\documentclass[fleqn,usenatbib]{mnras}

\usepackage[T1]{fontenc}
\usepackage{ae,aecompl}

\usepackage{graphicx}	% Including figure files
\usepackage{amsmath}	% Advanced maths commands
\usepackage{amssymb}	% Extra maths symbols

\title[The FP of ETGs at $z\sim1.3$]
{The Fundamental Plane of cluster spheroidal galaxies at z$\sim1.3$. Evidence for mass-dependent evolution.}

\author[P. Saracco et al.]{
P. Saracco,$^{1}$\thanks{E-mail: paolo.saracco@inaf.it}
A. Gargiulo,$^{2}$
F. La Barbera$^{3}$,
M. Annunziatella$^{4}$
D. Marchesini$^{4}$,
\\
\\
$^{1}$INAF - Osservatorio Astronomico di Brera, via Brera 28, 20121 Milano, Italy\\
$^{2}$INAF - Istituto di Astrofisica e Fisica Cosmica, IASF, via A. Corti
 12, 20133 Milano, Italy\\
$^{3}$INAF - Osservatorio Astronomico di Capodimonte, sal. Moiariello 16, 
80131 Napoli, Italy\\
$^{4}$Tufts University, Physics and Astronomy Department, 574 Boston Ave, Medford, 02155 MA, USA\\
}

\date{Accepted 2019 October 28. Received 2019 October 7; in original form 2019 May 13}

\pubyear{2019}

\begin{document}
\label{firstpage}
\pagerange{\pageref{firstpage}--\pageref{lastpage}}
\maketitle

\begin{abstract}
We present spectroscopic observations obtained at the {\it Large Binocular
Telescope} in the field of the cluster XLSSJ0223-0436 at $z=1.22$.
We confirm 12 spheroids cluster members and determine stellar velocity dispersion 
for 7 of them.
We combine these data with those in the literature for clusters 
RXJ0848+4453 at $z=1.27$ (8 galaxies) and XMMJ2235-2557 at $z=1.39$ 
(7 galaxies) to determine the Fundamental Plane of cluster spheroids.  
We find that the FP at $z\sim1.3$ is offset and 
{ rotated ($\sim3\sigma$)} with respect to the local FP.
The offset corresponds to a mean evolution  $\Delta$\rm{log}(M$_{dyn}$/L$_B$)=(-0.5$\pm$0.1)$z$.
High-redshift galaxies follow a steeper mass-dependent M$_{dyn}$/L$_B$-M$_{dyn}$ relation
than local ones.
Assuming $\Delta$ log$(M_{dyn}/L_B)$=$\Delta$ log$(M^*/L_B)$, 
higher-mass galaxies (log(M$_{dyn}$/M$_\odot$)$\geq$11.5) have a higher-formation 
redshift ($z_f\geq$6.5) than lower-mass ones ($z_f\leq$2 for 
log(M$_{dyn}$/M$_\odot$$\leq$10)), with a median $z_f\simeq2.5$ for the whole sample.
Also, galaxies with higher stellar mass density host stellar populations formed earlier 
than those in lower density galaxies.
At fixed IMF, M$_{dyn}$/M$^*$ varies systematically with mass and mass density.
It follows that the evolution of the stellar populations (M$^*/L_B$)
accounts for the observed evolution of M$_{dyn}/L_B$ 
for M$_{dyn}$$>10^{11}$ M$_\odot$ galaxies,  
while accounts for $\sim$85\% of the evolution at M$_{dyn}$$<10^{11}$ M$_\odot$.
We find no evidence in favour of structural evolution of individual galaxies, while we
find evidences that spheroids later added to the population
may account for the observed discrepancy between $\Delta$\rm{log}(M$_{dyn}/L_B)$ 
and $\Delta$ log$(M^*/L_B)$ at masses $<10^{11}$ M$_\odot$.
Thus, the evolution of the FP of cluster spheroids is consistent 
with the mass-dependent and mass density-dependent evolution of their stellar populations 
superimposed to a minor contribution of spheroids joining the population at later times.
\end{abstract}


%%%%%%%%%%%%%%%%%%%%%%%%%%%%%%%%%%%%%%%%%%%%%%%%%%

%%%%%%%%%%%%%%%%% BODY OF PAPER %%%%%%%%%%%%%%%%%%

\section{Introduction}
The process of stellar mass growth in galaxies
has led to lock more than 60\% of stars in the local Universe in spheroidal galaxies
\citep[e.g.][]{fukugita98}. 
The recent deep and wide multi-wavelength
surveys have shown that
massive (M$^*$$>$$10^{11}$~M$_{\odot}$) spheroids were already present
in the first 2 Gyr of cosmic time \cite[$z>$3; e.g.][]{glazebrook17}
and that they were characterized by 
an old stellar population when compared to the age of the 
Universe at that redshift \cite[e.g.][]{kriek16}. 
The mechanisms through which spheroids assembled and shaped their stellar mass, quenched their 
star formation and evolved along the cosmic time are still unclear and represent 
challenging problems in modern astrophysics. 

To reproduce the properties 
of spheroidal galaxies and to match the observed tight scaling relations,
simulations suggest that an early intense burst of star formation followed 
by a quick quenching are required
\citep{ciotti07,Naab07,oser12,porter14a,brooks16}.
Observations of local spheroids show that a minor fraction of spheroids
may accrete newly stellar mass through secondary events of 
star formation \cite[e.g.][]{thomas10}.
Therefore, studying spheroidal galaxies at high-redshift provides stringent
information of the main episode of their stellar mass growth. 

{ Environment seems to promote some mechanisms with respect to others, affecting
the morphological mix  of the galaxies but not their properties.}
For instance, the spheroidal galaxy fraction reaches
even 70-80\% in the core of some clusters, twice the one
observed in the field \citep[e.g.][]{dressler80,holden07,vanderwel07}.
There seems to be a lack of massive (M$^*$$>$2$\times$$10^{11}$~M$_{\odot}$)
and hence large (R$_e$$>$4-5 kpc) spheroids in the field with respect 
to cluster \citep[e.g.][]{saracco17}, even if they do not
differ from each other
at fixed stellar mass \citep[e.g.][]{huertas13b,newman14,kelkar15,saracco17}.
{ Thus, observations of cluster environment represents an efficient way to 
collect representative samples of spheroidal galaxies to study.}

Some studies based on the Fundamental Plane (FP) of spheroidal galaxies at intermediate redshift
(up to $z\sim0.9$) show that their evolution is consistent with a simple aging 
of their stellar populations \citep[e.g.][]{jorgensen06,vandokkum07}.
%both for field and cluster galaxies.
Other studies suggest that a structural evolution of individual galaxies
(a variation of R$_e$ and $\sigma_e$) besides their simple aging is required
to account for the FP evolution \citep[e.g.][]{saglia10,saglia16,beifiori17} even if progenitor
bias could affect significantly this interpretation \citep{prichard17}.
Moreover, while some authors find a steepening of the FP slope with redshift 
\citep{jorgensen06,jorgensen13,diserego05}, other authors do not detect any variation
\citep[e.g.][]{holden10}.

A steeper FP at higher redshift could be explained in terms of  a mass-dependent
evolution of galaxies, with higher-mass galaxies forming their stars at higher 
redshifts than lower-mass ones.
Also other properties may cause a differential evolution of 
galaxies and hence of the FP.
For instance,  
stellar metallicity and, at a less extent, age are found to correlate with 
velocity dispersion and dynamical mass in local  
\citep[e.g.][]{jorgensen99,trager00,thomas05,thomas10,gallazzi05,gallazzi14,harrison11,mcdermid15}  
and high-redshift spheroids \citep[e.g.][]{jorgensen17,saracco19}.
These correlations imply a mass dependent evolution of the luminosity of galaxies and, 
consequently, a variation of the FP slope.
Stellar metallicity is also correlated with the stellar mass density 
\citep[e.g.][]{saracco19}, 
and stellar population properties 
are found to correlate with some structural 
properties of galaxies \citep[e.g.][]{zahid17,vandesande18}.

Whether the FP slope changes along cosmic time is still a matter of debate.
Analogously, whether and how the above relationships affect the FP and its evolution
is not well understood.
In this paper we try to address these issues starting from the determination
of the FP of cluster spheroids at $z\sim1.3$ and establishing its evolution.
To this end, we have obtained new spectroscopic observations of 
candidate members of the cluster XLSSJ0223-0436 at $z\sim1.22$ deriving
stellar velocity dispersion for 7 spheroids cluster members.
Combining these data with those in the literature for the clusters 
RXJ0848+4453 at $z=1.27$ \citep[8 galaxies;][]{jorgensen14} 
and XMMJ2235-2557 at $z=1.39$ \citep[7 galaxies;][]{beifiori17}, 
{ we constructed an homogeneous and unbiased sample of 22 cluster spheroids 
to determine, for the first time, the FP at $\sim1.3$ and its evolution both in 
term of offset and slope over the last 9 Gyr}.

The paper is organized as follows.
Section 2 describes the observations, the data reduction and the redshift measurement.
Section 3 describes the velocity dispersion measurement and the dynamical mass derivation.
Section 4, describes the high-redshift and the local reference samples used to
derive the FP at $\sim1.3$ and in the local Universe.
In Section 5 the evolution of the M$_{dyn}$/L$_B$ is derived and used to constrain 
the formation redshift of cluster spheroids.
Section 6 discusses the evolution that spheroids likely experience since $z\sim1.3$ that
account for the FP evolution derived in the previous sections.
Section 7 presents a summary of results and the conclusions.

Throughout this paper we use a cosmology with
$H_0=70$ Km s$^{-1}$ Mpc$^{-1}$, $\Omega_m=0.3$, and $\Omega_\Lambda=0.7$
and assume a Chabrier \citep{chabrier03} initial stellar mass function (IMF).
All the magnitudes are in the Vega system, unless otherwise specified.

\section{Observations, data reduction and redshift measurement}
\label{sec:observations}
 % Target sample
\begin{table*}
\begin{minipage}[t]{1\textwidth}
\caption{\label{tab:sample}List of observed galaxies in the XLSSJ0223 field and
spectroscopic redshift measurements.}
\centerline{
\begin{tabular}{rccccccc}
\hline
\hline
  ID &    RA      &        Dec   &  F850LP & $i_{775}-z_{850}$&   Flag$^a$& $z^b_{spec}$ & Em$^c$  \\
     &   [h:m:s]& [d:p:s]&   [mag] & [mag] & & &\\
\hline
 4&      02:23:09.381&  -04:35:17.91&  21.63$\pm$0.01  & 0.89$\pm$0.02 & 0  & 0.8992&	  ...	  \\
 377&    02:23:09.428&  -04:37:07.82&  22.78$\pm$0.03  & 1.06$\pm$0.04 & 0  & 1.2154&  [OII]     \\  
 474&    02:23:07.909&  -04:36:45.16&  23.32$\pm$0.03  & 0.86$\pm$0.04 & -1  & (0.9005)&	   ...     \\  
 537&    02:23:08.485&  -04:37:18.07&  22.57$\pm$0.02  & 0.99$\pm$0.03 & 1  & 1.3153&  [OII]	   \\  
 651&    02:23:05.759&  -04:36:10.27&  21.62$\pm$0.01  & 1.09$\pm$0.02 & 1  & 1.2192&  [NeV]	   \\  
 962&    02:23:05.420&  -04:36:36.26&  22.23$\pm$0.02  & 1.02$\pm$0.03 & 1  & (1.2231)&  ...	   \\  
 972&    02:23:04.718&  -04:36:13.47&  22.61$\pm$0.03  & 0.92$\pm$0.03 & 1  & 1.2153&  ...	   \\  
 983&    02:23:04.843&  -04:36:19.87&  22.85$\pm$0.02  & 0.94$\pm$0.03 & 1  & 1.2149&  ...	   \\  
1076&    02:23:02.851&  -04:35:40.02&  21.52$\pm$0.02  & 1.06$\pm$0.03 & 0  & 1.0255&	[OII]   \\  
1117&    02:23:07.938&  -04:34:46.03&  24.09$\pm$0.05  & 0.70$\pm$0.07 & -1  & 0.5286&  [OII]  \\
1142&    02:23:03.262&  -04:36:14.60&  21.30$\pm$0.01  & 1.00$\pm$0.02 & 1  & 1.2204&  ...	   \\  
1147&    02:23:03.352&  -04:36:15.64&  22.56$\pm$0.02  & 0.82$\pm$0.03 & -1  & 0.7615&   Em	   \\
1175&    02:23:03.242&  -04:36:18.50&  21.90$\pm$0.02  & 1.08$\pm$0.02 & 1  & 1.2201&  {[OII]}	   \\  
1193&    02:23:02.799&  -04:36:05.92&  23.01$\pm$0.03  & 1.07$\pm$0.05 & 0  & 1.3890&  ...     \\  
1302&    02:23:00.882&  -04:35:39.85&  23.29$\pm$0.03  & 1.00$\pm$0.04 & 1  & ...   &  ...	   \\  
1317&    02:23:02.042&  -04:36:32.17&  21.37$\pm$0.01  & 0.96$\pm$0.02 & -1  & 0.7330&	   ...   \\
1320&    02:22:59.820&  -04:35:12.25&  23.52$\pm$0.05  & 0.82$\pm$0.07 & 0  & (1.3342)&     ...     \\  
1370&    02:23:02.021&  -04:36:43.26&  23.10$\pm$0.03  & 0.96$\pm$0.04 & 1  & 1.2249&  ...	   \\  
1442&    02:22:57.980 & -04:36:22.31&  21.88$\pm$0.01  & 0.91$\pm$0.02 & 0  & 1.2250&  {[OII]}     \\  
1448&    02:22:58.869&  -04:36:49.89&  23.94$\pm$0.05  & 1.01$\pm$0.07 & 1  & 1.2080&  [OII]	   \\  
1630&    02:23:00.929&  -04:36:50.19&  21.48$\pm$0.01  & 1.04$\pm$0.02 & 1  & 1.2109&  [OII]	   \\  
1711&    02:22:59.990&  -04:36:02.53&  20.92$\pm$0.01  & 1.02$\pm$0.02 & 1  & 1.2097&	...	   \\  
\hline
\end{tabular}
}
{$^a$ Target priority: 1 - the target belongs to the sample of spheroids cluster 
member candidates  \citep{saracco17}; 
0 - the target is a spheroidal galaxy with photometric redshift compatible
with the redshift of the cluster; -1 - the target is a filler galaxy best matching
the remaining slit positions on the mask.\\
$^b$ Bracketed values are uncertain redshift, based only on the fit of the overall 
shape of the continuum; dots indicate undetermined redshift.\\
$^c$ Emission lines detected: dots - no emission line; Em - more than one emission 
line.}
\end{minipage}
\end{table*}
\subsection{Spectroscopic observations and data reduction}
Spectroscopic observations in the field of the cluster XLSSJ0223-0436 at $z=1.22$ 
were performed  with the Multi-Object Double Spectrograph (MODS, 1 and 2) 
\citep{pogge10}) mounted at the Large Binocular Telescope (LBT).
%Observations have been carried out  with 
Filter GG495 coupled with the grim G670L sampling the wavelength range 
0.5$\mu$m$<\lambda<1.0\mu$m  at 0.85 \AA/pix and a slit width of 1.2'' 
were adopted.
The resulting spectral resolution is R$\simeq 1150$ (FWHM$\simeq$7.4 \AA).
The 22 targeted galaxies within 1 arcmin radius from the cluster center, 
were arranged into two masks and are listed in Tab. \ref{tab:sample}.
The selection of the parent sample of spheroids cluster member candidates 
from which the main targets were selected is described in \cite{saracco17}
and \cite{saracco19}.

{ Observations for each mask were collected with MODS1 and MODS2
cameras from October 2015 to January 2016.
The seeing was in the range
FWHM=0.7-1.1 arcsec.
Observations consisted in a sequence of exposures of 900 sec each 
taken at dithered (ABBA) positions offset by $\sim5$".
A total of 32 and 35 exposures were collected for the two
masks respectively, summing up to an effective integration time of about 
8 hours.}
The same bright star was included in the two masks in order to accurately
measure the offsets in the observing sequence and the
correction for telluric absorption lines.
A detailed description of the observations and of the data reduction
is given in \cite{saracco19}.

The spectra thus obtained were used to derive redshifts and, for
the 7 main targets having sufficient S/N, velocity dispersion.

\subsection{Redshift measurement}
Redshift and velocity dispersion measurements were performed
by fitting the observed spectra with MILES simple stellar population (SSP)
models \citep{vazdekis10} having a spectral resolution of 2.5\AA\ 
\cite[][]{beifiori11}, higher than the rest-frame resolution
($\sim$3.3\AA) of our spectra.
The spectral fitting was performed using the penalized
PiXel-Fitting method (pPXF, \cite{cappellari04,cappellari17}).
\begin{figure*}
	\includegraphics[width=10truecm]{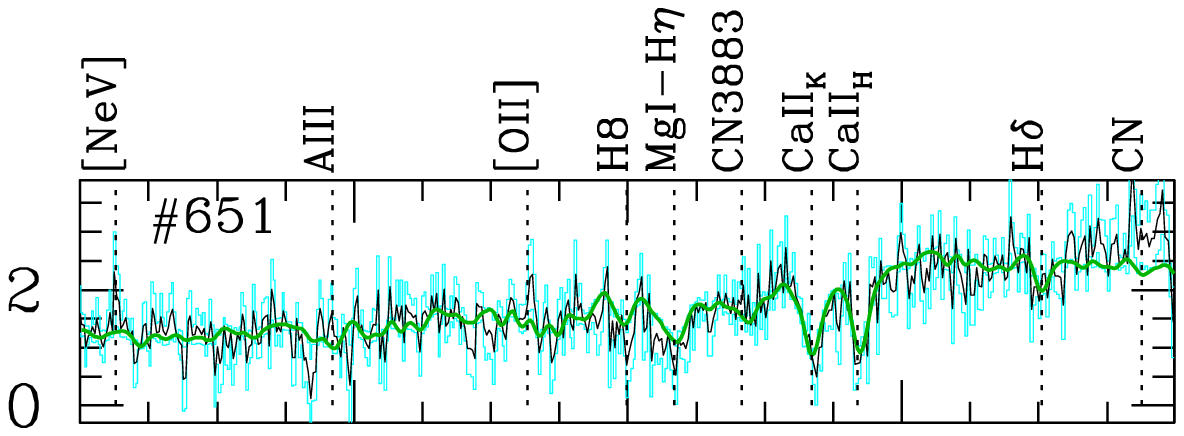}
  	\includegraphics[width=2truecm]{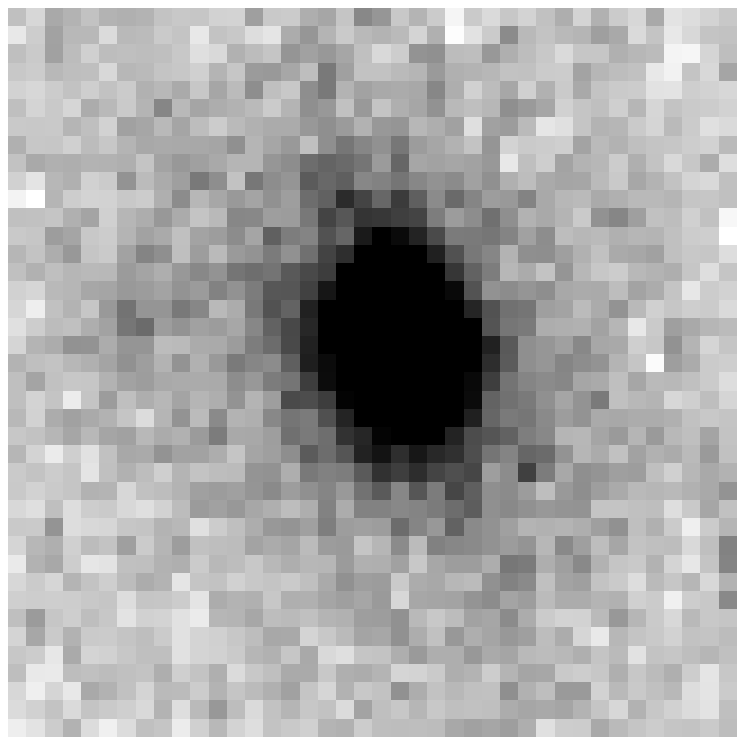}
	\includegraphics[width=10truecm]{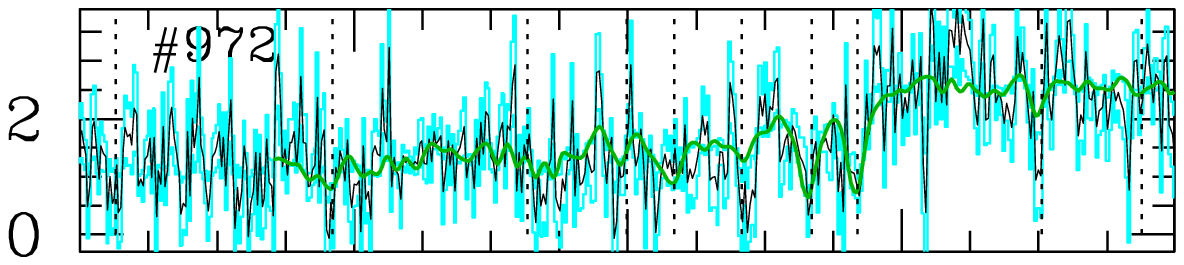}
	\includegraphics[width=2truecm]{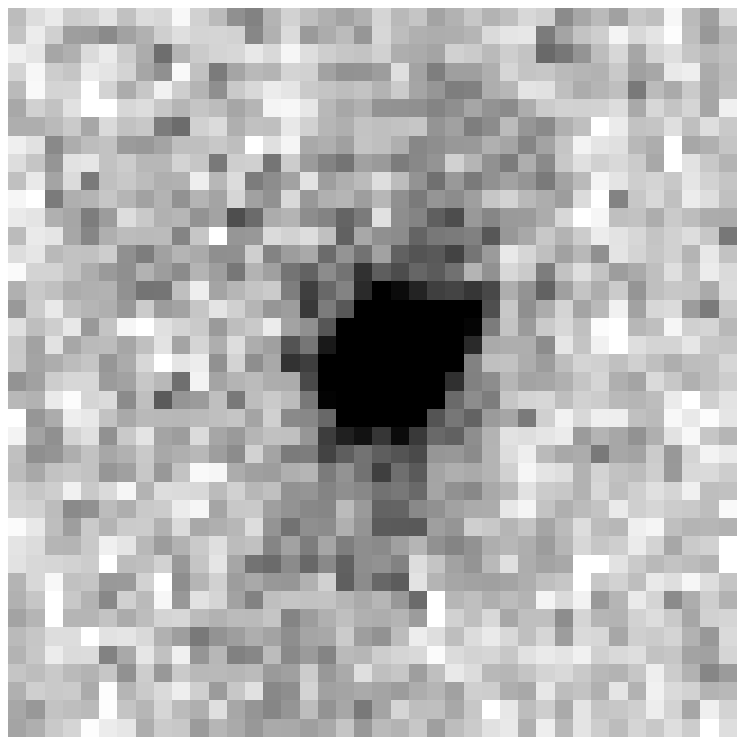}
	\includegraphics[width=10truecm]{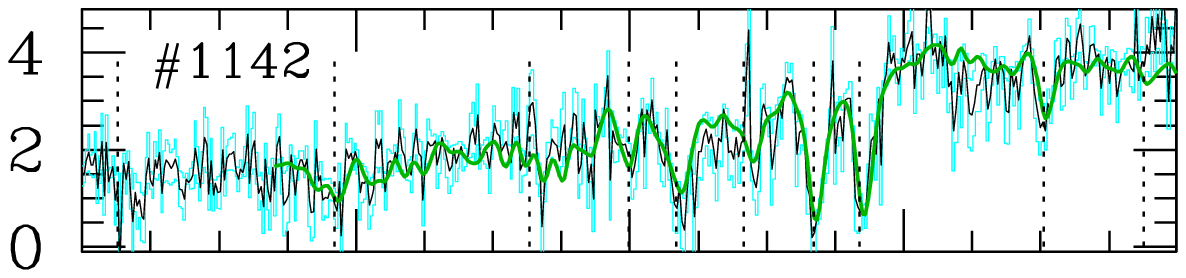}
	\includegraphics[width=2truecm]{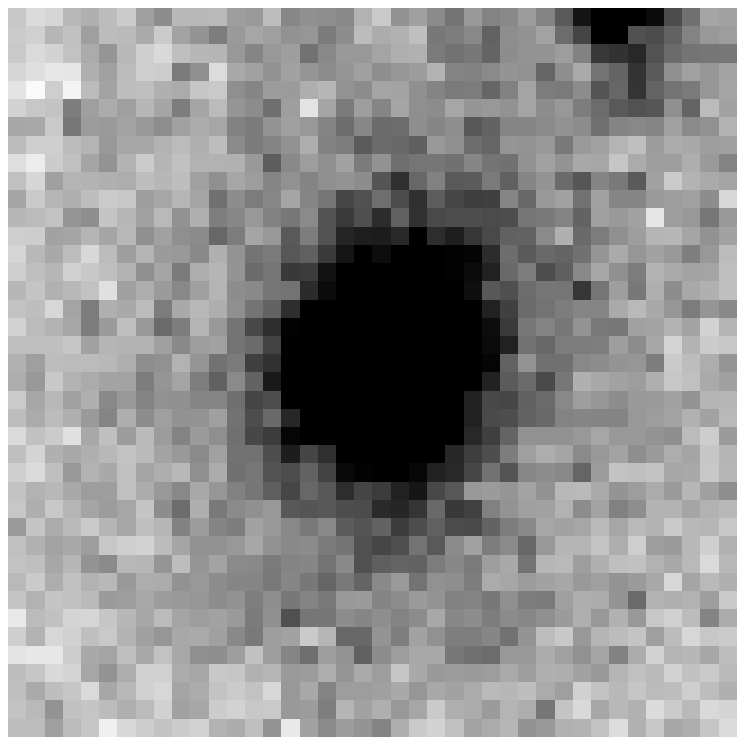}
	\includegraphics[width=10truecm]{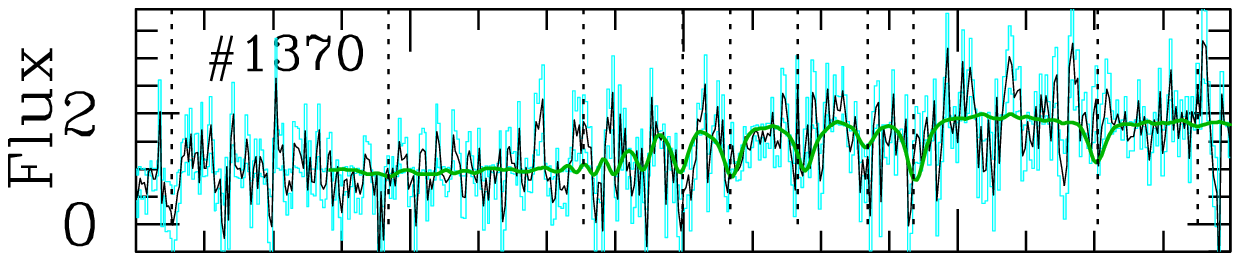}
	\includegraphics[width=2truecm]{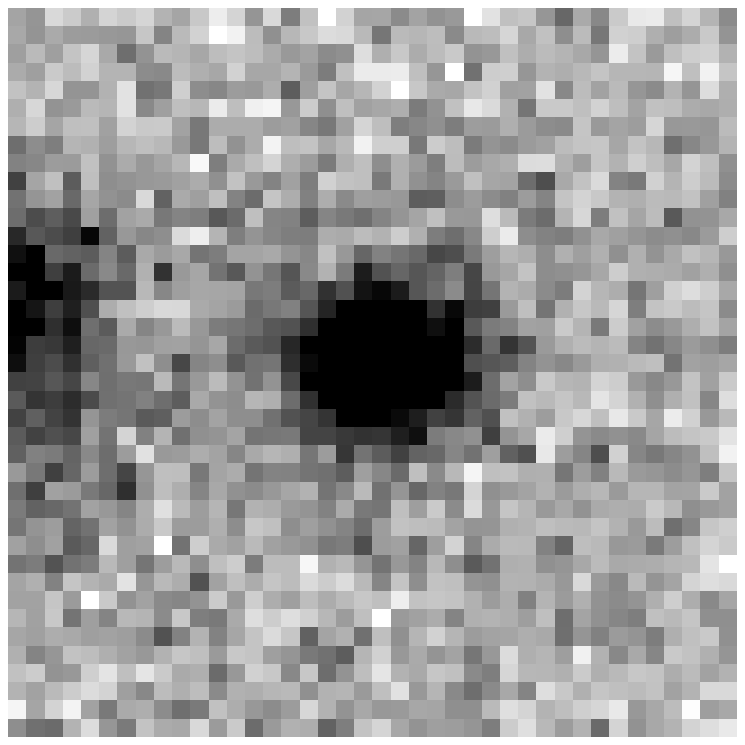}
	\includegraphics[width=10truecm]{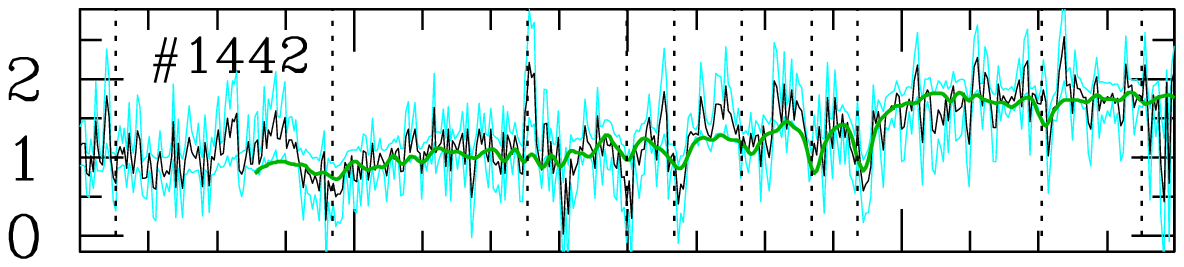}
	\includegraphics[width=2truecm]{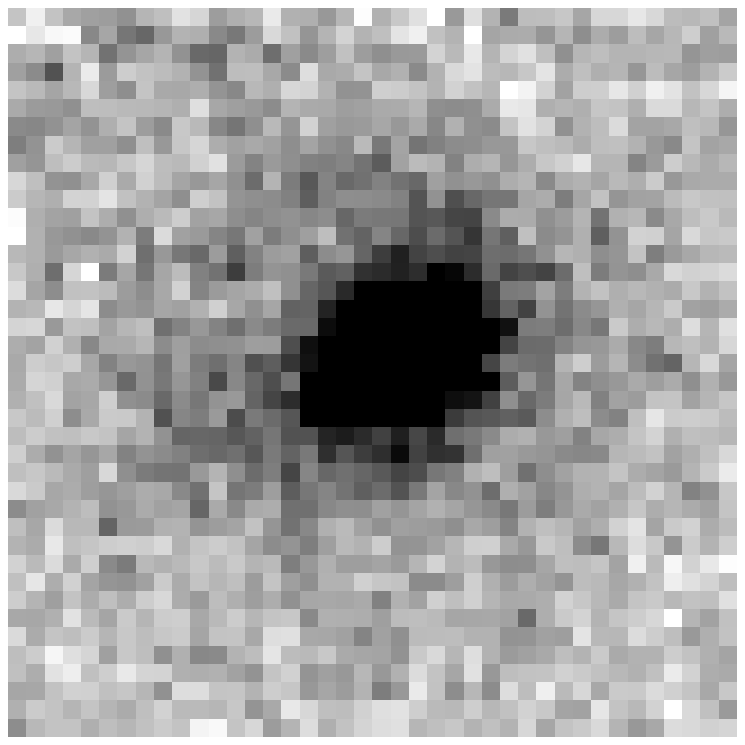}
	\includegraphics[width=10truecm]{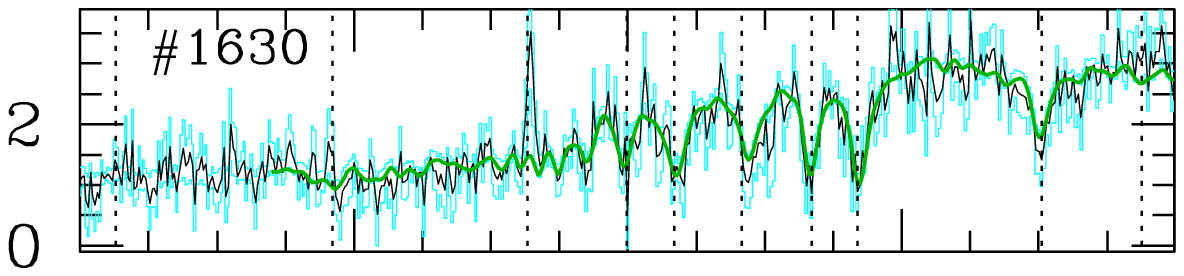}
	\includegraphics[width=2truecm]{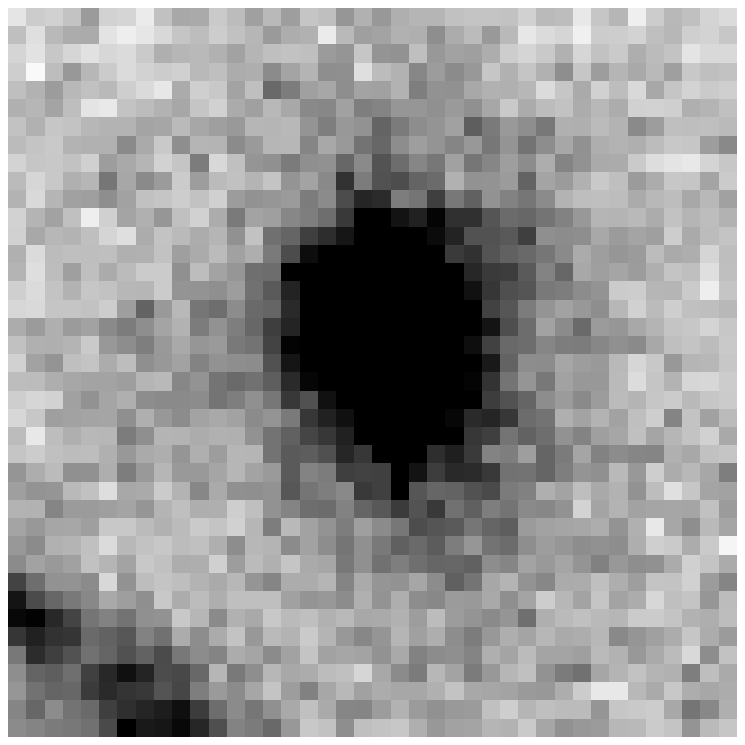}
	\includegraphics[width=10truecm]{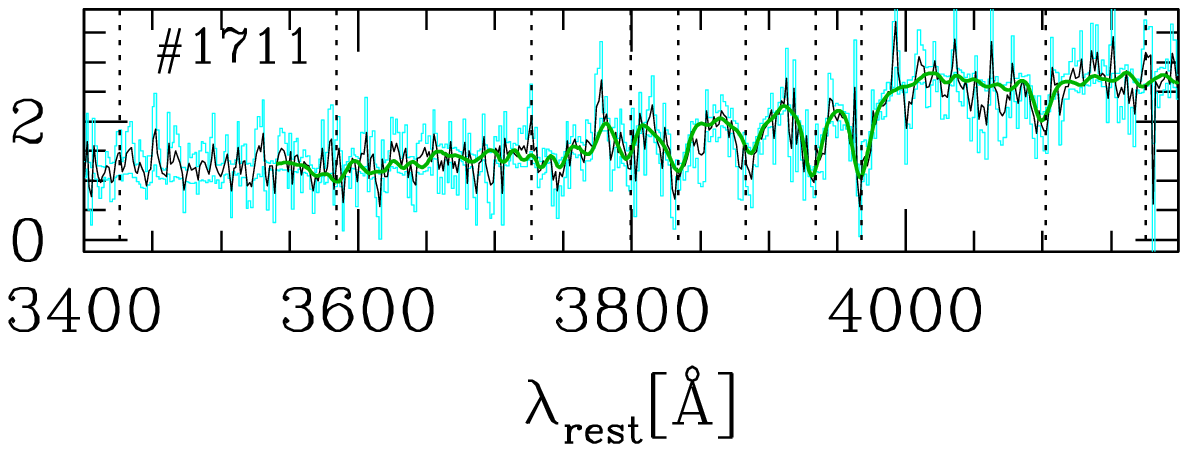}
	\includegraphics[width=2truecm]{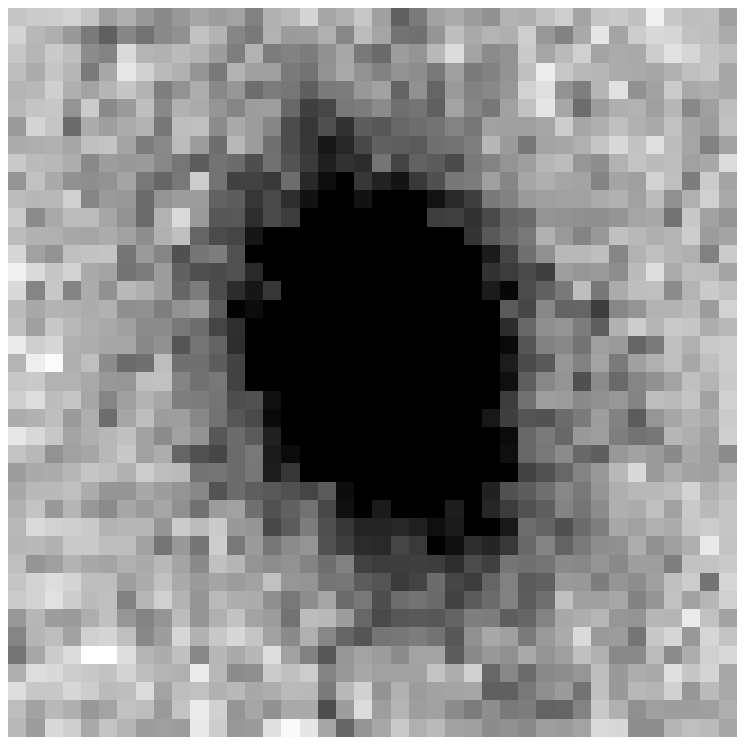}
   \caption{LBT-MODS spectra of the seven spheroidal galaxies cluster members for which
 velocity dispersion were measured. 
 These spectra have S/N=[6-18] per \AA\ in the galaxy rest-frame, in the range 
   3900$<\lambda_{rest}<$4100\AA. 
The black curve is the observed spectrum binned to 3.4 \AA/pixel as normalized by pPXF
(flux is in arbitrary units).
The green curve is the best-fitting MILES model resulting from the pPXF spectral fitting. 
{ Cyan histogram is the error ($\pm\sigma$) scaled by the same normalization factor of the spectrum 
to preserve the original S/N.} 
   The dotted vertical lines mark the main spectral features labeled on top
   of the figure. To the right of each spectrum, the ACS-F850LP image
   (2$\times$2 arcsec) of the galaxy is shown.}
   \label{fig:7spectra}
\end{figure*}
\renewcommand{\thefigure}{\arabic{figure} (Cont.)}
\addtocounter{figure}{-1}

\begin{figure*}
	\includegraphics[width=11truecm]{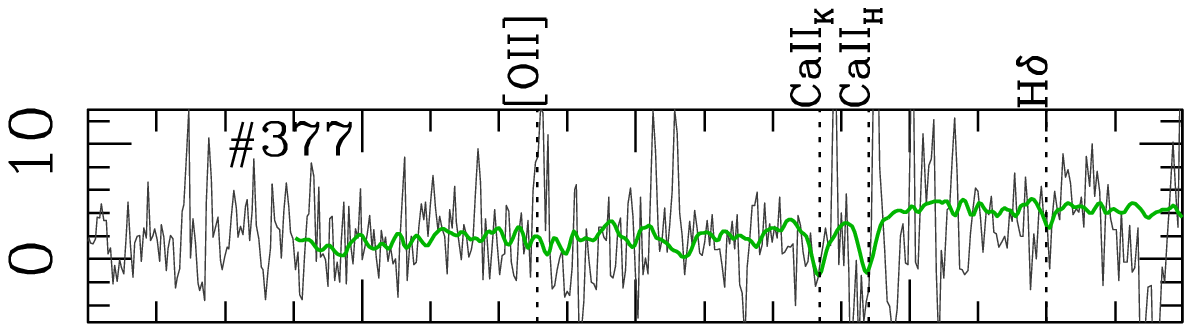}
	\includegraphics[width=2.truecm]{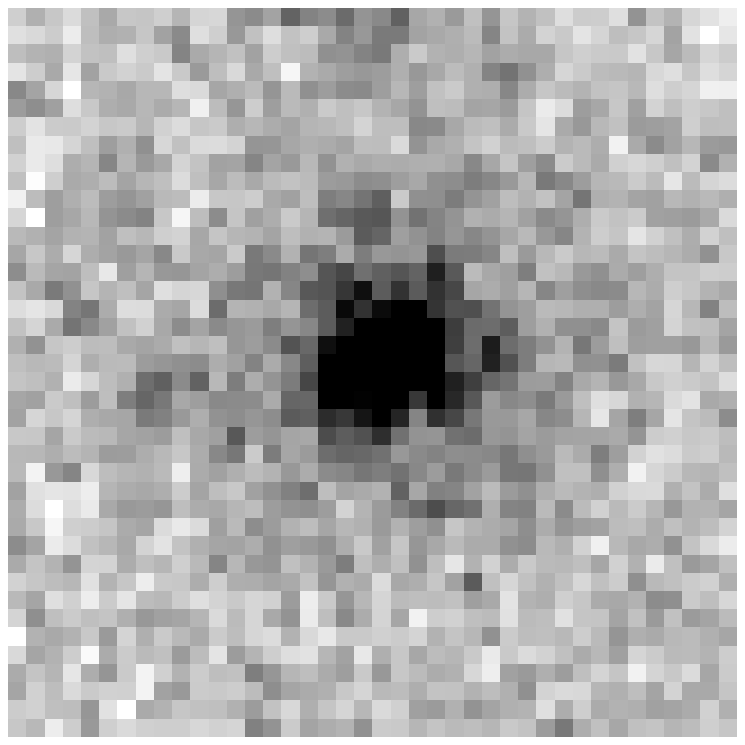}	
        \includegraphics[width=11truecm]{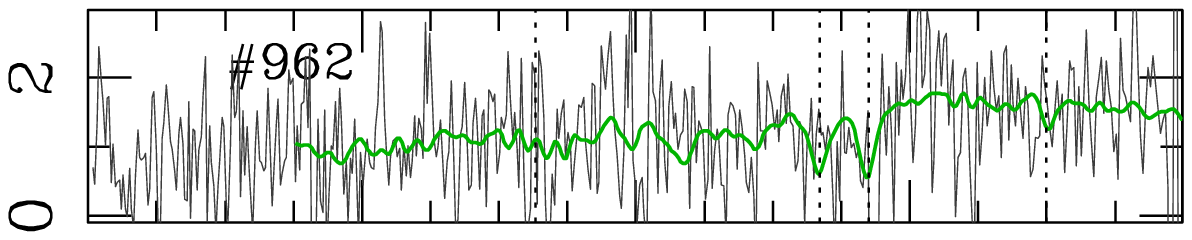}	
	\includegraphics[width=2.truecm]{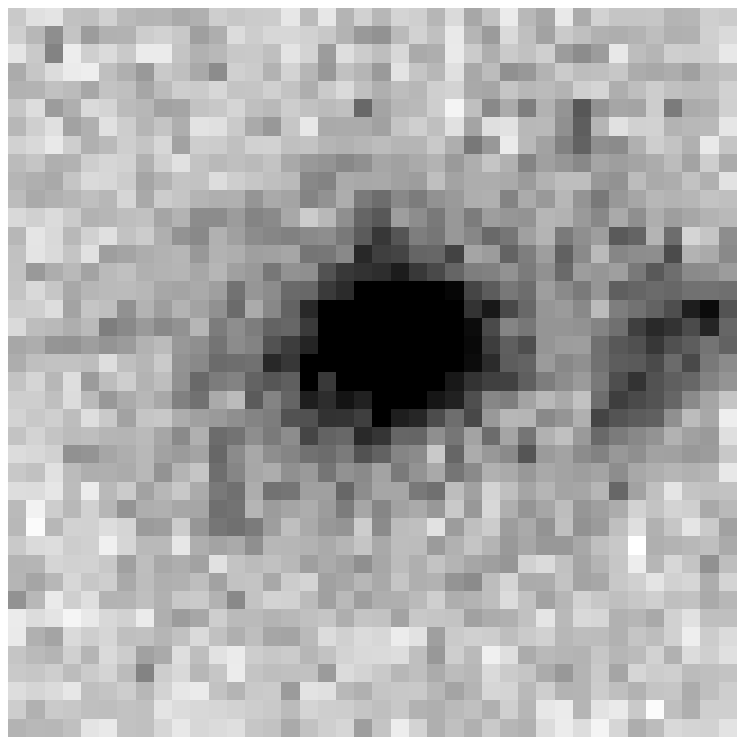}	
        \includegraphics[width=11truecm]{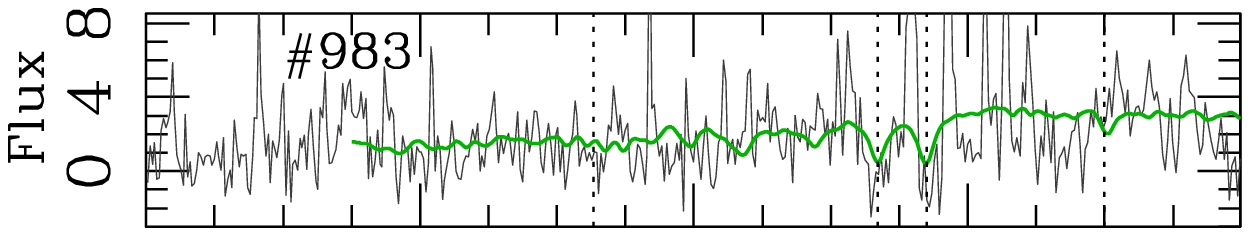}	
	\includegraphics[width=2.truecm]{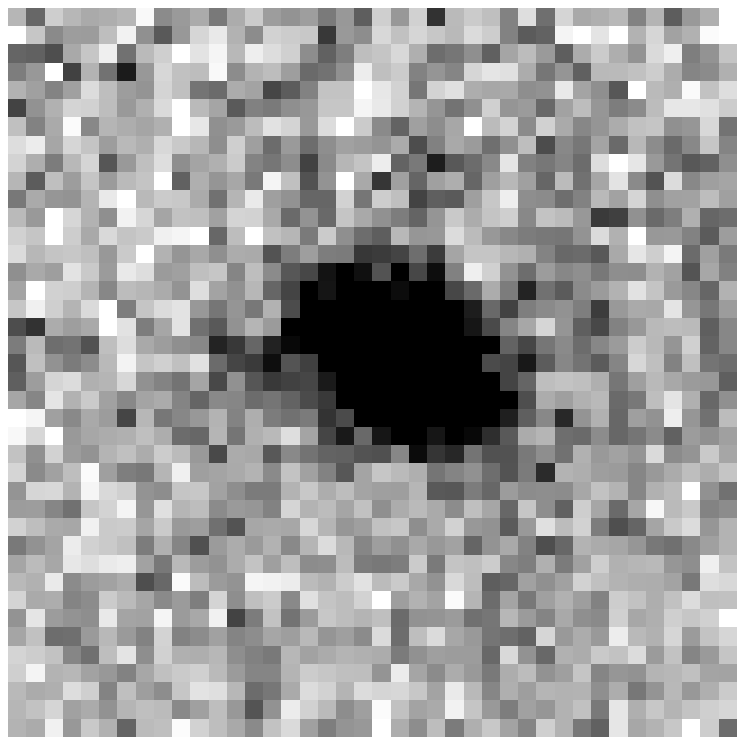}	
        \includegraphics[width=11truecm]{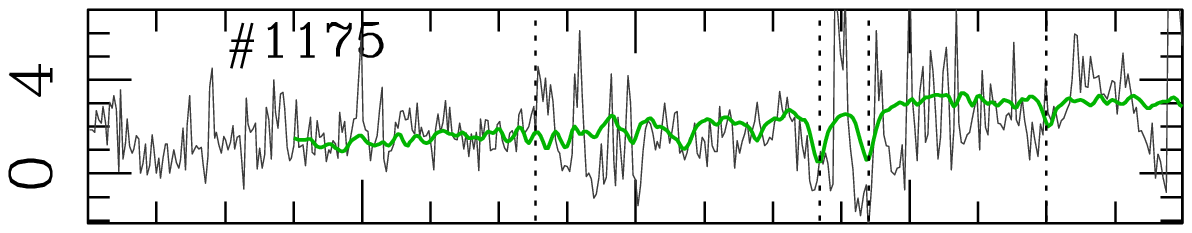}	
	\includegraphics[width=2.truecm]{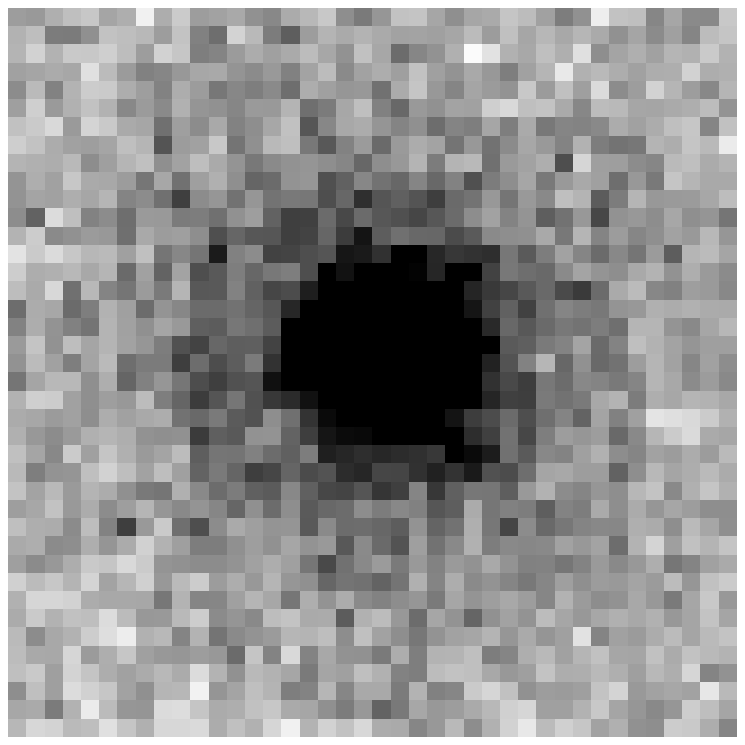}
        \includegraphics[width=11truecm]{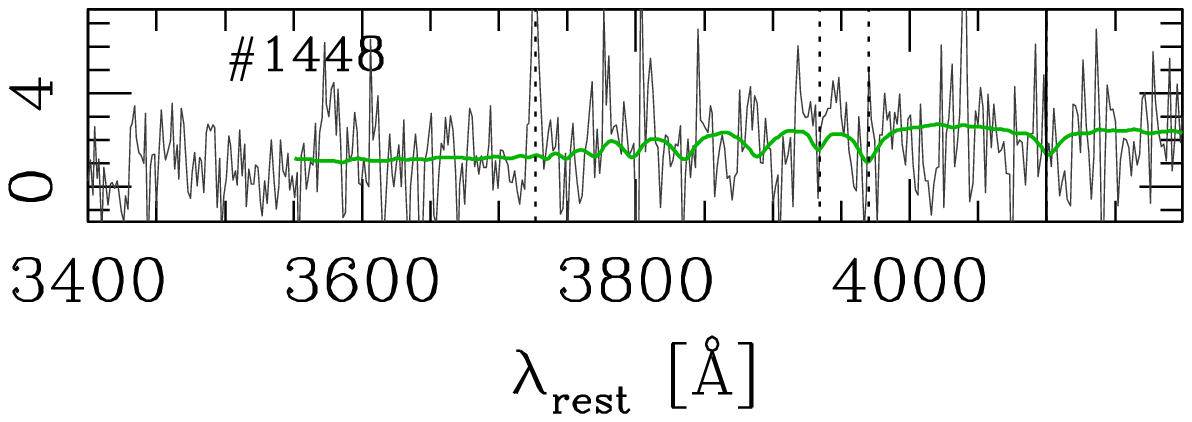}	
	\includegraphics[width=2.truecm]{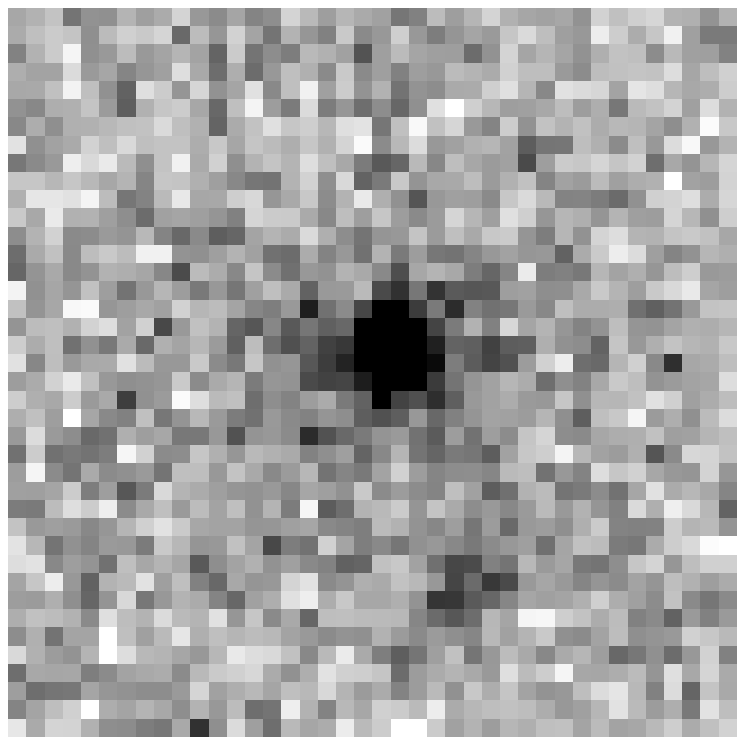}	
   \caption{\label{fig:6spectra} MODS spectra for the remaining six spheroidal cluster member galaxies 
   for which only spectroscopic redshift was obtained. }
\end{figure*}
\renewcommand{\thefigure}{\arabic{figure}}

We derived redshifts for 21 out of the 22 target galaxies. 
In Tab. \ref{tab:sample} we report the values of the estimated redshifts.
We considered secure the redshifts for those galaxies (18) clearly showing 
one or more spectral (absorption and/or emission) features and/or 
the 4000 \AA\ break, while we considered insecure the redshifts 
derived only from the fitting to the overall shape of the continuum
with no clear features. 
In these latter cases, (galaxies \#474, \#962 and \#1320) the values are shown
in brackets in  Tab. \ref{tab:sample}.
Out of the 21 galaxies with redshift measurements, 12 spheroidal galaxies
are confirmed members of the cluster.
For seven of them,  the S/N measured in the 
wavelength range 3900-4100\AA, turned out to be sufficient (S/N$\sim$3-8 per \AA, 
6-18 in the rest-frame) to derive stellar velocity dispersion. 
Fig. \ref{fig:7spectra} shows the spectra of the 7 spheroidal galaxies with higher 
S/N, while Fig. \ref{fig:6spectra} shows the spectra of the remaining 5 spheroidal 
galaxies cluster members. 
In the figures, the black curve is the 8-hours MODS spectrum binned
to 3.4 \AA/pix, the green curve is the best fitting template.
For each galaxy the $2\times2$ arcsec ACS-F850LP image is also shown.

\section{Velocity dispersion and dynamical mass}

\begin{table*}
\caption{\label{tab:sigmas} Structural and kinematic parameters of galaxies.
Column 1: ID; Column 2: stellar velocity dispersion; Column 3: S\'ersic index;
Column 4: { circularized} effective radius for S\'ersic profile; Columns 5 and 6: surface brightness and 
velocity dispersion within the S\'ersic effective radius;
Columns 7,8 and 9: same as columns 4,5 and 6 but for de Vaucouleurs profile;
Column 10: dynamical mass (eq. \ref{eq:mdyn}); Column11: stellar mass
\citep[from][]{saracco17}. { It is worth noting that $\sigma_{e/8}=1.132\times\sigma_e$,
according to eq. \ref{eq:sigscale}. The extraction radius slightly varies from
galaxy to galaxy in the range 0.9\arcsec-1.1\arcsec.}
}
\centerline{
\begin{tabular}{rcccccccccc}
\hline
\hline
  ID &$\sigma_*$&$n_{Ser}$& R$_e^{F850}$ & log$\langle I_e\rangle$ & $\sigma_e$&R$_{e,Dev}$&log$\langle I_{e,Dev}\rangle$&$\sigma_{e,Dev}$&log$\mathcal{M}_{dyn}$&
  log$\mathcal{M}_*$   \\
     &[km/s]       &  &[kpc] &[L$_\odot$ pc$^{-2}$]  & [km/s]& [kpc]	  &[L$_\odot$ pc$^{-2}$]& [km/s]  &[M$_\odot$] & [M$_\odot$] \\
\hline
 651&   215$\pm28$   &     2.9$\pm0.3$&      3.5$\pm$0.2 & 3.04$\pm$0.02&   220 & 5.1$\pm$0.2 & 2.75$\pm$0.02& 215 &  11.29$\pm$0.1 &10.94  \\  
 972&   207$\pm47$   &     6.5$\pm0.6$&      2.4$\pm$1.5 & 3.02$\pm$0.20&   216 & 1.4$\pm$0.1 & 3.79$\pm$0.10& 224 &  11.11$\pm$0.1 &10.54  \\  
1142&   221$\pm38$   &     5.4$\pm0.3$&      9.1$\pm$1.1 & 2.40$\pm$0.06&   213 & 5.1$\pm$0.2 & 3.33$\pm$0.06& 221 &  11.68$\pm$0.1 &11.50  \\  
1370&   102$\pm90$   &     3.8$\pm0.5$&      0.8$\pm$0.2 & 3.72$\pm$0.10&   115 & 1.1$\pm$0.2 & 3.80$\pm$0.10& 112 &  10.10$\pm$0.5 &10.08  \\  
1442&   178$\pm69$   &     6.5$\pm0.7$&      2.5$\pm$0.4 & 3.26$\pm$0.10&   210 & 1.4$\pm$0.1 & 3.76$\pm$0.06& 222 &  11.11$\pm$0.4 &10.90   \\  
1630&   193$\pm50$   &     5.3$\pm0.3$&      3.3$\pm$0.3 & 3.12$\pm$0.04&   198 & 2.3$\pm$0.1 & 3.74$\pm$0.04& 202 &  11.18$\pm$0.1 &10.75  \\  
1711&   247$\pm23$   &     6.0$\pm0.2$&      4.7$\pm$0.3 & 3.08$\pm$0.04&   247 & 2.4$\pm$0.1 & 3.94$\pm$0.04& 258 &  11.52$\pm$0.2 &11.20  \\  
\hline
\end{tabular}
}
\end{table*}
The galaxy stellar velocity dispersion $\sigma_*$ is given by the relation
$\sigma^2_*=\sigma_{obs}^2-\sigma_{inst}^2$
where $\sigma_{obs}$ is the velocity dispersion measured on the observed 
spectrum and $\sigma_{inst}=111$ km/s is the instrumental broadening 
resulting from the instrumental resolution 
(R$\simeq$1150). 
The uncertainties on $\sigma_*$ were derived 
by repeating the measurement for a set of 100 spectra obtained by
summing to the best-fitting template, the real 1D residual background extracted 
from the final 2D spectra, randomly shuffled in the wavelength range
considered for measurement.
We adopted, as uncertainty, the median absolute deviation 
(MAD) resulting from the distribution of the measurements.

The independence of the velocity dispersion measurements of the 
template library used, was tested by repeating the fitting to the observed 
spectra with the set of synthetic stars used by \cite{gargiulo16}, selected 
from the high-resolution (R$\sim20000$) spectral library of \cite{munari05}
and with a set of SSPs of \cite{bruzual03}.
These additional sets of templates returned values always within 8\% from
the reference values with no systematic.  
The robustness of the measurements against the wavelength range considered,
was tested by repeating
the fitting considering different wavelength ranges of the spectrum.
In all the cases considered, we obtained measurements well within the one sigma error.
In Tab. \ref{tab:sigmas} we report the values of the velocity dispersion $\sigma_*$ 
 and their errors.
Since the measurements refer to 1.2\arcsec\ slit width and spectra were extracted
within 0.9\arcsec-1.1\arcsec, we scaled measurements 
to a radius  $r=0.6$\arcsec.
In Tab. \ref{tab:sigmas}, we report also the velocity dispersion $\sigma_e$ within the 
effective radius R$_e$ obtained following the relation 
\begin{equation}
\label{eq:sigscale}
 \sigma_*/\sigma_e=(r/R_e)^{-0.065}
\end{equation}
\cite[][see also \cite{jorgensen95}]{cappellari06}.

%\subsection{Dynamical masses}

The dynamical masses $\mathcal{M}_{dyn}$ of the galaxies have been derived from 
the velocity dispersions $\sigma_e$ and the effective radii R$_e$ according to the
relation
\begin{equation}
\label{eq:mdyn}
 \mathcal{M}_{dyn}=k_n {\sigma^2_e R_e \over G}.
\end{equation}
with $k_n=5$.
G is the gravitational constant and $k_n$ is the virial coefficient which takes
into account the distribution of both luminous and
dark matter (DM) and the projection effects \cite[see e.g.][]{bertin02, lanzoni03}.

\cite{cappellari06}, using a sample of local spheroidal galaxies, show that 
a value $k_n\simeq5$ provides, on average, the best approximation
to the real value at $z\sim0$.
Therefore, also for homogeneity with other data in the literature, we adopted
this value of $k_n$ in eq. \ref{eq:mdyn}.

\section{Fundamental Plane determination}
In the three dimensional logarithmic space defined by the effective radius R$_e$
[kpc], the central stellar velocity dispersion $\sigma_0$ [km/s] and the surface 
brightness <I$_e$> [L$_\odot$ pc$^{-2}$], usually evaluated { within} the half light radius,
elliptical galaxies 
%are not homogeneously distributed.
%They 
are observed to be arranged along a plane known as Fundamental Plane 
\cite[FP, e.g.][]{djorgovski87,dressler87,jorgensen96,pahre98}.
The FP is usually expressed in the form:
\begin{equation}
\label{eq:fp}
{\rm log R}_e = \alpha~ {\rm log} \sigma_0 + \beta~ {\rm log \langle I\rangle}_e + \gamma
\end{equation}
where $\alpha$ and $\beta$ are the slopes and $\gamma$ the offset.
The central velocity dispersion $\sigma_0$ is often corrected to an aperture of R$_e$/8.
The coefficients of the FP only weakly depend on wavelength
\citep[e.g.][]{bernardi03, labarbera10a},
%On the contrary, the values of the coefficients 
while they can significantly differ according
to the fitting procedure adopted, as discussed by, e.g.,
\cite{saglia01,bernardi03, labarbera10a, sheth12, cappellari13}.

In this section, we compare the FP of cluster spheroidal galaxies at $z\sim1.3$ with 
the one at $z\sim0$ to constrain the evolution subtended in the last 9 Gyr.
To perform a reliable comparison, we use the same fitting procedure for
the high-redshift and the local reference samples.
\begin{figure}	
	\includegraphics[width=8truecm]{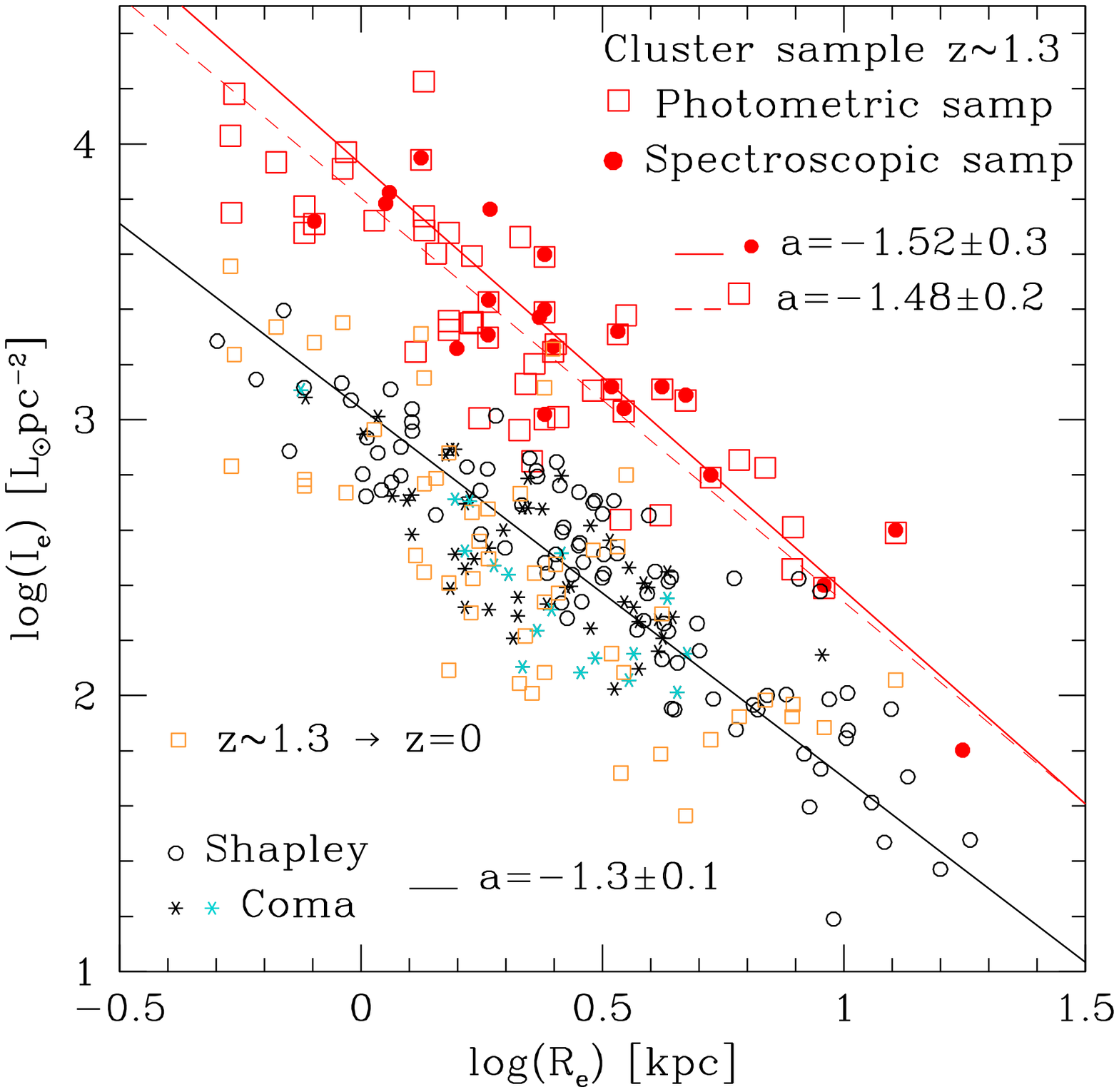}
    \caption{ \label{fig:kor} Kormendy relation of spheroidal galaxies in cluster.  
     Skeletal symbols are the Coma cluster galaxies  { (cyan symbols are galaxies younger than 7 Gyr,
     see text)}
     \citep{jorgensen95,jorgensen95a};
     open circles are the galaxies in the Shapley sample \citep{gargiulo09}; 
     filled red symbols are the 22 spheroids of the cluster sample 
at $z\sim1.3$
    (\S 4.1); large red open squares are the photometric sample of 55 spheroids 
   cluster member candidates at $z\sim1.3$ \citep{saracco17}, to which the 16 
    spectroscopic members belong to; small open orange squares are the spheroids
  passively evolved to $z=0$.
The best fitting relation 
    $\rm{log}(I_e)=a\times \rm{log}(R_e) + b$ obtained through orthogonal fit is shown.
}
\end{figure}

\subsection{High-redshift sample}
We combined the data of the 
cluster XLSSJ0223, with those collected by 
\cite{jorgensen14} for the cluster RXJ0848 (LinxW) at $z=1.27$ and the velocity dispersion 
measurements obtained 
by \cite{beifiori17} for a sample of galaxies in the cluster XMMJ2235 at $z=1.39$.
%A detailed description of the selection of galaxies in these two clusters 
%and of their parameters is given in appendix A.
For the cluster RXJ0848, following \cite{jorgensen14}, we considered 
%in the analysis of the FP only 
the 8 spheroidal galaxies belonging to sample \#5 in 
their Tab. 7.
%, even if we plot also 
%the five galaxies of their sample \#4.
The parameters of these galaxies are summarized in Tab. \ref{tab:rxj0848}. 
{For the cluster XMMJ2235, out of the 9 galaxies observed by \cite{beifiori17},
we considered the 7 galaxies having spheroidal morphology,} 
six of which previously studied in detail by \cite{ciocca17}.
The parameters of these galaxies are summarized in Tab. \ref{tab:xmm2235}.

The high-redshift sample is thus composed of 22 cluster spheroidal galaxies 
in the redshift range 1.2$<$$z$$<$1.4.
The parameters of all the galaxies in these three clusters  have been homogeneously 
derived using \texttt{Galfit} on ACS-F850LP images, sampling the rest-frame B-band 
at the redshift of the clusters.
{ In particular, photometry and structural parameters of galaxies have been derived 
and studied by \cite{saracco14} for cluster RXJ0848,
by \cite{ciocca17} for cluster XMMJ2235 and by \cite{saracco17} for cluster XLSS0223.
\cite{jorgensen14} find a good agreement between their measurements for cluster  RXJ0848 and
those by \cite{saracco14}.
Analogously, \cite{ciocca17} find a good agreement between their measurements for cluster 
XMMJ2235 and those by \cite{chan16} for the same cluster.
(See also Appendix for further information.)}

\subsection{Local reference samples}
As reference local samples, we considered the Coma cluster sample of \cite{jorgensen95} 
and the Shapley sample of \cite{gargiulo09}. 
The Coma cluster sample is composed of the 74 spheroidal galaxies with stellar 
velocity dispersion measurements \citep{jorgensen95} and structural parameters 
derived from r$^{1/4}$ profile fitting in the r-Gunn filter \citep{jorgensen95a}.
We derived the corrections to apply to the parameters in the r-band to obtain those
in the B-band, using the sample of 28 ellipticals in the cluster for which 
measurements were performed also in the B-band \citep{jorgensen95a}.
The effective radius in the B-band is obtained by scaling the effective radius in
the r-band by the mean ratio $\langle r_{e,B}/r_{e,r}\rangle=1.05\pm0.08$.
The B-band surface brightness in physical units, log<I$_e$>=-0.4$\mu_e$+10.8 
[L$_\odot$ pc$^{-2}$], has been derived from the surface brightness 
$\mu_e=r+\langle B - r\rangle+5{\rm log}(r_{e,B})+2.5{\rm log}(2\pi)$,
where  $\langle B - r\rangle=1.14\pm0.07$ is the mean color of
the 28 ellipticals. 
The Shapley sample is composed of 141 spheroidal galaxies whose 
structural parameters were derived through Sersic profile fitting in the R-band 
\citep{gargiulo09}.
The effective radii in the B-band were obtained by applying the same scaling
factor derived for the Coma galaxies (given the small different wavelength between $r$ and R), 
while the B-band surface brightness was
derived by applying to the R-band magnitude, the mean color correction B-R=1.57
of local elliptical galaxies \citep[e.g.,][]{fukugita95}.

\subsection{The Fundamental Plane at z$\sim1.3$}
\begin{figure*}
	\hskip -0.5truecm\includegraphics[width=8.3truecm]{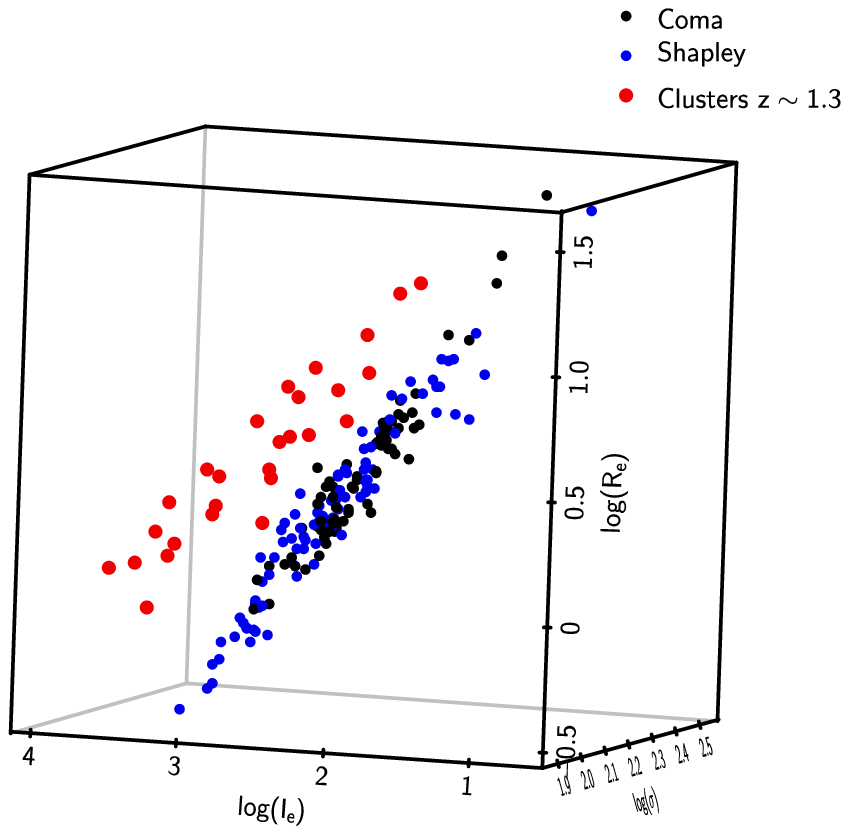}
	\hskip 0.8truecm \includegraphics[width=8.3truecm]{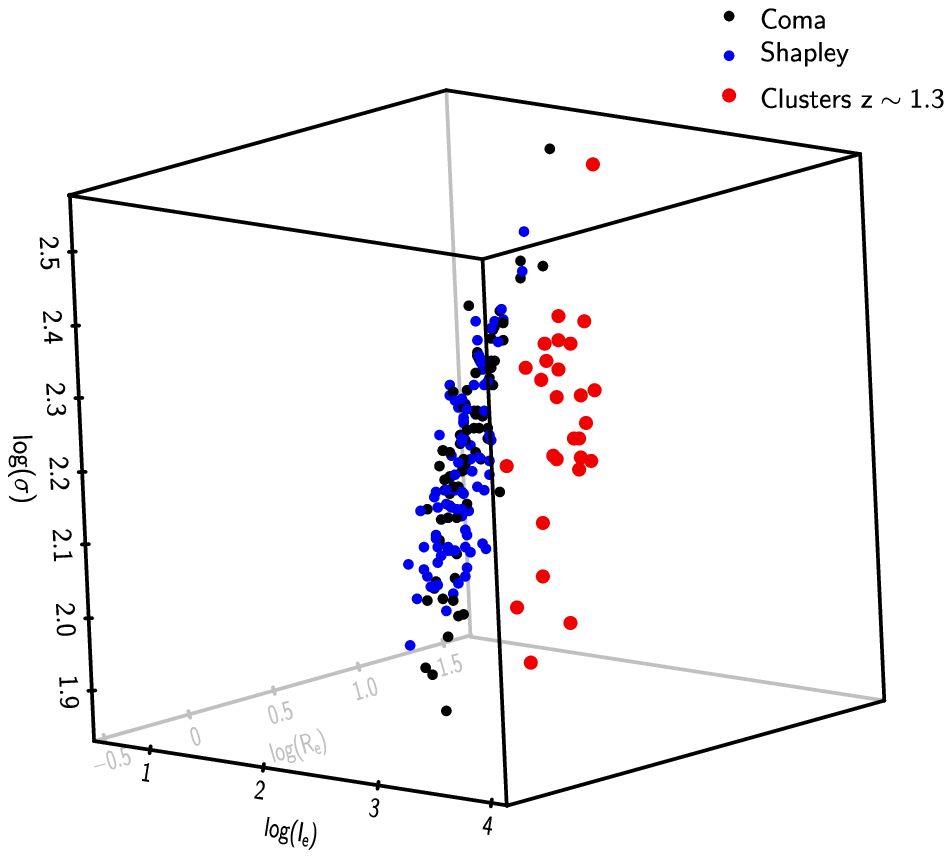}
\caption{\label{fig:fp_cube} 3D views in the log($\sigma$), log<I$_e$>, log(R$_e$) space of the 
low- and high-z cluster data. Black dots are the Coma data, blue dots are the Shapley data and 
red dots are the data of the three clusters at $z\sim1.3$.
 }
\end{figure*}	

\begin{figure*}	
	\includegraphics[width=8truecm]{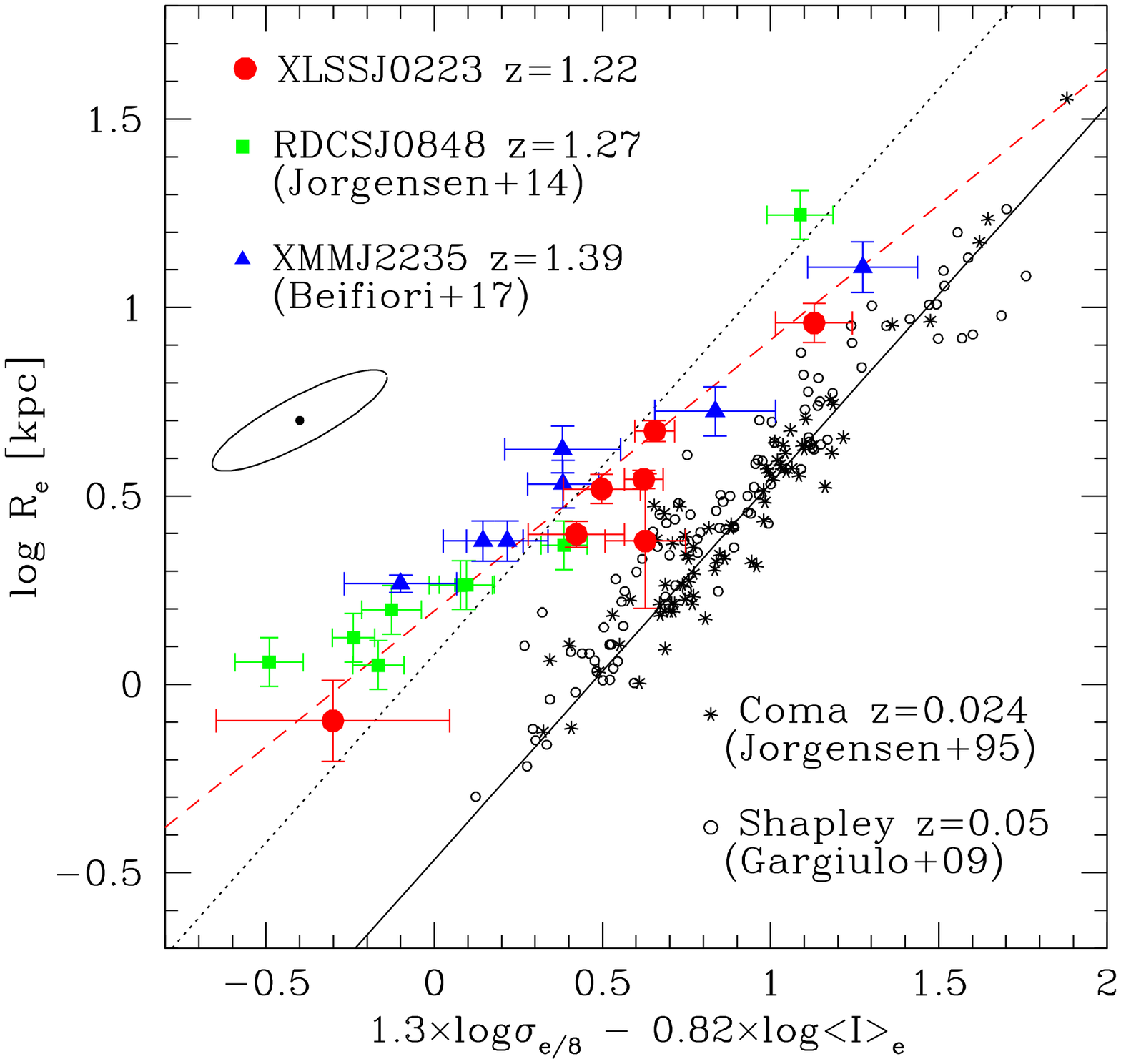}
	\includegraphics[width=8truecm]{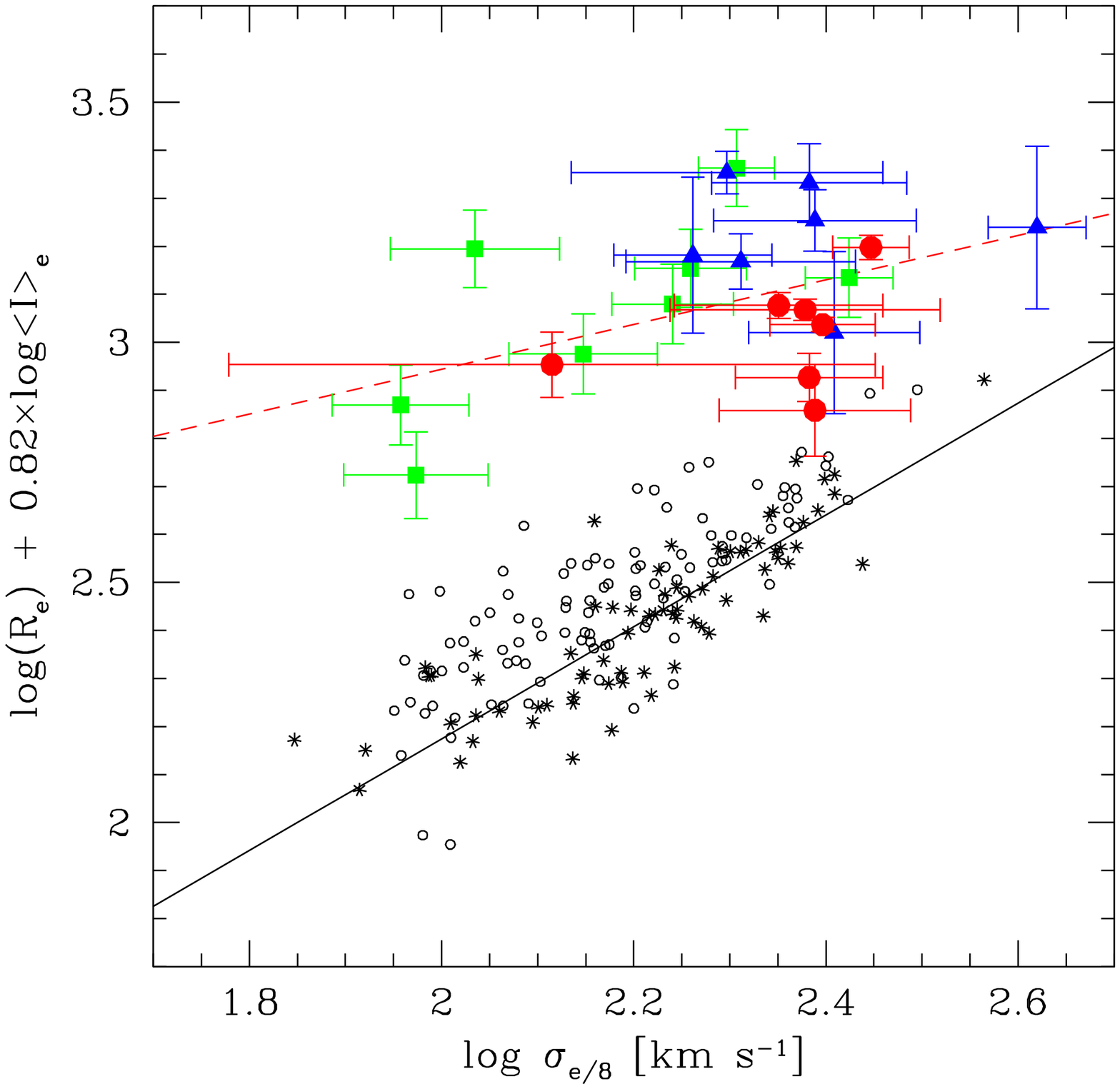}
    \caption{ \label{fig:fpp} 
Edge-on views of the B-band Fundamental Plane of spheroidal galaxies in cluster.  
     Skeletal symbols are the 74 galaxies of the Coma cluster sample
     \citep[][see details in the text]{jorgensen95,jorgensen95a}; 
     open circles are the 92 galaxies in the Shapley sample with $\sigma$$>$100
      km/s \citep{gargiulo09}; 
     filled symbols are the 7 galaxies
     in the cluster XLSS0223 (red circles, this work), the 8 galaxies in the cluster RDCS0848
     \citep[green symbols;][]{jorgensen14} and the 7 galaxies in the cluster XMMJ2235 
     \citep[blue triangles;][]{beifiori17}.
     The black solid lines are the best fitting relation to the Coma cluster sample.
     The black dotted line in the left panel is the best fit to the Coma data offset
     to the mean zeropoint of the high-z data. 
     The red-dashed lines are the best fit to the high-z data when the
     values 1.3log($\sigma$) and 0.82log<I$_e$> are considered. 
     { Note that the error bars on the left plot should not be crosses but rather tilted
ellipses due to the correlated errors between R$_e$ and I$_e$, as shown by
the ellipse representing the typical (mean) contour level at 1$\sigma$.}
%     Thus, they do not represent the real best fitting plane.
}
\end{figure*}
Before proceeding, we assessed possible selection effects in the high-redshift  
spectroscopic sample that may affect the analysis.
To this end, we compared the spectroscopic sample of 22 spheroids cluster
members with the photometric sample of 55 spheroids member candidates of the three clusters at $z\sim1.3$ 
\citep{saracco17} to which the 16 spectroscopic members belong.
Fig. \ref{fig:kor} shows the distribution of the galaxies on the 
log<I$_e$> vs log(R$_e$) plane, that is the Kormendy relation.
The high-redshift spectroscopic sample (red filled points) follows the  
same [log<I$_e$>;log(R$_e$)] distribution of the photometric sample (red open squares). 
This is confirmed by the 2-dimensional Kolmogorov-Smirnov (KS) test that gives a 
probability P=0.42 that the two samples are extracted from the same parent population. 
Thus, no obvious bias affecting size and/or surface brightness is present in the 
high-redshift spectroscopic sample.

We derived the best fitting relation 
 $\rm{log}(I_e)=a\times log(R_e) + b$ by using orthogonal fit.
We obtained $a=-1.5\pm0.3$ and $b=3.9\pm0.2$ for the
high-redshift sample and $a=-1.3\pm0.1$ and $b=3.04\pm0.06$ for the low redshift sample.
{ 
Given the correlated errors between <I$_e$> and R$_e$
%However, since <I$_e$> depends on R$_e$, the errors on these two quantities  
%are correlated among them.
%To properly take into account for this, 
we repeated the fitting also using the routine \texttt{linmix\_err} \citep{kelly07} 
that applies a Bayesian method to account for correlated measurement errors in linear regression.
The resulting best fitting relations are summarized in Tab. \ref{tab:kelly}.
No significant differences are found with respect to the relations obtained
with the orthogonal fit.
We notice that the orthogonal fit has been proved to be  robust
with respect to outliers \cite[e.g.,][]{jorgensen96} treating all the variables simmetrically, 
and has been adopted in many previous studies \cite[see e.g.][]{labarbera10}. 
For this reason, while comparing our results to those obtained with the Kelly's method, we adopt
the orthogonal fit as our reference fitting throughout the present work.}

{ We tried to account for the progenitor bias \citep[e.g.][]{vandokkum01,carollo13}
identifing galaxies in the Coma
sample potentially too young to be descendants of high redshift galaxies. 
To this end, we used the estimates based on line-strength indices by \cite{harrison10} 
which provide ages for 43 galaxies ($\sim$60\%) of the Coma sample here considered. 
Seventeen galaxies out of the 43 ($\sim$40\%) are younger than 7 Gyr.
Thus, other $\sim$10 galaxies younger than this age could be expected among
the remaining galaxies.\footnote{ It is worth noting that the light-weighted age
derived from H$_\beta$ index, as in the case of \cite{harrison10}, 
is strongly affected by recent episodes of star formation.
Hence, on average, it is biased toward ages younger than the true age of 
the bulk of stars.
Therefore, 40\% should be considered an upper limit to the
fraction of young galaxies.}
We choose 7 Gyr as limiting age instead of the lookback time $t_{LB}(z=1.3)\sim8.5$ Gyr,
to leave room to spheroids that may have experienced a 
secondary burst of star formation at later times.
These young galaxies are highlighted in cyan in Fig. \ref{fig:kor}.
The best fitting relation obtained without considering  young galaxies
($a=-1.4\pm0.3$, $b=2.9\pm0.2$) does not significantly differ from the one obtained
considering the whole sample.
}
We further discuss the Kormendy relation below.

Fig. \ref{fig:fp_cube} shows two different views of the data in the 3D space 
log$\sigma$, log<I$_e$>, log(R$_e$).
To homogenize the low-z samples, we considered for the Shapley sample
only galaxies (92) with $\sigma>$100 km/s.
The two different views show that the high-z data (red points) 
are offset with respect to the local data (blue and black points), and define 
a plane with different slope.
  
In Fig. \ref{fig:fpp}, we plot the so-called long and short edge-on projetions 
of the FP \citep[see, e.g.,][]{jorgensen95}.
We considered the velocity dispersion  $\sigma_{e/8}$ scaled to R$_e/8$ and 
parameters derived from Sersic profile fitting for all the samples, with the exception 
of the Coma cluster sample whose parameters are derived from Devaucoulers profile.
The figures show the two edge-on views of the FP for the spheroidal galaxies in 
the three clusters at $z\sim 1.3$ and for the local reference samples, 
assuming the values $\alpha=1.3$ and $\beta=0.82$ in eq. \ref{eq:fp} \citep{jorgensen06}.
Also in this case, it is visible the different slopes of the plane defined by the 
high redshift cluster sample  with respect to local ones.

A rotation of the FP at high-z is still a controversial issue.
While some authors find evidence of an evolution of the parameters $\alpha$
and $\beta$ since $z=0.8-0.9$ 
\cite[e.g.,][]{treu05a,treu05b,diserego05,jorgensen06,jorgensen13,woodrum17},
other authors do not detect significant variation at similar redshift
\cite[e.g.,][]{holden10}.  
Different slopes imply a differential evolution of the population 
of elliptical galaxies along the plane:
the mean properties of the population change over time along the plane 
in a different way according to the position on the plane, i.e., according 
to the physical parameters mass and/or size, and/or 
to the stellar population properties. 

We constrained the evolution of the FP applying the same fitting procedure 
to the low-redshift and the high-redshift data.
The best fitting coefficients $\alpha$, $\beta$ and $\gamma$ of eq. \ref{eq:fp}
have been determined using the code \texttt{lts\_planefit} 
\citep{cappellari13}
%The code 
that performs a robust linear regression with errors in all variables.
We first established the FP in the local Universe as reference. 
For the Coma cluster galaxies, whose parameters were derived from de Vaucoulers profile, 
we obtained $\alpha=1.27(\pm0.07)$, $\beta=-0.87(\pm0.03)$
and $\gamma=0.46(\pm0.01)$ with an rms of 0.07, 
not significantly different from the relation found by \cite{jorgensen06} in the B-band.
{ By excluding galaxies younger than 7 Gyr we obtained
$\alpha=1.34(\pm0.07)$, $\beta=-0.89(\pm0.03)$ and $\gamma=0.48(\pm0.01)$}.
For the Shapley galaxies, whose parameters were derived from Sersic profile, we obtained 
$\alpha=1.12(\pm0.09)$, $\beta=-0.78(\pm0.02)$
and $\gamma=0.46(\pm0.01)$ with an rms of 0.09,
not significantly different from the relation found by \cite{gargiulo09}.
The best-fitting relations obtained for the local samples are shown in 
Fig. \ref{fig:fit_fpl}.
By fitting the 22 cluster spheroids at $z\sim1.3$ with structural parameters
derived from S\'ersic profile fitting, we obtain 
\begin{equation}
\label{eq:fpz}
{\rm log R}_e = 1.3(\pm0.4){\rm log}\sigma_{e/8} 
-0.49(\pm0.08) {\rm log \langle I\rangle}_e + 0.4
(\pm0.1)
\end{equation}
with an rms of 0.05.
The best fit is shown in Fig. \ref{fig:fit_fph}.
Considering r$^{1/4}$ profile fitting we obtain $\alpha=1.0(\pm0.2)$ and $\beta=-0.34(\pm0.03)$.
We verified that excluding from the fitting galaxy \#1370, whose velocity dispersion 
has a large uncertainty (see Tab. \ref{tab:sigmas}),  
does not affect significantly the FP best-fitting coefficients in Eq. \ref{eq:fpz}.

{ The variation of $\alpha$ is not statistically significant, given the large 
uncertainty affecting this parameter. 
On the contrary, the variation of $\beta$, $\Delta\beta$$>$$0.29(\pm0.08)$, seems
to be significant, a conclusion reached also when r$^{1/4}$ profile is considered.  
{ We checked whether correlated errors between R$_e$ and <I$_e$>
can affect this result by repeating the fitting of the FP to 
100 simulated samples of 22 galaxies, 
even if we already verified  that correlated errors does not affect the 
relation between these two parameters
(Fig. \ref{fig:kor} and Tab. \ref{tab:kelly}).
We considered r$^{1/4}$ measurements to directly compare the result with 
the Coma FP parameter ($\beta=-0.87\pm0.03$).
The values of R$_e$ and L$_B$ of each galaxy have been varied by 
adding a shift randomly chosen from gaussian distributions with sigmas 
dR$_e$ and dL$_B$ respectively, where dR$_e$ and dL$_B$
are the errors on the two parameters.
Then, a new value of I$_e$ has been derived from these perturbed values
and assigned to the galaxy (see also appendix).

In Fig. \ref{fig:sim}, the distribution of the best fitting $\alpha$ and $\beta$ values obtained 
for the 100 simulated samples are shown.
%An example of best-fitting FP obtained for one of the 100 realizations is shown
%in Fig. \ref{fig:fp_sim}.
We confirm that the effect of the correlated errors is very weak ($\sim$10\%) in agreement with 
what already shown in \cite{labarbera10a}.
The simulations provide a median value $\langle\alpha\rangle=0.93\pm0.14$ 
and $\langle\beta\rangle=-0.35\pm0.04$ confirming a variation of this latter
parameter at the $\sim3\sigma$ level.
}
Thus, in agreement with, e.g.,
\cite{treu05a,treu05b,diserego05,jorgensen06,jorgensen13,woodrum17}, but 
at variance with others, e.g. \cite{holden10},  
our analysis shows a rotation of the FP at $z\sim1.3$ with respect to 
the local plane. 

\begin{figure}
\begin{center}
	\includegraphics[width=8.5truecm]{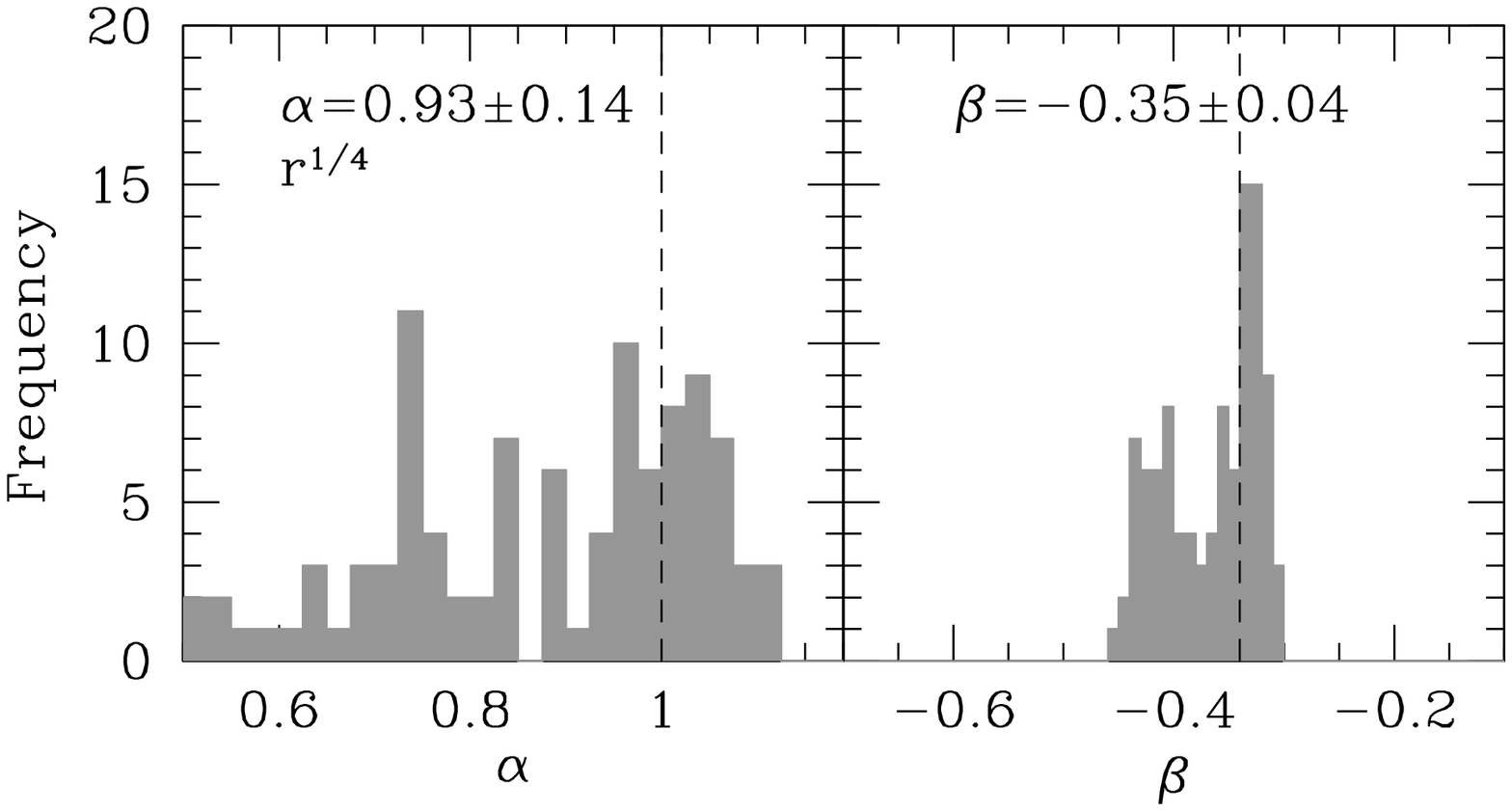}
\caption{\label{fig:sim} 
Distribution of the best fitting $\alpha$ and $\beta$ values obtained for the 100
simulated samples in case of r$^{1/4}$ profile.
The dashed line represents the best-fitting values $\alpha$=1.0$\pm$0.2 and 
$\beta$=-0.34$\pm$0.03 obtained for the true high-z cluster sample (r$^{1/4}$ profile).
The median value and the median absolute deviation of the distributions are
$\langle\alpha\rangle=0.93\pm0.14$ and
$\langle\beta\rangle=-0.35\pm0.04$.}
\end{center}
\end{figure}

\section{The M/L ratio\label{sec:msul}}
\begin{figure*}
	\includegraphics[width=7.truecm]{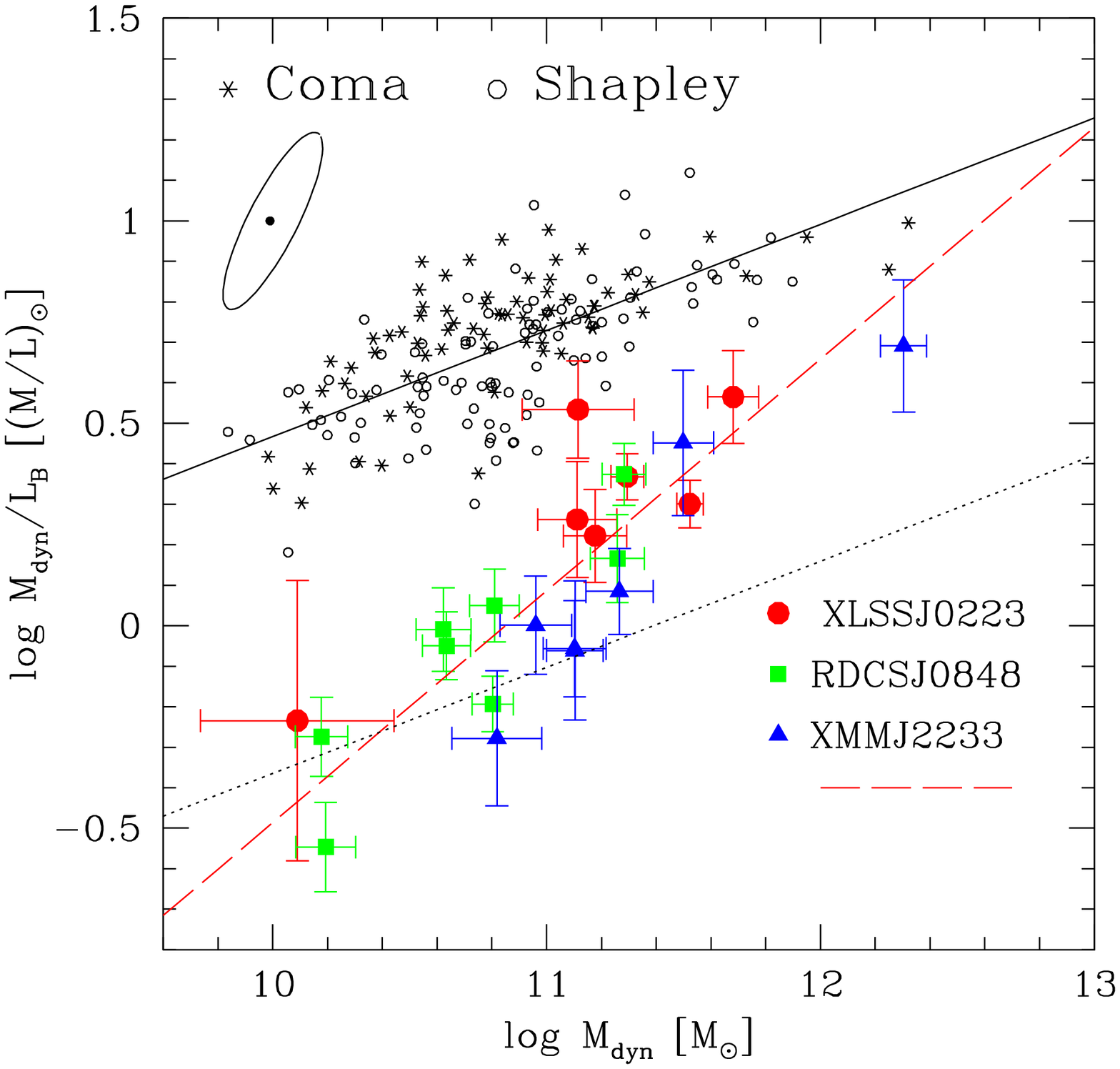}
	\includegraphics[width=7.truecm]{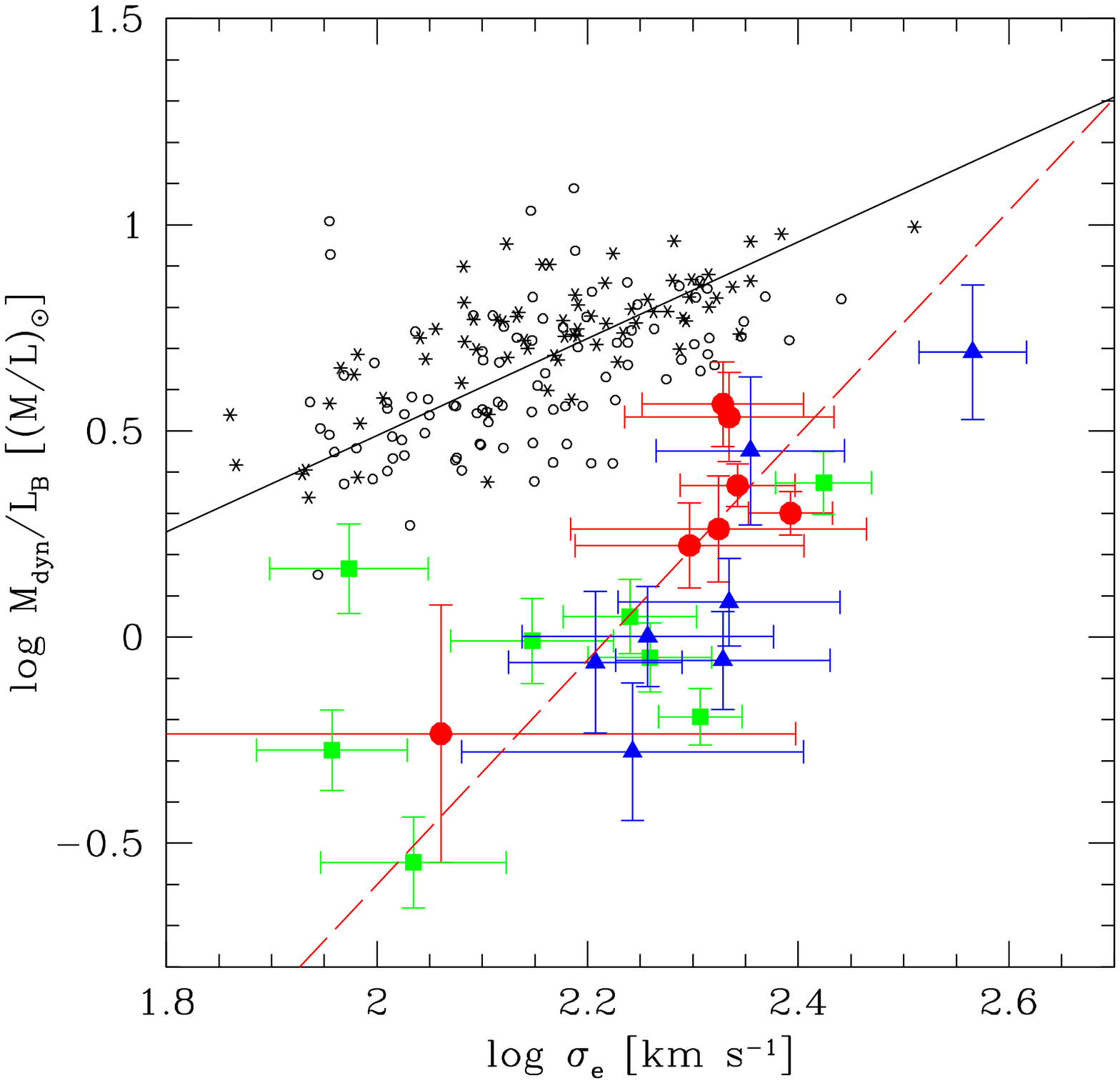}
\caption{\label{fig:MsuL} 
Dynamical mass to light ratio M$_{dyn}$/L$_B$ vs dynamical mass M$_{dyn}$ (left)
and velocity dispersion $\sigma_e$ (right) for high-redshift and low-redshift cluster 
spheroids.
Lines are the best fitting relations log(M/L)=a~log(M) + b and 
log(M/L)=a$_\sigma$~log($\sigma$) + b$_\sigma$ obtained with orthogonal fit
to the low-redshift data (black solid line, a=0.26$\pm0.03$ and b=-2.13$\pm0.06$;
a$_\sigma=1.2\pm0.1$ and b$_\sigma=-1.82\pm0.05$)
and to the high-redshift data (red dashed-line, a=0.6$\pm0.1$ and b=-6.2$\pm0.5$;
a$_\sigma$=2.7$\pm0.5$ and b$_\sigma$=-6.1$\pm0.4$).
{ The best fitting relations obtained considering
the correlated errors between the variables are reported in Tab. \ref{tab:kelly}.} 
Dotted line in the left panel is the local relation offset to the median 
log(M$_{dyn}$/L$_B$) value of high-redshift data.
{ Error bars should not be crosses but rather tilted ellipses, due to the correlated errors between 
M$_{dyn}$/L$_B$, M$_{dyn}$ and $\sigma_e$,  as shown by the ellipse plotted in the left panel representing
the typical (mean) 1$\sigma$ confidence contour.
}}
\end{figure*}

In Fig. \ref{fig:MsuL}, galaxies are plotted on the M$_{dyn}$/L$_B$ vs M$_{dyn}$ 
plane (left) and M$_{dyn}$/L$_B$ vs $\sigma_e$ (right) plane,
another way to represent the FP projections, 
where the mass M$_{dyn}$ is given by eq. \ref{eq:mdyn}.
As expected, the best fitting relation (orthogonal fit)
\footnote{ It is worth to note that the 
error on the b parameter has been obtained by fitting  
the relation y=a log(x/<x>)+b  \cite[e.g.][]{tremaine02}.}
of high-redshift spheroids
\begin{equation}
\label{eq:msul_z}
{\rm log}(M_{dyn}/L_B)=(0.6\pm0.1){\rm log}(M_{dyn})-(6.2\pm0.5)
\end{equation}
is steeper than the local one
\begin{equation}
\label{eq:msul_0}
{\rm log}(M_{dyn}/L_B)=(0.26\pm0.03){\rm log}(M_{dyn})-(2.13\pm0.06)
\end{equation}
as also previously found by other authors
\citep[e.g.,][]{treu05a,treu05b,diserego05,jorgensen06,saglia10,jorgensen13,
vandesande15,woodrum17}.
%{ It is worth to note that the 
%error on the b parameter has been obtained by fitting  
%the relation y=a log(x/<x>)+b  \cite[e.g.][]{tremaine02}}.
{ In Tab. \ref{tab:kelly} we report the relations obtained considering
the correlated errors among  M$_{dyn}$/L$_B$, M$_{dyn}$ and $\sigma_e$ in the fitting.
Also in this case, no significant differences are found with respect
to the relations obtained through orthogonal fitting.}
The different slope obtained for high and low redshift samples implies that M$_{dyn}$/L$_B$ 
has changed over time in a different way for lower mass and higher 
mass galaxies.

Trends visible in  Fig. \ref{fig:MsuL} can be due to a differential evolution of individual 
galaxies, to different properties of the spheroids joining the population at later times, 
or to a mix of these two causes. 
The relationships between stellar population properties (such as metallicity
and age) and mass \citep[e.g.][]{trager00,thomas05,gallazzi14,mcdermid15,jorgensen17,saracco19}
imply a mass dependent evolution of the M$_{dyn}$/L$_B$ ratio, even in the case 
that structural parameters of individual galaxies ($\sigma$ and R$_e$, and hence M$_{dyn}$) 
do not change over time. 
These correlations imply a mass dependent evolution of L$_B$ and, consequently, 
of M$_{dyn}$/L$_B$.
Moreover, metallicity and age seem to be also correlated to the
stellar mass density of galaxies \citep[e.g.][]{saracco11,saracco19,tacchella17}
that seems to be connected to the duration of the star formation \citep{gargiulo09}.
Therefore, we also expect a differential evolution of M$_{dyn}$/L$_B$ versus 
mass density as a function of redshift.

\subsection{The evolution of M$_{dyn}$/L$_B$}
We first consider the case of no evolution of the FP slope to compare our 
results with those in the literature, then we consider the
change of the slope.

The offset from the local relationship shown in Figure \ref{fig:MsuL}, 
is usually used to measure the mean M$_{dyn}$/L$_B$ evolution, under the assumption that
$\alpha$ and $\beta$ do not change with redshift.
In this case, from eq. \ref{eq:fp},  the offset of each galaxy is 
$\Delta\gamma_i=\gamma_{i,z}-{\gamma}_{0}=\beta(log (I_{e,0}/I_{ei,z}))$, that can be
rewritten as $\Delta$\rm{log}(M$_{dyn,i}$/L$_{B,i}$)=$\Delta\gamma_i/\beta$.
%where $\gamma_z=log(R_e)-\alpha~log(\sigma_{e/8})-\beta~log(I_e)$.
The mean (median) value of the individual $\Delta$\rm{log}(M$_{dyn,i}$/L$_{B,i}$)
values will result in the mean (median) M$_{dyn}$/L$_B$ variation of the 
population of galaxies. 
It is worth to note that this $\Delta$\rm{log}(M$_{dyn}$/L$_{B}$) derived from the FP, 
is not necessarily the same shown in Fig. \ref{fig:MsuL} where M$_{dyn}$ comes from the virial 
theorem (eq. \ref{eq:mdyn}).

We assumed, as reference, the coefficients 
of the local FP  derived in \S 4.3,
[$\alpha=1.27$, $\beta=-0.87$] for Coma and [$\alpha=1.12$, $\beta=-0.78$] 
for the Shapley sample.
We derived the offset $\Delta\gamma_i$ by computing for each individual galaxy
at $z\sim1.3$ the quantity 
$\gamma_{i,z}=log(R_{e,i})-\alpha~log(\sigma_{e,i/8})-\beta~log(I_{e,i})$  
\citep[e.g.][]{jorgensen06,saglia10,labarbera10a}.
Fig. \ref{fig:MsuL_mass}, shows $\Delta$\rm{log}(M$_{dyn}$/L$_B$) 
for the high-redshift spheroids.
The resulting median variation with respect to Coma is
$\Delta$\rm{log}(M$_{dyn}$/L$_B$)=-0.7$\pm$0.1
that does not differ significantly from $\Delta$\rm{log}(M$_{dyn}$/L$_B$)=-0.6$\pm$0.1
obtained considering Shapley as reference.
These values agree with the luminosity evolution $\Delta$\rm{log}(I$_e$)=$-0.72\pm0.1$
resulting from the Kormendy relation shown in Fig. \ref{fig:kor}
and also found by \citet[][$\Delta$\rm{log}(I$_e$)=$-0.77$]{jorgensen14}.

The resulting evolution with redshift of M$_{dyn}$/L$_B$ of cluster spheroids is
$\Delta$\rm{log}(M$_{dyn}$/L$_B$)=$(-0.5\pm0.1)z$, (filled black triangle), 
in agreement with many previous estimates 
\cite[e.g.][]{vandokkum03,vandokkum07,vandermarel07,saglia10,holden10,jorgensen14,beifiori17,prichard17}.
 We do not detect significant differences in the evolution of M$_{dyn}$/L$_B$
of individual clusters. 
We obtained $\Delta$\rm{log}(M$_{dyn}$/L$_B$)=$(-0.45\pm0.1)z$, 
$\Delta$\rm{log}(M$_{dyn}$/L$_B$)=$(-0.54\pm0.1)z$ and
$\Delta$\rm{log}(M$_{dyn}$/L$_B$)=$(-0.51\pm0.1)z$ for XLSSJ0223, RDCS0848 and XMMJ2235
respectively.
This result supports our approach of analyzing simulatenously all galaxies in the three clusters, when deriving the FP relation at $z\sim1.3$ (Sec.~4.3).

Fig. \ref{fig:MsuL_mass}, shows quantitatively the dependence of
the M/L variation with redshift on dynamical mass.
We divided the sample between low-mass galaxies having log(M$_{dyn}$)$<$11.1 M$_\odot$
and high-mass galaxies  with log(M$_{dyn}$)$\ge$11.1 M$_\odot$, 
being log(M$_{dyn}$)$=$11.1 M$_\odot$ the median value of the whole sample.
High-mass galaxies (median mass log(M$_{dyn}$)$=$11.27 M$_\odot$) 
%galaxies with log(M$_{dyn}$)$>$11.1 M$_\odot$
%(the median M$_{dyn}$ of the sample) 
are characterized by a milder evolution 
$\Delta$\rm{log}(M$_{dyn}$/L$_B$)=$(-0.38\pm0.07)z$ than low-mass galaxies 
(median mass log(M$_{dyn}$)$=$10.63 M$_\odot$),
whose variation is
$\Delta$\rm{log}(M$_{dyn}$/L$_B$)=$(-0.59\pm0.08)z$.

From eq. \ref{eq:msul_z} and \ref{eq:msul_0}, one obtains the differential evolution of 
the M$_{dyn}$/L$_B$ between low- and high-redshift galaxies as a function of galaxy mass:
\begin{equation}
\label{eq:msul_diff}
\Delta log(M_{dyn}/L_B)=(-0.34\pm0.1) log(M_{dyn})+(3.8\pm0.5)
\end{equation}
{ The differential evolution derived considering the best fitting relations obtained
with \texttt{linmix\_err} does not differ significantly (see Tab. \ref{tab:kelly}).}
This trend is consistent with that found by \cite{jorgensen06} for cluster spheroids 
at $z\sim0.9$, and reflects the fact that the M$_{dyn}$/L$_B$-mass relation is steeper 
for high- relative to low-redshift galaxies, 
consistent with the redshift evolution of the FP slopes.

\subsection{The formation redshift of cluster spheroids}
\label{sub:interp}

\begin{figure}
	\includegraphics[width=8.truecm]{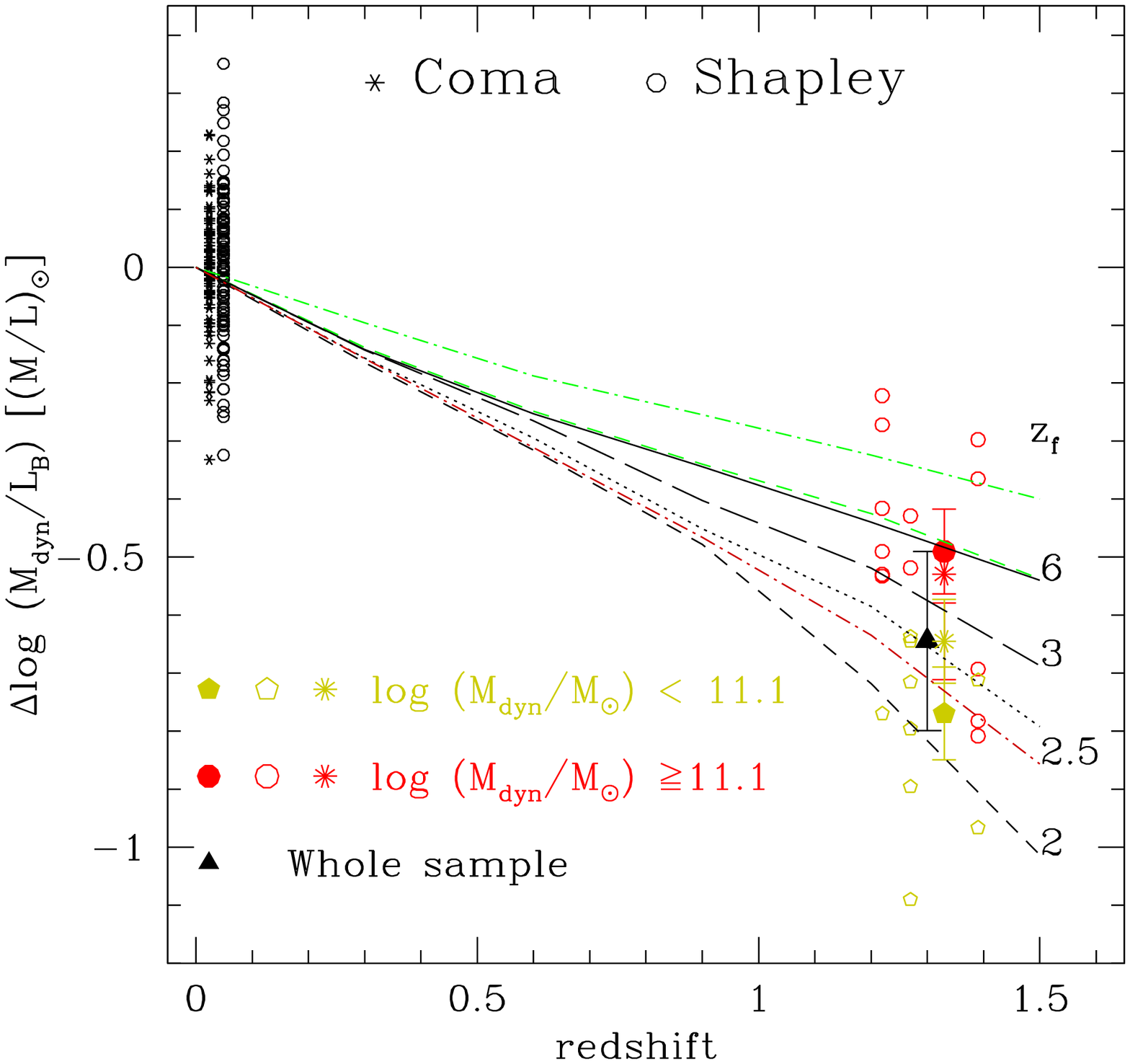}
\caption{\label{fig:MsuL_mass} 
Offset of log(M$_{dyn}$/L$_B$) from the local value for the high-redshift
spheroids coded by dynamical mass
as in the legend of the figure.
Filled black triangle is the median value $\Delta$\rm{log}(M$_{dyn}$/L$_B$)=-0.6$\pm$0.1  
of the high-redshift sample.
Coloured filled symbols (offset by 0.03 in $z$ for clarity) are the median 
$\Delta$\rm{log}(M$_{dyn}$/L$_B$) values of the M$_{dyn}$-selected sub-samples. 
The skeletal symbols are the median $\Delta$\rm{log}(M$^*$/L$_B$) values
for the M$_{dyn}$-selected.
The black curves are the expected $\Delta$\rm{log}(M$^*$/L$_B$) from BC03 
SSPs \citep{bruzual03} with Chabrier IMF and solar metallicity for different 
formation redshift $z_f$.
The dark-red curve was obtained for $z_f=2.5$ as the dotted line, but for 
metallicity Z=2.5Z$_\odot$. 
The green curves were obtained for $z_f=6$,  Salpeter IMF (dashed curve;
i.e. a single power law with index $s=2.35$) and for a single power-law IMF with 
slope $s=3.5$ (dot-dashed curve).
}
\end{figure}
If the variation of the M$_{dyn}$/L$_B$ was only due to the evolution
of the stellar populations over time and M$_{dyn}$/M$^*$=cost, 
then $\Delta$\rm{log}(M$_{dyn}$/L$_B$)=$\Delta$\rm{log}(M$^*$/L$_B$). 
In this case, using simple stellar population (SSP) models, this variation can be 
translated into the epoch of the last major episode 
of star formation, i.e. the formation redshift $z_f$ of the bulk of the stellar mass.

In Fig. \ref{fig:MsuL_mass}, the predicted $\Delta$\rm{log}(M$^*$/L$_B$) 
values for the SSP models of \cite{bruzual03} (BC03) are shown for different
values of $z_f$ and Chabrier initial mass function (IMF) and 
compared to the observed $\Delta$\rm{log}(M$_{dyn}$/L$_B$) of high-redshift galaxies.
Black curves are obtained with solar metallicity, while   
the dark-red dashed curve (obtained for $z_f=2.5$, as the black dotted curve) is 
obtained for Z=2.5Z$_\odot$, to show the effect of the metallicity.
For comparison, the green-dashed curve ($z_f=6$, Z$_\odot$) was obtained with
a Salpeter IMF (i.e. a single power-law IMF functional form, with slope $s=2.35$) while 
the dot-dashed curve refers to
the bottom-heavier IMF with $s=3.5$ (i.e. a distribution with a strong contribution from low-mass stars).

Under the assumptions above, the resulting median formation redshift of the 
stellar populations for the whole population of high-redshift spheroids 
(black filled triangle)
is $z_f\simeq2.5$ for Z=Z$_\odot$ ($z_f\simeq2.7$ for Z=2.5Z$_\odot$).
This agrees with the formation redshift obtained from stellar absorption line strengths 
and full spectral fitting for the 7 spheroids in the cluster XLSSJ0223  \citep{saracco19}
and with what found by \cite{beifiori17} for cluster XMMJ2235.

Fig. \ref{fig:MsuL_mass} shows also that stellar populations in high-mass 
(log(M$_{dyn}>$11.1 M$_\odot$) spheroids form earlier ($z_f\simeq6$) than those 
in lower-mass spheroids ($z_f\sim2.2$), in agreement with many previous works
\citep[e.g.][]{thomas10}. 
A better estimate of $z_f$ as a function of galaxy mass can be obtained from 
eq. \ref{eq:msul_diff}, which reflects the evolution with redshift of the FP slopes.
Interpreting the variation of M$_{dyn}$/L$_B$ of eq. \ref{eq:msul_diff} 
as a difference in the epoch of the last major star formation episode, we obtain
\begin{equation}
\label{eq:age_mdyn}
 {\rm log}(age(z_f))=(-0.39\pm0.1)\times {\rm log}(M_{dyn})+(4.4\pm0.6)
\end{equation}
where we used the best-fitting relation
log(M$^*$/L$_B$)=0.86log(age) obtained for BC03 models (Chabrier IMF, Z$_\odot$). 
This implies that a galaxy with mass log(M$_{dyn}$)$\ge$11.5 M$_\odot$ has experienced 
its last burst when the Universe was $\le$0.82 Gyr old, i.e. at $z_f$$\ge$6.5, while a galaxy 
with log(M$_{dyn}$)$\le$10 M$_\odot$ at $z_f$$\le$2.

The same trend is qualitatively obtained considering M$^*$/L$_B$ instead of M$_{dyn}$/L$_B$.
This is shown by the skeletal symbols in Fig. \ref{fig:MsuL_mass} that
represent the median $\Delta$\rm{log}(M$^*$/L$_B$) values of low-M$_{dyn}$ and high-M$_{dyn}$
spheroids. 
{ The median $\Delta${\rm{log}}(M$^*$/L$_B$) values have been obtained by summing 
to $\Delta${\rm{log}}(M$_{dyn}$/L$_B$) the median
log(M$_{dyn}$/M$^*$)=log(M$_{dyn}$/L$_B$)-log(M$^*$/L$_B$) values obtained for 
low-M$_{dyn}$ and high-M$_{dyn}$ galaxies.}
The M$^*$/L$_B$-based formation redshifts of the stellar populations 
in low- and high-mass spheroids differ, to some extent, with respect to those obtained considering  
$\Delta$M$_{dyn}$/L$_B$.
This shows that the assumption $\Delta$\rm{log}(M$_{dyn}$/L$_B$)=$\Delta$\rm{log}(M$^*$/L$_B$) does not 
match the real case and suggests a variation of M$_{dyn}$/M$^*$ among the population of spheroids.

\begin{figure*}
	\includegraphics[width=7.truecm]{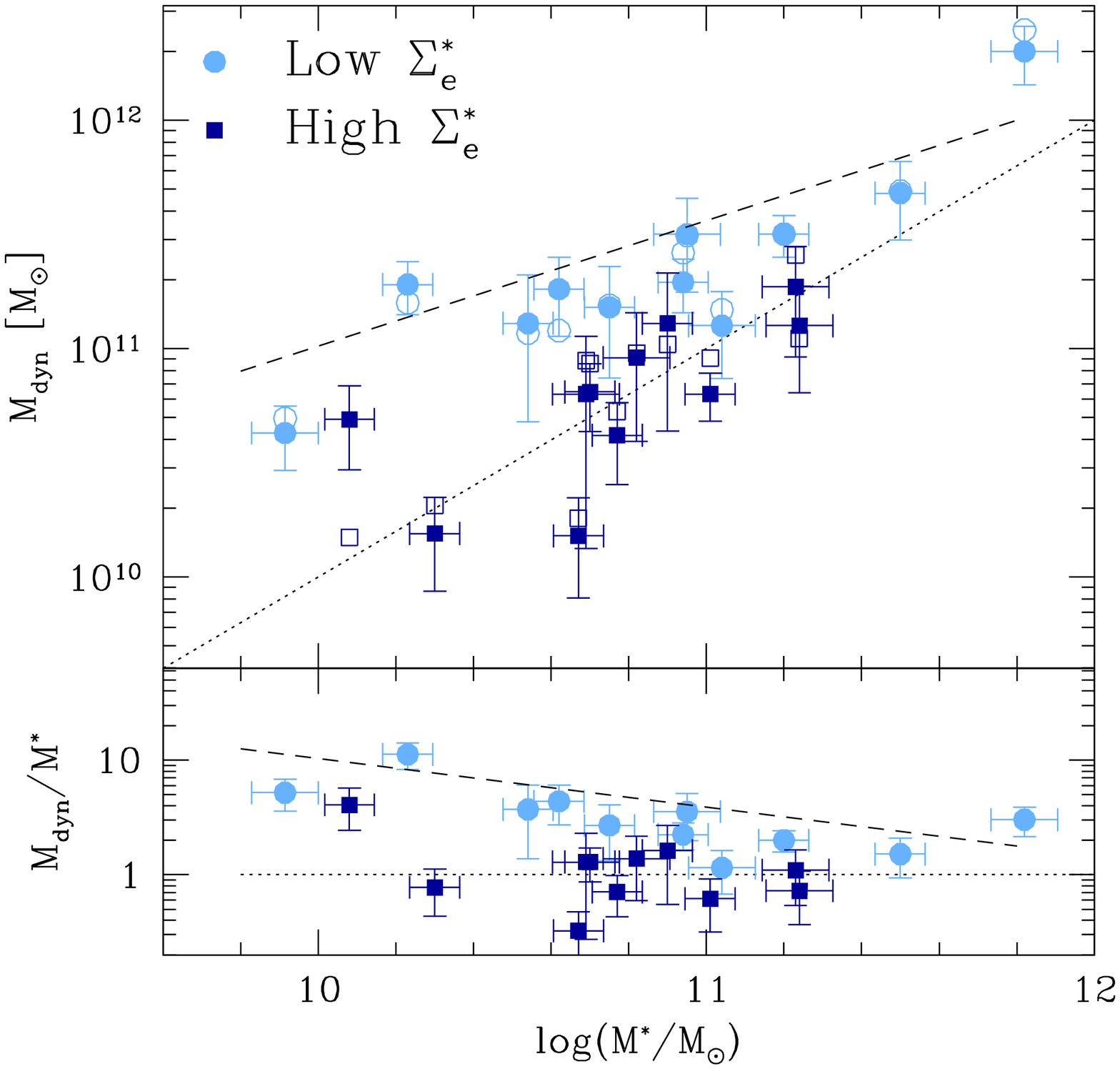}
	\includegraphics[width=7.truecm]{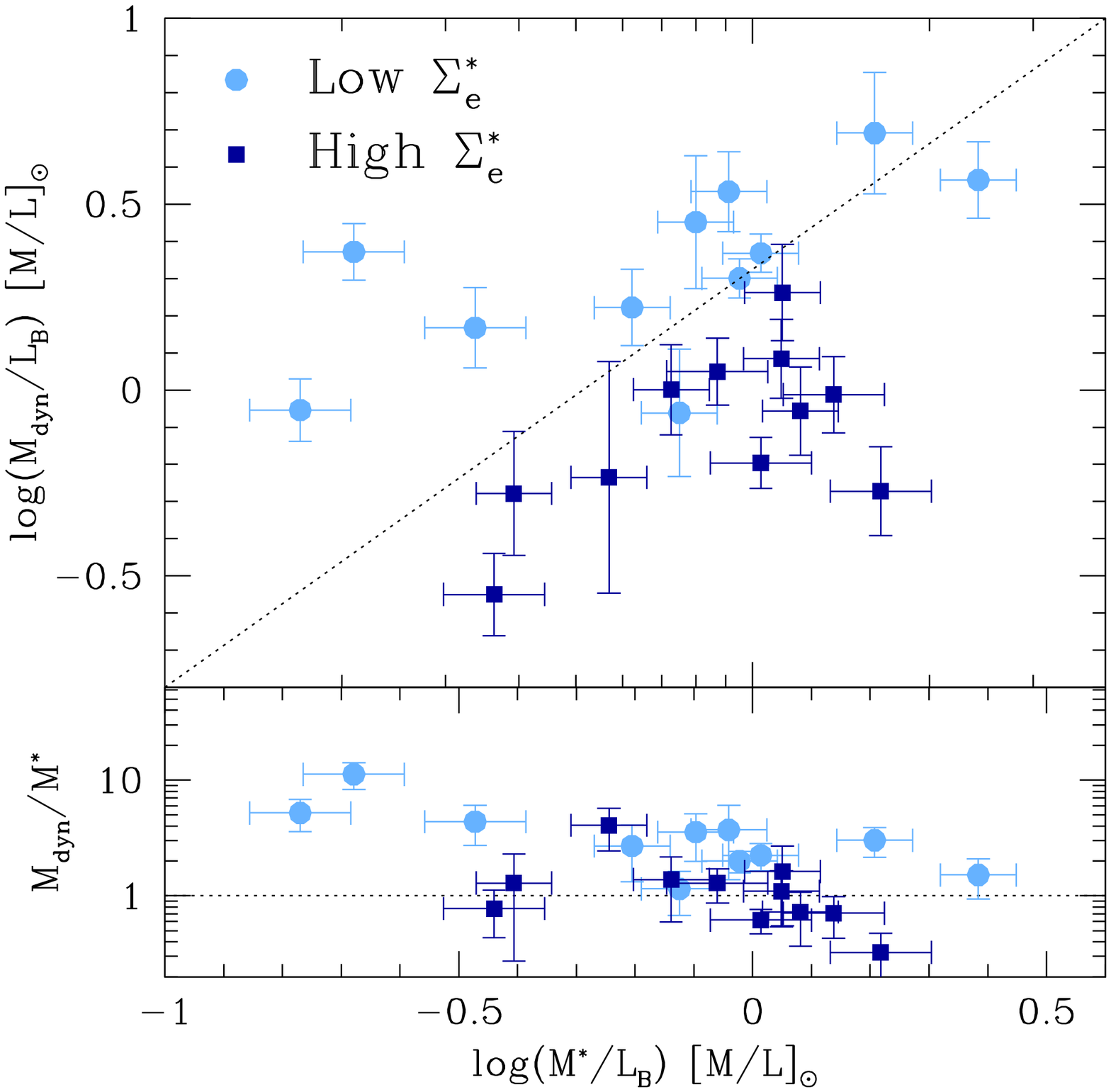}
\caption{\label{fig:mdyn_mstar} 
Left - Relation between dynamical mass M$_{dyn}$, as derived from eq. \ref{eq:mdyn},
and stellar mass M$^*$ for the 22 cluster spheroids at $z\sim1.3$. 
Galaxies are coded according to their stellar mass density $\Sigma_e^*$
(see \S 5):
light-blue filled circles are low-density galaxies, with stellar mass density lower than
the density median value of the sample; navy filled squares
are high-density galaxies. 
The dotted line is the 1:1 relation, the dashed line is the upper relation 
enclosing 90\% of the data.
The open symbols represent the dynamical mass obtained using the virial coefficient
$\beta(n)=8.87-0.831n+0.0241n^2$ \citep{cappellari06}, where $n$ is the Sersic index.
The lower panel shows the dynamical to stellar mass ratio as a function of stellar mass.
{ It is worth to remind that error bars on $M_{dyn}/M^*$ would be tilted ellipses because
of the correlation with $M^*$.}
The median value of the ratio is $\langle M_{dyn}/M^*\rangle$=1.6$\pm$0.2.
Symbols are as in the upper panel.
Right - Relation between the dynamical mass-to-light ratio $M_{dyn}/L_B$ versus 
$M*/L_B$. 
}
\end{figure*}	

Fig. \ref{fig:mdyn_mstar} (left) shows the relation between M$_{dyn}$ and M$^*$ for the
high-redshift spheroids coded according to their stellar mass density 
$\Sigma_e^*=M^*/(2\pi R^2_e)$: low-density spheroids (light blue pentagons), 
have $\Sigma_e^*$$<$$\Sigma_{med}^*$$\simeq$1800 M$_\odot$ pc$^{-2}$ 
(the median $\Sigma_e^*$ of the sample), 
high-density spheroids (dark blue points) have $\Sigma_e^*$$>$$\Sigma_{med}^*$.
The overall best-fitting relation is log(M$_{dyn}$)=(1.09$\pm$0.2)log(M$^*$)-(0.7$\pm$1.3).
The median value of the whole sample is $\langle M_{dyn}/M^*\rangle$=1.6$\pm$0.2.
{ We note that one galaxy, \#1264 belonging to the cluster RXJ0848 has M$_{dyn}$/M$^*$$<$1.
This galaxy has the lowest velocity dispersion ($\sim90$ km/s) with a small error and a relatively 
high stellar mass (see Tab. \ref{tab:rxj0848}). 
We cannot exclude a sistematicity in one of these measurements.}
Two trends are visible: at fixed stellar mass, lower density spheroids have systematically 
higher M$_{dyn}$, hence larger M$_{dyn}$/M$^*$; the mean M$_{dyn}$/M$^*$ of low-$\Sigma^*$ 
galaxies ($\langle$M$_{dyn}$/M$^*\rangle_{l\Sigma}$=3.0$\pm$1) 
increases for decreasing stellar masses and tends to the 
nearly constant value $\langle$M$_{dyn}$/M$^*\rangle_{h\Sigma}$=1.1$\pm$0.3 
of high-$\Sigma^*$ galaxies at large stellar masses.
{These trends do not depend on the virial mass estimator adopted (eq. \ref{eq:mdyn}),
as shown by the open symbols representing M$_{dyn}$=$\beta(n)$$\sigma_e^2$R$_e$/G,
where $\beta(n)=8.87-0.831n+0.0241n^2$ \citep{cappellari06} and $n$ is the S\'ersic index}.

Some evidences have been accumulated in favour of a variation
of the IMF with velocity dispersion \citep[e.g.][]{labarbera13}, 
or mass density \citep[at constant dark matter (DM) fraction, e.g.][]{gargiulo15} 
or mass of galaxies  \cite[see, the Introduction in][for a concise review]{demasi19}.
Some of these studies consider the IMF normalization ratio  M$_{dyn}^*$/M$^*$, often called $\Gamma$,
as a comparison between stellar mass M$^*$, obtained for a reference IMF (e.g. Chabrier), 
and the true mass of the galaxy, M$_{dyn}^*$, corrected for the DM contribution.
M$_{dyn}^*$/M$^*$ is found to increase with velocity dispersion, both
in low- and high-redshift samples \citep[e.g.][]{cappellari12,tortora13,gargiulo15,ferreras13}.
We find that M$_{dyn}$/M$^*$ varies with  velocity dispersion
also for cluster spheroids at $z\sim1.3$, as shown in Fig. \ref{fig:mdyn_sigma}.
%where M$_{dyn}$/M$^*$ is plotted as a function of the stellar velocity dispersion.
We note that the variation holds both for low-density and high-density galaxies,
in agreement with what previously found by \cite{gargiulo15} for field spheroids at high redshift.
{ For comparison, the M$_{dyn}$/M$^*$-$\sigma$ relation of a sample
of local spheroids extracted from the SPIDER sample \cite[gray symbols;][]{labarbera10}
is also shown.
We notice that the relation between M$_{dyn}$ and M$_{dyn}^*$, as well the trend of M$_{dyn}^*$/M$^*$-$\sigma$, 
i.e. the variation of IMF on $\sigma$ and its possible
evolution is out of the scope 
of the present work and it is treated in detail in other papers 
\cite[e.g.,][and references therein]{cappellari12,cappellari13,tortora13,ferreras13,gargiulo15,labarbera15,lyubenova16}}.
 \begin{figure}
\begin{center}
	\includegraphics[width=7.truecm]{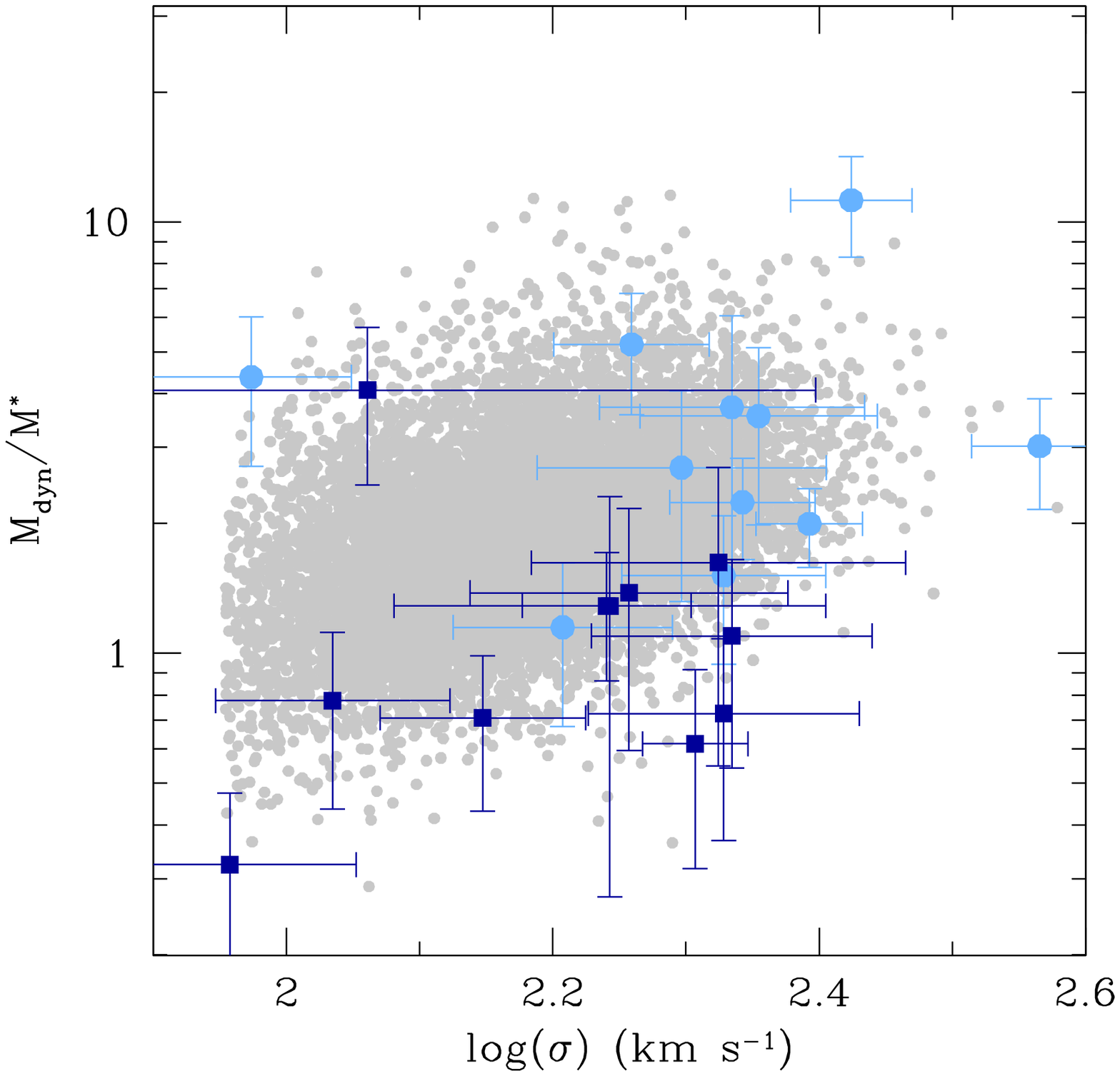}
\caption{\label{fig:mdyn_sigma} 
M$_{dyn}$/M$^*$ is plotted versus 
the stellar velocity dispersion for the 22 spheroidal galaxies in cluster
at $z\sim1.3$ (blue and light-blue symbols) coded according the their stellar mass density
{ and for a sample
of local spheroids (gray symbols) extracted from the SPIDER sample \citep{labarbera10}. 
Also in this case, errors would be tilted ellipses (see e.g. Fig. \ref{fig:fpp})
because of the correlation between M$_{dyn}$} and $\sigma$.}
\end{center}
\end{figure}

Assuming a Salpeter IMF for low density spheroids, while keeping the Chabrier IMF for high density ones, 
would not change the trend of M$_{dyn}$/M$^*$ with mass while the offset
between  $\langle$M$_{dyn}$/M$^*\rangle$ values of high- and low-density spheroids (a factor $\sim$3, see above)
would reduce by a factor of ~1.7 \citep[e.g.][]{longhetti09}, remaining still
higher for low-density galaxies. 
As to the formation redshift, the effect of assuming an heavier IMF 
is that to decrease the $z_f$ derived from the observed 
$\Delta$\rm{log}(M$_{dyn}$/L$_B$), as shown by the green curves in 
Fig. \ref{fig:MsuL_mass} obtained for $z_f=6$ with a Salpeter
IMF (dashed curve, slope $s=2.35$) and a significantly bottom-heavy IMF with 
slope $s=3.5$ (dot-dashed curve) \cite[see also Fig. 10 in][]{renzini06}.
The right panel of Fig. \ref{fig:mdyn_mstar} shows explicitly that, for a given
M$^*$/L$_B$, lower-density spheroids have higher  M$_{dyn}$/L$_B$ and vice versa.

%[E' importante questo paragrafo (e le figure) o no?]

These trends are independent of redshift and of the environment, as shown by 
Fig. \ref{fig:msul_spd} that shows the same quantities of Fig. \ref{fig:mdyn_mstar}
for a sample of local ($z\sim0.07$) spheroids. 
They have been selected from the SPIDER sample 
\citep{labarbera10}
%%We selected galaxies 
with $\sigma_e>150$ km/s (the range where 17 out of our 22 spheroids lie)
and ages older than 7 Gyr, to account for the progenitor bias.\footnote{We allowed for an age
younger than the lookback time $t_{LB}\sim8.5$ Gyr, to account for the progenitor 
bias, yet leaving room to spheroids that may have experienced a secondary minor 
burst of star formation at later times. SPIDER sample includes principally 
spheroidal galaxies in the field and in groups.}
The SPIDER galaxies are coded according to their stellar mass density $\Sigma^*$
(upper panel) and to their M$_{dyn}$/L$_B$ (lower panel).
High-redshift spheroids (red filled points) are superimposed to the local spheroids
and the size of their symbols scale according to their  $\Sigma^*$ and M$_{dyn}$/L$_B$.

%[.....]
\begin{figure}
\begin{center}
	\includegraphics[width=8.0truecm]{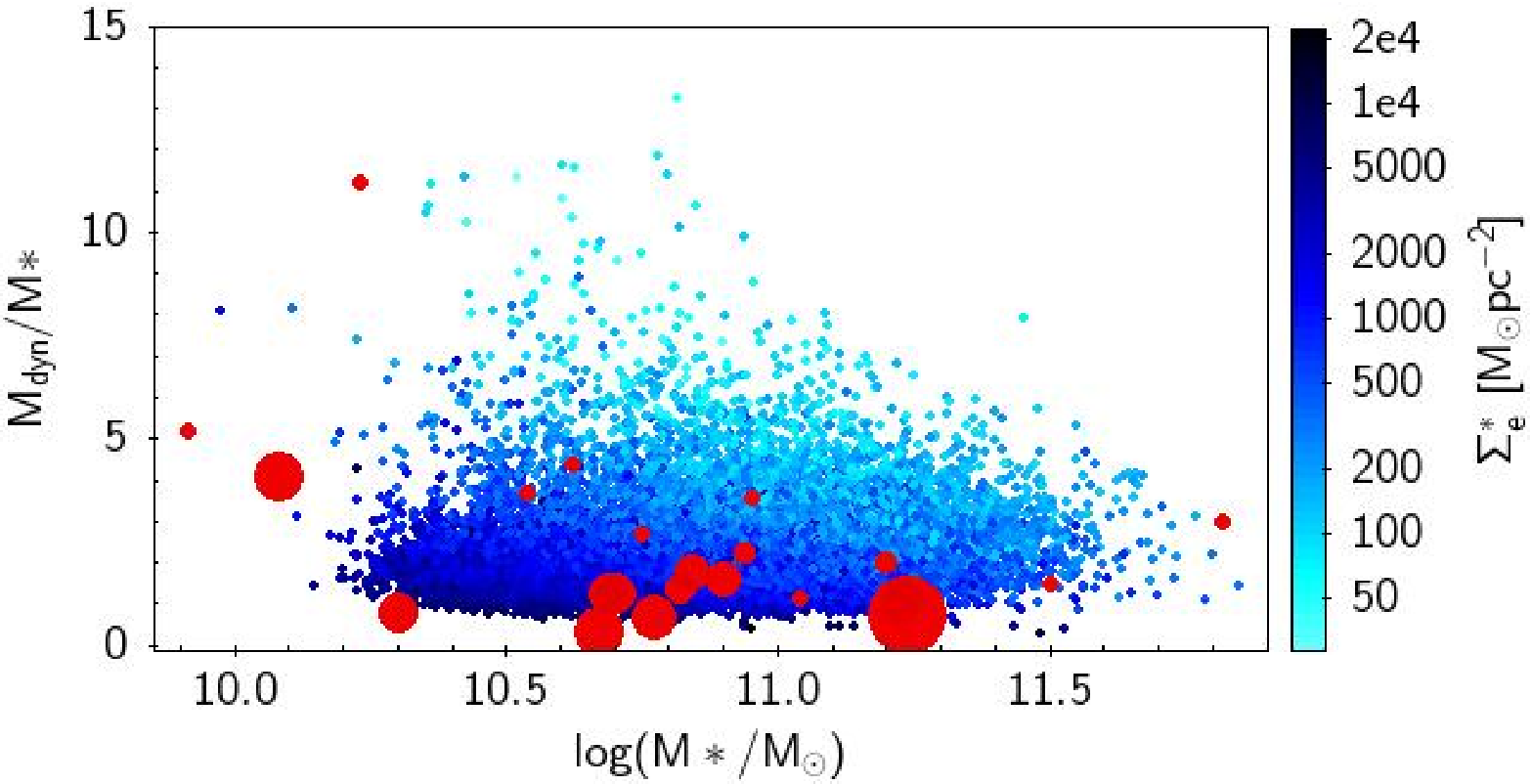}
	\includegraphics[width=7.5truecm]{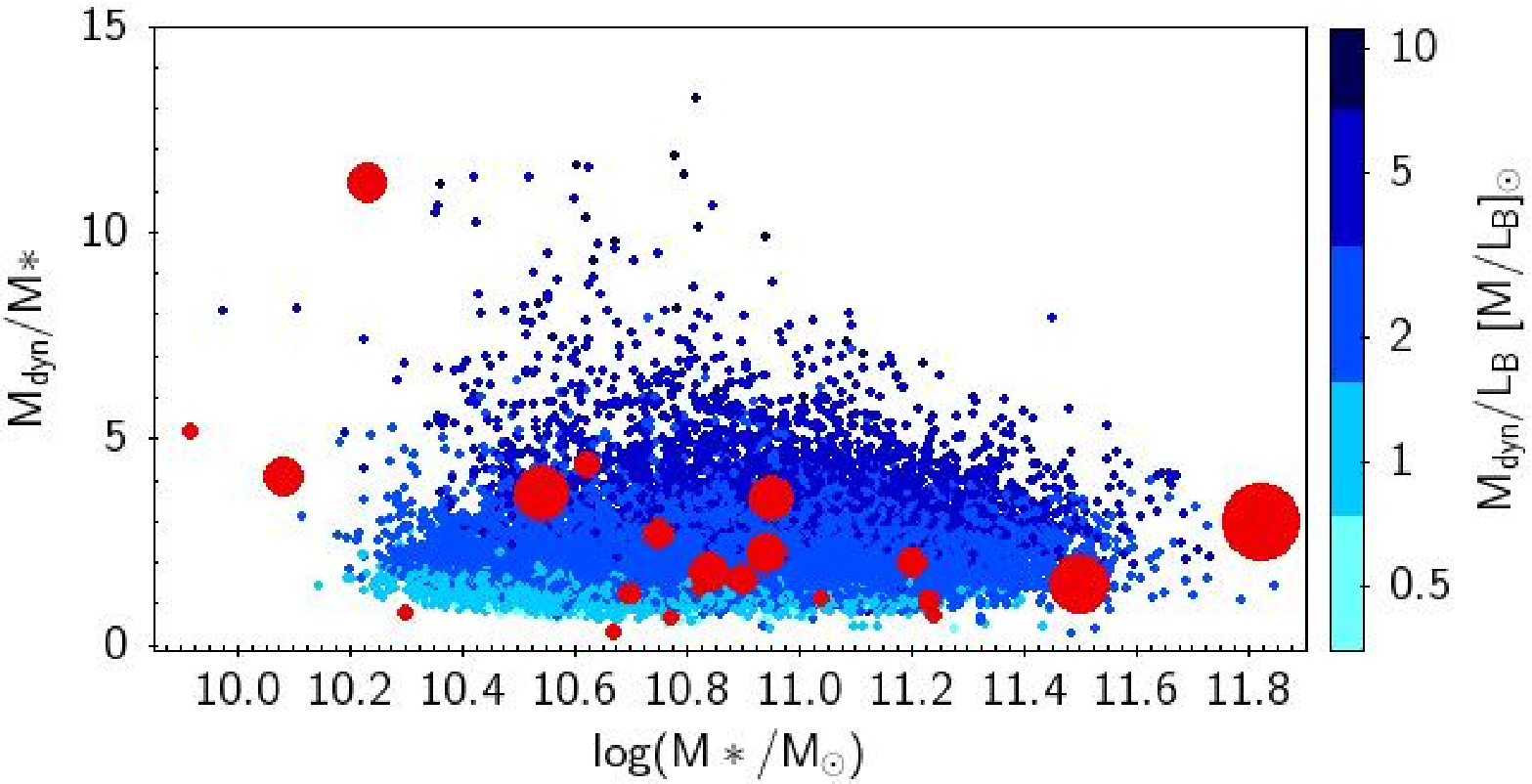}
\caption{\label{fig:msul_spd} 
Dynamical to stellar mass ratio as a function of stellar mass for the SPIDER sample
(blue dots) and high-redshift cluster spheroids (red dots).
The SPIDER galaxies are coded according to their stellar mass density (upper panel)
and M$_{dyn}$/L$_B$ ratio (lower panel).
The size of the symbols of high-redshift spheroids (red dots)
scale according to their $\Sigma^*$ and M$_{dyn}$/L$_B$. }
\end{center}
\end{figure}

Therefore, M$_{dyn}$/M$^*$ varies systematically both with stellar mass and mass density,
as shown in figures \ref{fig:mdyn_mstar} and \ref{fig:msul_spd}.
In Fig. \ref{fig:msul_dens} we compare the observed $\Delta$\rm{log}(M$^*$/L$_B$) values 
of high-density and low-density spheroids to those predicted by SSPs:
high-density spheroids experience 
their last major burst of star formation earlier ($z_f$$\sim$4.5) than low-density spheroids
($z_f$$<$2.5),
in agreement to the rather well established evidence that denser galaxies
host older stellar populations  
\citep[e.g.][]{gargiulo09, gargiulo17, saracco09, saracco11, valentinuzzi10, poggianti13, 
fagioli16,williams17,diaz19}.
Therefore, the formation redshift of cluster spheroids depends on their 
(dynamical/stellar) mass and it is correlated to the stellar mass density.

\begin{figure}
	\includegraphics[width=8.truecm]{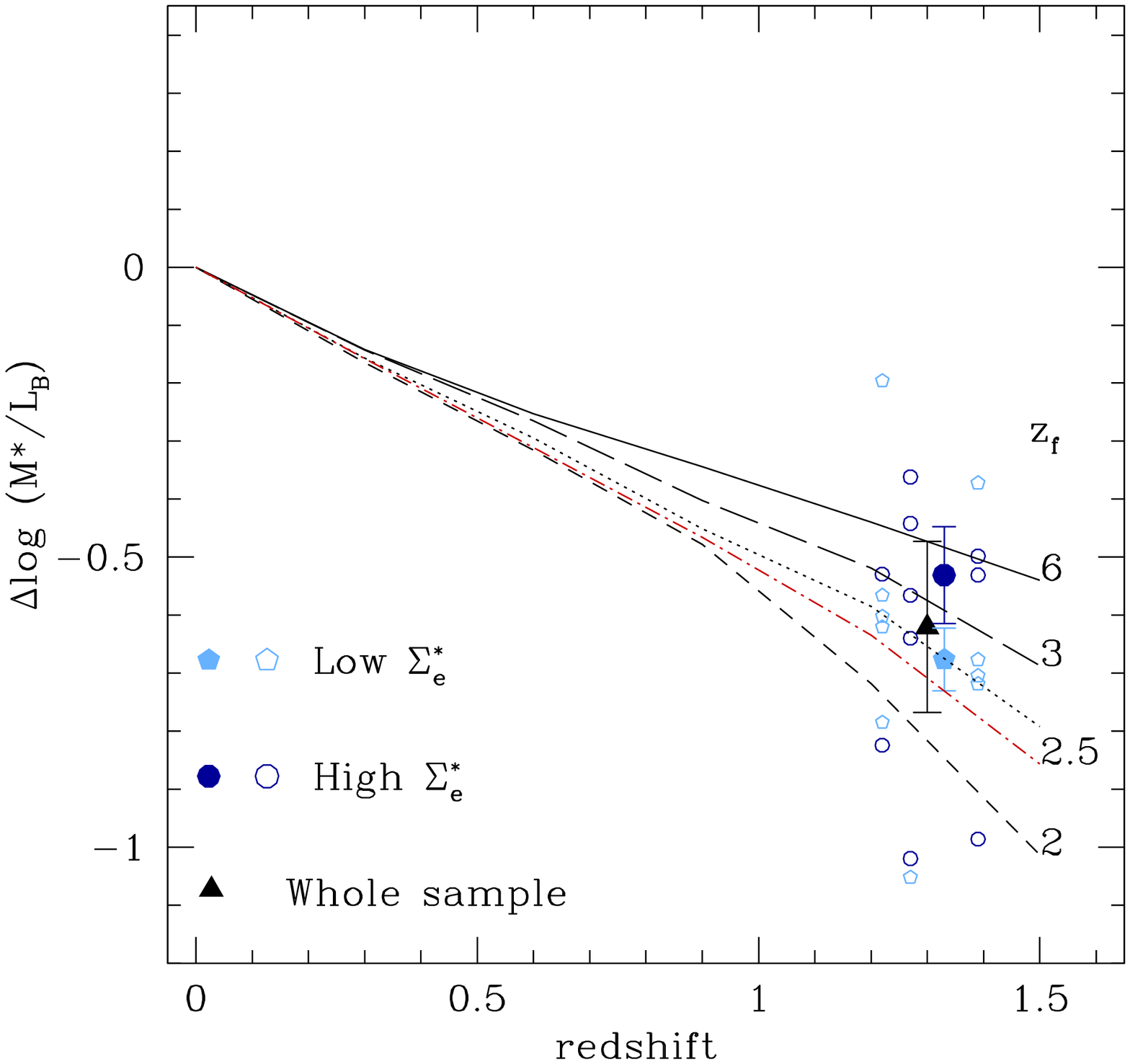}
\caption{\label{fig:msul_dens} 
Offset $\Delta$\rm{log}(M$*$/L$_B$) for the high-redshift spheroids coded by stellar mass density
$\Sigma^*_e$.
{ It has been obtained by summing to the $\Delta${\rm log}(M$_{dyn}$/L$_B$) of high-density
and low-density galaxies, referred to the median log(M$_{dyn}$/L$_B$) value of local galaxies
(like the models), 
the median log(M$_{dyn}$/M$^*$)=log(M$_{dyn}$/L$_B$)-log(M$*$/L$_B$)
computed for high-density and low-density galaxies.}
Filled black triangle is the median value $\Delta$\rm{log}(M$^*$/L$_B$)$\simeq$-0.6$\pm$0.1  
of the high-redshift sample.
Coloured filled symbols (offset by 0.03 in $z$ for clarity) are the median 
$\Delta$\rm{log} values of the two subsamples. 
The curves are as in Fig. \ref{fig:MsuL_mass}.
 }
\end{figure}

\section{The evolution of cluster spheroids}
The different slope of the M$_{dyn}$/L$_B$-M$_{dyn}$ relation at low and high redshift
(Fig. \ref{fig:MsuL}), 
implies that M$_{dyn}$/L$_B$  has varied over time in a different way for low-mass 
and high-mass galaxies according to eq. \ref{eq:msul_diff}, at least since $z\sim1.4$.
Fig. \ref{fig:MsuL_mass} shows that the variation of log(M$^*$/L$_B$) (skeletal symbol) 
is consistent with the observed variation of log(M$_{dyn}$/L$_B$) for high-mass galaxies while
it is not sufficient to account for the variation of low-mass galaxies.
Assuming that high- and low-mass galaxies have the same stellar IMF,
the evolution of the stellar populations account for $\sim$85\% of the observed evolution of 
M$_{dyn}$/L$_B$ for M$_{dyn}<10^{11}$ M$_\odot$ spheroids and, hence, that 15($\pm$8)\% 
of this evolution must be due to other causes.
{ It is worth to mention that an heavier IMF for low-mass galaxies, 
like Salpeter IMF, would not account
for the different behaviour of 
$\Delta$\rm{log}(M$_{dyn}$/L$_B$) vs $\Delta$\rm{log}(M$^*$/L$_B$)} (see Fig. \ref{fig:MsuL_mass}).
Moreover, low-mas galaxies would be  expected to have a lighter IMF than
 high-mass systems.
 
Two other components may affect the evolution of M$_{dyn}$/L$_B$, 
DM and galaxies that join the population of spheroids at later times.
\begin{figure}
	\includegraphics[width=8.5truecm]{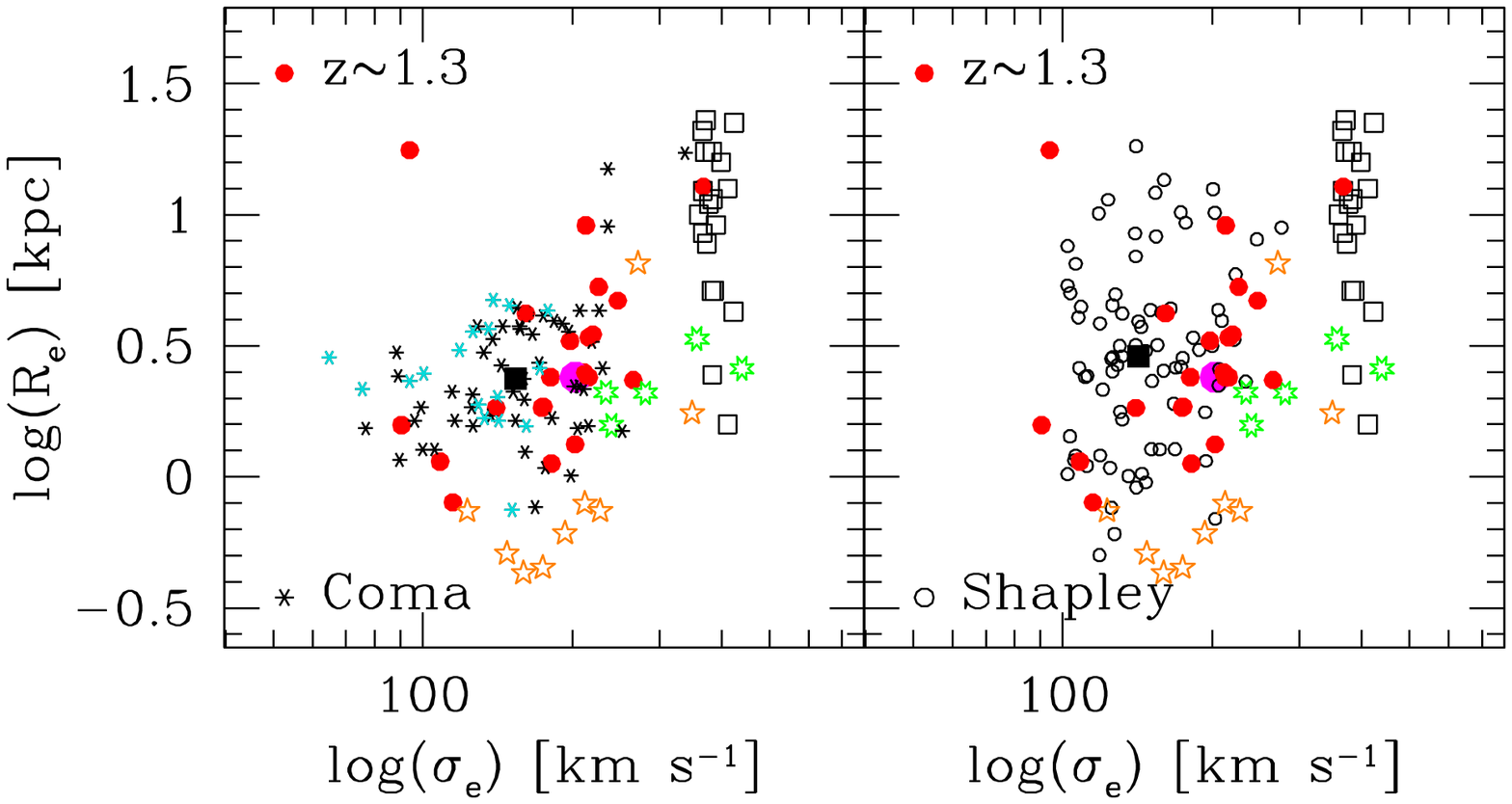}
	\includegraphics[width=8.5truecm]{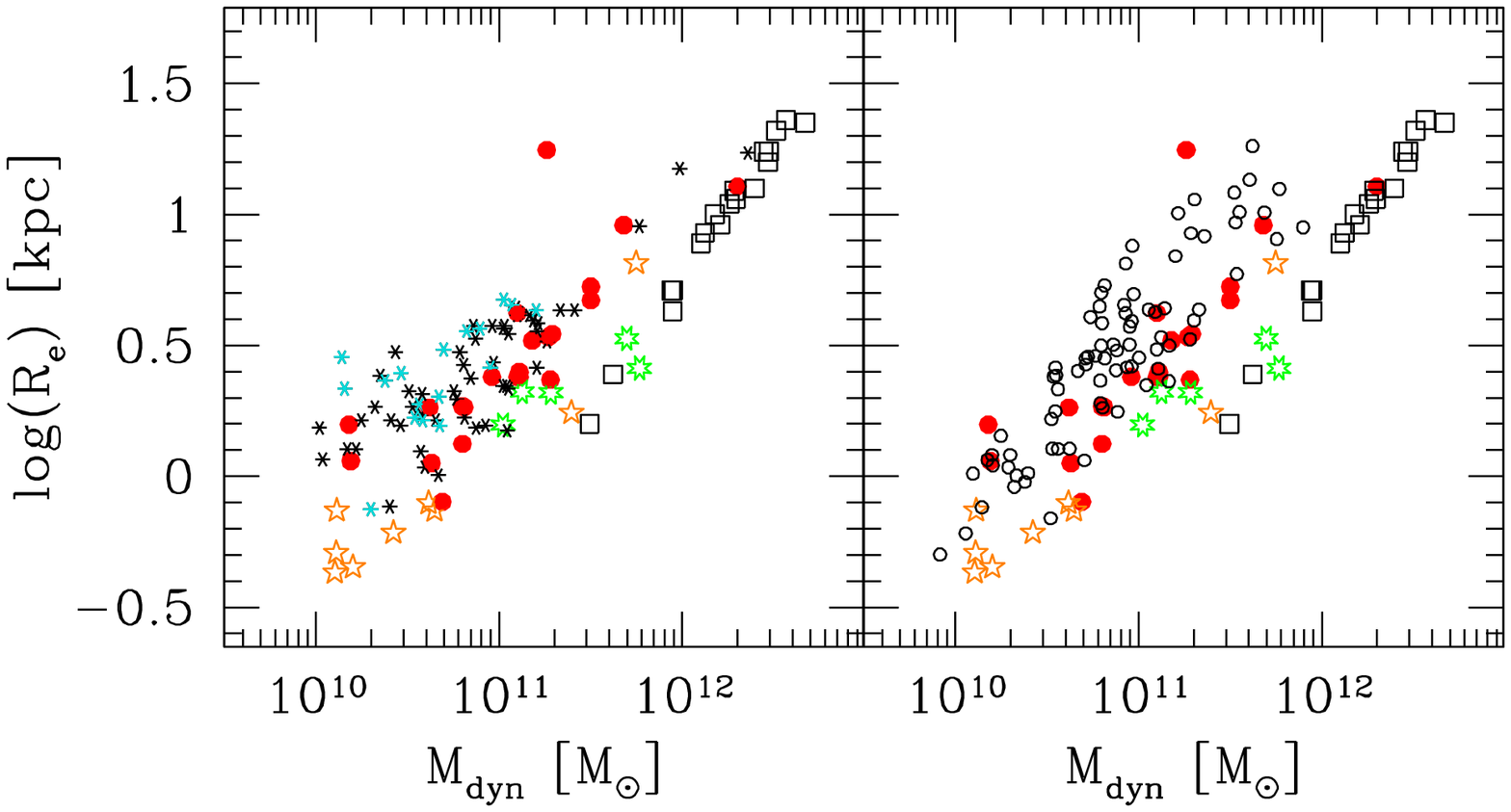}	
\caption{\label{fig:sigma_re}   
Upper panel - Effective radius versus velocity dispersion relation  
of high-redshift spheroids (red filled symbols) compared to the relation for the Coma 
cluster sample (skeletal symbols, left) and the Shapley sample (open circles, right).
{ Cyan skeletal symbols are galaxies younger than 7 Gyr in the Coma cluster (see \S 4.3).} 
The large magenta point is the median value of the high-redshift sample, the large
filled square is the median value of the local samples.
The open squares are the 23 local massive (0.3-5$\times$10$^{12}$ M$_\odot$)
early-type galaxies with $\sigma>350$ km/ s$^{-1}$ of \citet{bernardi06, bernardi08}; 
the orange stars are the extremely 
compact galaxies identified by \citet{damjanov13} at $z\sim0.3$;
the green stars are the ultramassive dense early-types at $z\sim1.4$ by
\citet{gargiulo16}.
Lower panel - Effective radius versus dynamical mass relation. 
Symbols are as in the upper panel.}
\end{figure}	
Trends of M$_{dyn}$/M$^*$ shown in figures \ref{fig:mdyn_mstar} and \ref{fig:msul_spd} 
suggest that
denser galaxies host a lower fraction of DM (affecting M$_{dyn}$) 
than lower density ones.\footnote{It is worth to note that the apertures
used for spectroscopic observations sample regions $>>$2R$_e$ for all the galaxies, 
with the exception of the largest galaxy of each cluster. 
Hence, the observed difference between dense and less dense galaxies cannot be
ascribed to size aperture effects.}
However, a different fraction of DM does not produce a mass-dependent evolution of 
M$_{dyn}$/L$_B$  (eq. \ref{eq:msul_diff}, Fig. \ref{fig:MsuL}).
To account for such an evolution,
the DM fraction should increase with time more in higher density (lower M$_{dyn}$) 
galaxies than 
in lower (higher M$_{dyn}$) ones.
However, if this were the case, spheroidal galaxies in the local universe
should not display the same M$_{dyn}$/M$^*$ trends of
high-redshift spheroids, { that is, higher density galaxies should not 
have M$_{dyn}$/M$^*$ values lower than low density ones}, 
contrary to what it is shown in Fig. \ref{fig:msul_spd}.
Moreover, a variation of the DM fraction can be realized through merging, 
a mechanism largely disfavoured in cluster environment given the large relative 
velocities of galaxies \citep[e.g.][]{harrison11,treu03}.
Actually, Figures \ref{fig:kor}, \ref{fig:sigma_re} and \ref{fig:re_z} do not show 
evidences in favour of merging for cluster spheroidal galaxies.

Fig. \ref{fig:kor} shows that passive aging would bring high-z spheroids onto the
local Kormendy relation. 
Hence, structural evolution of individual galaxies is not required and would move 
them away from the local relation
\citep{saracco14,jorgensen14}.
Figure \ref{fig:sigma_re} shows the effective radius as a function of 
velocity dispersion (upper panel) and of dynamical mass
(lower panel) for high-redshift cluster spheroids 
(red filled circles) and for the local cluster samples.
For comparison, the sample of local BCGs with very large velocity dispersion 
($\sigma>350$ km/ s$^{-1}$, open squares) of \cite{bernardi08}, the sample of extremely 
compact galaxies identified by \cite{damjanov13} (orange stars) at $z\sim0.3$
and the sample of ultramassive dense early-types at $z\sim1.4$ by \cite{gargiulo16}
(green starred symbols)
are also shown.
High-redshift cluster spheroids occupy a region of the $\sigma_e$-R$_e$ and
M$_{dyn}$-R$_e$ planes 
well populated by local cluster spheroids, and all of them have a local counterpart 
\citep[see also Fig.7 in][]{beifiori17}.
In case of merging, the region they occupy, especially at small radii and large velocity
dispersions and masses (where the most compact/densest galaxies lie), 
should not be populated by local spheroids, contrary to what happens.
This evidence contrasts with 
the scenario recently proposed by \cite{matharu19} in which compact cluster spheroids should 
disappear over time\footnote{
%This evidence contrasts with 
%the scenario recently proposed by \cite{matharu19} in which compact cluster spheroids should 
%disappear over time to justify the observed negligible difference between 
%the mean size of cluster and field spheroids 
%\citep[e.g.][but see also
%\cite{raichoor12,strazzullo13,andreon18}, for a different result]{weinmann09, maltby10, cappellari13, fernandez13, huertas13b, 
%cebrian14, kelkar15, sweet17,saracco17, mosleh18}.
The disappearance of compact galaxies would be due to merging with the BCG 
($\sim$40\% of them) and to tidal destruction (the remaining 60\%).
However, merging with BCG would not affect preferentially compact galaxies
while tidal forces would destroy preferentially less dense galaxies 
rather than compact ones.}
to justify the observed negligible difference between 
the mean size of cluster and field spheroids 
\citep[e.g.][but see also
\cite{raichoor12,strazzullo13,andreon18}, for a different result]{weinmann09, maltby10, cappellari13, fernandez13, huertas13b, 
cebrian14, kelkar15, sweet17,saracco17, mosleh18}.
{ It is worth to note that galaxies younger than 7 Gyr  in the Coma cluster
(cyan symbols, see \S 4.3) are offset with respect to high redshift galaxies, 
having preferentially larger R$_e$ and lower $\sigma$ at fixed mass.}

{ For sake of completeness, in Fig. \ref{fig:size_mass}
the effective radius and the velocity dispersion are also shown 
as a function of stellar mass, relations extensively studied (especially R$_e$-M$^*$)
in many previous papers of cluster galaxies 
\citep[e.g.][and references therein]{valentinuzzi10,poggianti13,saracco14,jorgensen14,saracco17,andreon17,beifiori17}.
The comments on Fig. \ref{fig:sigma_re}  apply also to  Fig. \ref{fig:size_mass}.}
\begin{figure}
	\includegraphics[width=8.5truecm]{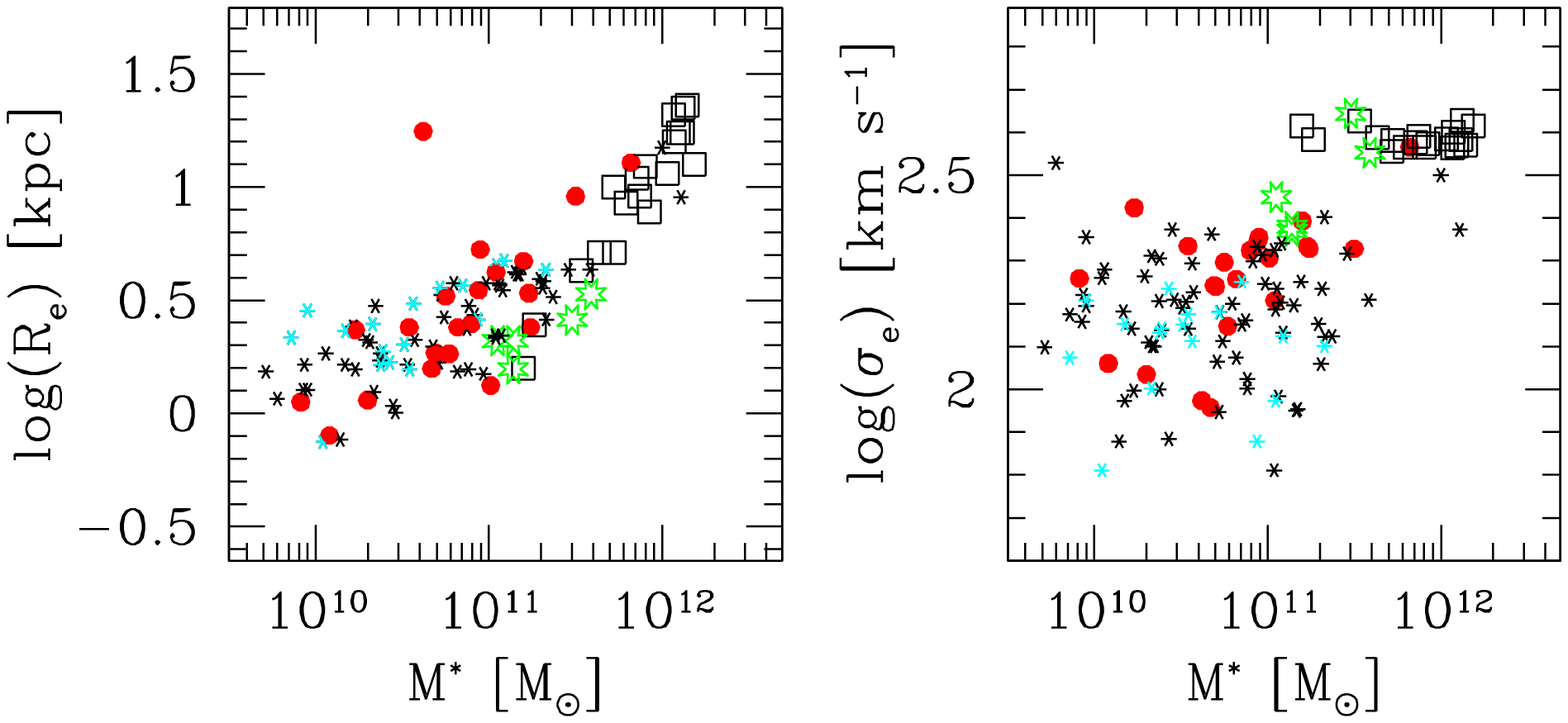}
\caption{\label{fig:size_mass}   
Effective radius (left) and velocity dispersion (right) versus stellar mass for  
high-redshift and low-redshift (Coma sample) spheroids.
Symbols are as in Fig. \ref{fig:sigma_re}.}
\end{figure}

To admit merging, densest galaxies should be replaced by those that join the
population at later times. 
However, these latter are expected, on average, less dense than those assembled earlier
because of the lower density of the universe and the lower star formation rate density.
Therefore, they will preferentially have larger radii and relatively lower velocity 
dispersions (and probably a higher DM fraction within R$_e$).
This agrees with the lack of high-redshift spheroids with low velocity dispersion 
($\sim$100-150 km s$^{-1}$) and large radii (R$_e$$>$1.5 kpc) 
with respect to local ones in Fig. \ref{fig:sigma_re}. 

In Fig. \ref{fig:re_z} the median effective radius R$_e$ and 
mass-normalized radius R$_e/m_*^a$ as a function
of redshift are shown for different samples of cluster 
spheroids and for two ranges of stellar masses.
Besides our data (red filled point) we plot data from the literature,
as reported in the caption of the figure and described in the appendix.
We considered 5$\times$10$^{10}$ M$_\odot$ as lower limit to keep the comparison
safe from the different mass limits of the samples.
The mass-normalized radius R$_e/m_*^a$
\citep[see e.g.][]{newman12, cimatti12,huertas13, delaye14, allen15,saracco17},
where $m_*=\mathcal{M}_*/10^{11}$ and $a=0.5$ 
is the slope of the size-mass relation  at $z\sim1.3$ 
\citep[][]{saracco17}, removes the correlation between
radius and mass.
Fig. \ref{fig:re_z} does not show evidence of evolution of the median R$_e$
since $z\sim1.5$ for masses larger than 10$^{11}$ M$_\odot$.
Considering the scatter of the data, an increase up to $\sim$40\% of the median 
value of R$_e$ could be present for galaxies with masses M$^*$$<$10$^{11}$ M$_\odot$ 
at redshift $z$$<$0.6 while the data do not show
significant variation over the redshift range 0.5$<z<$1.5 
\citep[see also][for negligible or null size evolution of cluster spheroids]{stott11,huertas13,jorgensen14,saracco14}.

Thus, the data do not support merging and, hence, a mass-dependent DM variation. 
On the contrary, the results shown by Fig. \ref{fig:sigma_re}, \ref{fig:size_mass} and
\ref{fig:re_z} support the scenario in which the spheroids joining the population at later 
times account for the missing fraction of evolution of M$_{dyn}$/L$_B$ of the population of spheroids 
at masses M$_{dyn}$$<$10$^{11}$ M$_\odot$.
Concluding, we find evidences that the observed evolution of the FP is mainly due
to the evolution of the stellar populations and that galaxies 
that join the population of spheroids at later times may contribute at masses 
$<$10$^{11}$ M$_\odot$ for about 15\% of the observed evolution.

\begin{figure}
\includegraphics[width=8.truecm]{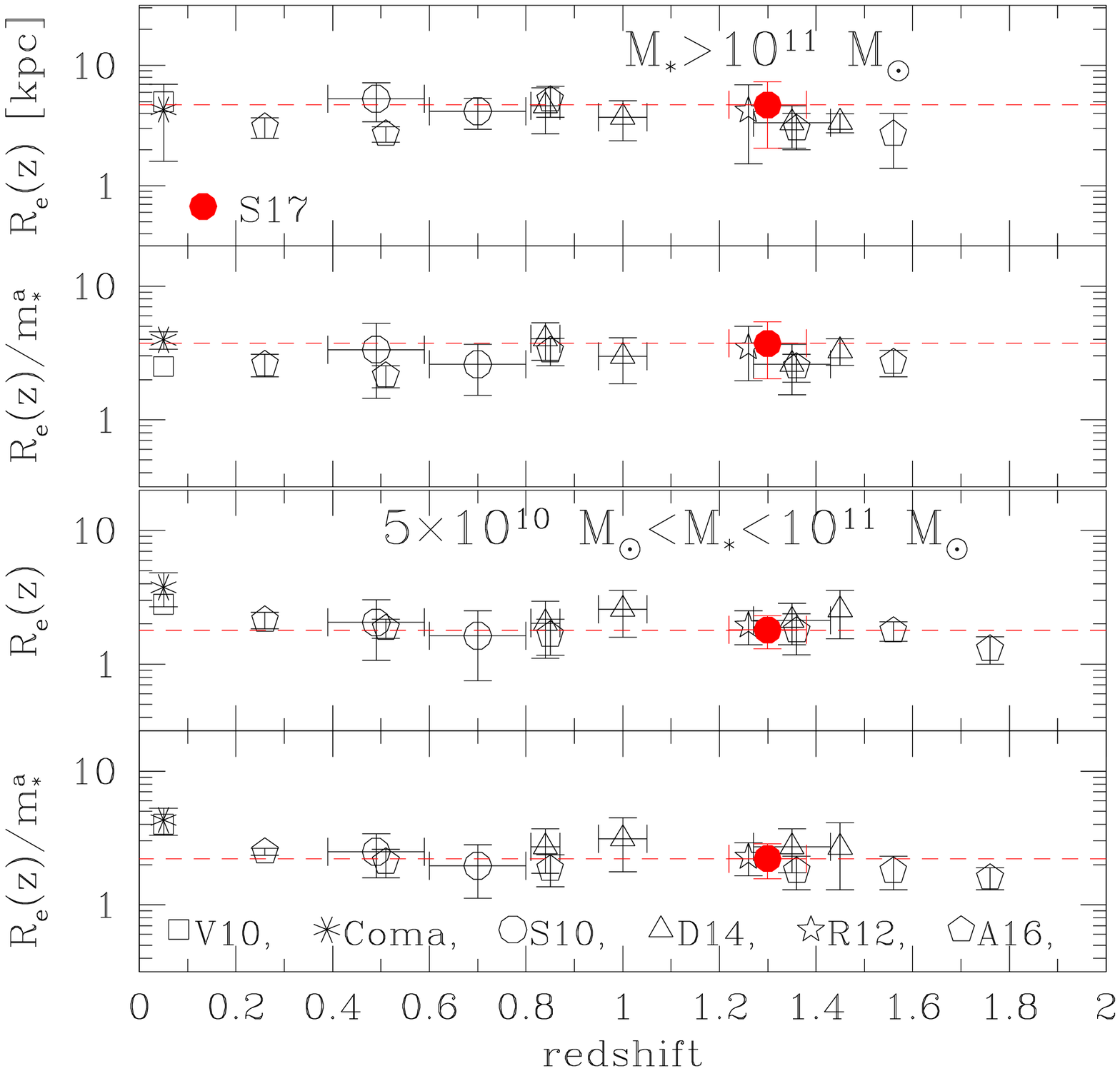}
 \caption{\label{fig:re_z} Median size-redshift relation.
 The median R$_e$ and R$_e/m_*^a$ of cluster early-type galaxies from
 different samples are shown as a function of redshift in the two
 ranges of stellar masses
 $\mathcal{M}_*>10^{11}$ M$_\odot$
 and
  $5\times10^{10}$ M$_\odot<\mathcal{M}_*<10^{11}$ M$_\odot$.
 The different symbols are as follow:
 filled red circle represents our photometric sample of (55) cluster
 spheroids at $z\simeq1.3$ from \citet[][S17]{saracco17};
 open triangles are the sample of cluster elliptical galaxies (313)
 at $0.84<z<1.45$ by \citet[][D14]{delaye14}; 
 open diamonds represent the red sequence early-types (158)
 in clusters at $0.2<z<1.8$ selected by 
\citet[][A14]{andreon16};
 the open star is the sample of cluster and group ellipticals (41)
 selected at $z\sim1.27$ by \citet[][R12]{raichoor12}; 
 the open circles are the cluster elliptical galaxies (188)
 at $0.4<z<0.8$ studied by \citet[][S10]{saglia10};
 the starred symbol is the Coma cluster sample,  
 the open square is the median size of local ellipticals
 from \citet[][V10]{valentinuzzi10} \citep[see also][]{poggianti13a}.
 The dashed red line represent the median value of our sample at $z\sim1.3$. }
\end{figure}

\section{Summary and Conclusions}
We obtained LBT-MODS spectroscopic observations for 22 galaxies in the field of the 
cluster XLSSJ0223-0436 at $z=1.22$.
We confirmed the membership for 12 spheroidal galaxies and determined the stellar velocity 
dispersion for 7 of them.
We combined these data with those in the literature for clusters 
RXJ0848+4453 at $z=1.27$ (8 galaxies) and XMMJ2235-2557 at $z=1.39$ (7 galaxies), 
to determine the FP of spheroidal galaxies in cluster at $z\sim1.3$.
The analysis related to the FP of high-redshift spheroids and the comparison with that of
local cluster spheroids have led to the following results.
\begin{itemize}
\item 
The FP of cluster spheroids at $z\sim1.3$ is offset
with respect to the FP of the local comparison samples, 
the Coma cluster sample and the Shapley sample, {and we detect
a rotation at $\sim3\sigma$}.
The rotation (the different slope) of the FP implies a differential 
evolution of the population of galaxies along cosmic time.
Once expressed as a relation between M$_{dyn}$/L$_B$ and the total mass 
M$_{dyn}$, high-redshift galaxies follow a steeper relation than found for local ones
implying a mass-dependent evolution of M$_{dyn}$/L$_B$ 
expressed by the relation 
$
\Delta log(M_{dyn}/L_B)=-0.34\Delta log(M_{dyn})+3.8.
$
\item
Assuming that $\Delta log(M_{dyn}/L_B)$=$\Delta log(M^*/L_B)$, the above relation 
implies a higher formation redshifts for high-mass galaxies ($z_f\ge$6.5 for 
log(M$_{dyn}$/M$_\odot$)$\ge$11.5) than for low-mass ones  ($z_f\le$2 for 
log(M$_{dyn}$/M$_\le$10).
Analogously, we find that galaxies with higher stellar mass density 
host stellar populations that formed earlier than those in lower mass density
galaxies.
\item
Assuming no variation of the IMF, at fixed stellar mass, lower density spheroids
have larger M$_{dyn}$/M$^*$ and vice versa, suggesting a different DM fraction; 
the mean M$_{dyn}$/M$^*$ of low-density spheroids tends to increase
for decreasing stellar masses while tends to the constant value of high-density
spheroids at large stellar masses.
As a consequence of this systematic variation of M$_{dyn}$/M$^*$,
%Assuming no variation of the IMF, 
the observed $\Delta log(M_{dyn}/L_B)$ 
for M$_{dyn}$$>10^{11}$ M$_\odot$ galaxies is accounted for by the evolution of
the stellar populations ($\Delta$\rm{log}(M$^*/L_B)$),
%variation of M$^*$/L$_B$, 
while for spheroids with masses M$_{dyn}$$<10^{11}$ M$_\odot$ the evolution of M$^*$/L$_B$ account 
for $\sim$85\% of the observed $\Delta$\rm{log}(M$_{dyn}/L_B)$.
\item
We find no evidence in favour of merging for cluster spheroids,
of structural evolution of individual galaxies that may produce a differential
DM variation to fill in the gap between the evolution of M$_{dyn}/L_B$ and M$^*$/L$_B$
for lower mass galaxies.
On the contrary, we find evidences that spheroids later added to the population
may affect the observed $\Delta$\rm{log}(M$_{dyn}/L_B)$ at masses  $<10^{11}$ M$_\odot$.
\end{itemize}
The above findings show that the evolution of the FP of cluster spheroids is consistent 
with the mass-dependent and mass density-dependent evolution of their stellar populations 
superimposed to a contribution from spheroids that join the population at later times.

\section*{Acknowledgements}
We thank the referee for the many useful comments and suggestions that improved the presentation of the results.
This work is based on observations carried out at the Large Binocular Telescope
(LBT) under program ID 2015\_2016\_28. 
The LBT is an international collaboration among institutions in the US, Italy 
and Germany. 
We acknowledge the support from the LBT-Italian Coordination Facility
for the execution of the observations, the data distribution and for 
support in data reduction. 
P.S. would like to thank M. Singer for the useful suggestions.

%%%%%%%%%%%%%%%%%%%%%%%%%%%%%%%%%%%%%%%%%%%%%%%%%%



\bibliographystyle{mnras}
\bibliography{paper_fp2_r2} % if your bibtex file is called example.bib

%%%%%%%%%%%%%%%%%%%%%%%%%%%%%%%%%%%%%%%%%%%%%%%%%%

%%%%%%%%%%%%%%%%% APPENDICES %%%%%%%%%%%%%%%%%%%%%

\appendix

\section{High-redshift cluster sample}
The high-redshift sample is composed of 22 cluster spheroidal galaxies 
in the redshift range 1.2$<$$z$$<$1.4,
7 of which belonging to cluster XLSSJ0223 at $z=1.22$
(this work, Tab. \ref{tab:sigmas}), 8 spheroids to cluster RXJ0848 (LinxW) at $z=1.27$
\citep{jorgensen14} and 7 spheroids to cluster XMMJ2235 at $z=1.39$ \citep{beifiori17}.
The parameters of these galaxies are summarized in Tab. \ref{tab:rxj0848}
and Tab. \ref{tab:xmm2235}, respectively.
{ Photometry and structural parameters of galaxies belonging to the 
three clusters considered were derived and extensively 
studied in previous papers, by \cite{saracco14} 
for cluster RXJ0848, \cite{ciocca17} for cluster XMMJ2235 and 
\cite{saracco17} for cluster XLSS0223. 
For RXJ0848, we used the measurements from \cite{jorgensen14},
given the good agreement between our \citep{saracco14} and their measurements
\cite[see][]{jorgensen14}.
In \cite{ciocca17} extensive simulations are presented in Appendix A1
assessing the robustness of the photometry and of structural parameters 
derived for the galaxies in cluster XMMJ2235.
In the same Appendix, simulations dedicated to estimate the errors 
on the above parameters are also described, while Appendix C is dedicated to a 
comparison with \cite{chan16} measurements showing a good agreement with their published 
measurements \citep[see Fig. C2 of][]{ciocca17}.}

\begin{table*}
\begin{minipage}[t]{1\textwidth}
\caption{Parameters of spheroidal galaxies in cluster RDCS0848}
\label{tab:rxj0848}
\centerline{
\begin{tabular}{rccccccccccc}
\hline
\hline
         ID   & ID$^c$ &    z	  &   Re	 &   log<I$_e$>	 &  $\sigma_e$    & R$_{e,Dev}$  & log<I$_e$>$_{Dev}$ &  $\sigma_{e,Dev}$&log$\mathcal{M}_{dyn}$&log$\mathcal{M}_*$\\ 
	      &        &  	  &	[kpc]  &[L$_\odot$ pc$^{-2}$]&[km s$^{-1}$]&[kpc]& [L$_\odot$ pc$^{-2}$]&[km s$^{-1}$]&[M$_\odot$] &[M$_\odot$]	       \\
\hline
         240  &        &  1.261   &  2.3$\pm$0.4 & 3.37$\pm$0.05 & 265$\pm$28 & 1.3    &     3.76    &      275 &	     11.3   & 	 10.2 \\
	1264  &        &  1.273   &  1.6$\pm$0.2 & 3.26$\pm$0.05 &  91$\pm$15 & 1.6    &     3.24    &        90&	     10.3   &	 10.7 \\
        1748  & 606    &  1.270   &  1.8$\pm$0.3 & 3.43$\pm$0.05 & 174$\pm$26 & 2.3    &     3.30    &      172 &	     10.8   &	 10.7 \\
	2111  &	       &  1.279   & 17.6$\pm$2.6 & 1.84$\pm$0.04 &  94$\pm$26 & 3.6    &     2.86    &      103 &            11.3   &       10.6\\
        2735  &        &  1.267   &  1.1$\pm$0.2 & 3.78$\pm$0.05 & 182$\pm$25 & 1.1    &     3.82    &      182 &	     10.6   &	  9.9 \\
        2943  & 1160   &  1.268   &  1.8$\pm$0.3 & 3.31$\pm$0.05 & 140$\pm$25 & 2.2    &     3.19    &      139 &	     10.6   &	 10.8 \\
        2989  &        &  1.264   &  1.1$\pm$0.2 & 3.82$\pm$0.05 & 108$\pm$22 & 1.3    &     3.76    &      108 &	     10.2   & 	 10.3 \\
        3090  & 3      &  1.278   &  1.3$\pm$0.2 & 3.95$\pm$0.05 & 203$\pm$19 & 1.8    &     3.78    &      199 &	     10.8   & 	 11.0 \\
\hline
%         807  &        &  1.270   &  1.6$\pm$0.2 & 3.07$\pm$0.05 & 167$\pm$44 & 2.6    &     2.79    &      162 &	     10.7   &	  9.4 \\
%        1362  &        &  1.273   &  6.4$\pm$1.0 & 2.58$\pm$0.06 & 219$\pm$25 &11.6    &     2.19    &      211 &	     11.6   & 	 10.1 \\
%        1763  & 4      &  1.275   &  6.9$\pm$1.0 & 2.84$\pm$0.06 & 187$\pm$18 &11.0    &     2.52    &      182 &	     11.5   & 	 11.3 \\
%        1888  & 2      &  1.277   &  6.1$\pm$0.9 & 2.86$\pm$0.06 & 149$\pm$15 & 3.9    &     3.15    &      153 &	     11.2   & 	 11.2 \\
%        2063  & 590    &  1.267   &  2.3$\pm$0.3 & 3.21$\pm$0.05 & 215$\pm$23 & 3.6    &     2.91    &      209 &	     11.1   & 	 10.8 \\
%\hline
 \end{tabular}
}
{$^c$ ID from \cite{saracco14}. \\
All the measurements are taken from \cite{jorgensen14}.
}
\end{minipage}
\end{table*}

\begin{table*}
\begin{minipage}[t]{1\textwidth}
\caption{Parameters of spheroidal galaxies in cluster XMMJ2235}
\label{tab:xmm2235}
\centerline{
\begin{tabular}{rccccccccccc}
\hline
\hline
         ID& ID$^a_C$&  z$^b$  &   n &    R$_e$	   & log<I$_e$>    &  $\sigma^b_e$  &  R$_{e,Dev}$&     log<I$_e$>$_{Dev}$&    $\sigma^b_{e,Dev}$ &log$\mathcal{M}_{dyn}$&log$\mathcal{M}_*$\\
	   &  	&     &     &	[kpc]&[L$_\odot$ pc$^{-2}$]&  [km s$^{-1}$]&   [kpc]&	   [L$_\odot$ pc$^{-2}$]&      [km s$^{-1}$]  &[M$_\odot$] &[M$_\odot$]	 \\
\hline
        352& 1782  & 1.375 &  3.6$\pm0.2$&   3.4$\pm$0.5 &  3.32$\pm$0.02 &  216$\pm$53  &  3.5$\pm$0.5&    3.28$\pm$0.02&    216$\pm$53&   11.27&      11.23\\
        407& 1790  & 1.385 &  4.4$\pm0.3$&   2.4$\pm$0.3 &  3.64$\pm$0.06 &  213$\pm$50  &  2.2$\pm$0.3&    3.68$\pm$0.06&    214$\pm$50&   11.10&       11.24\\
        220& 1284  & 1.390 &  4.3$\pm0.2$&   2.4$\pm$0.3 &  3.44$\pm$0.02 &  181$\pm$50  &  2.3$\pm$0.3&    3.48$\pm$0.02&    181$\pm$50&   10.96&       10.82\\
         36& 2054  & 1.392 &  4.4$\pm0.3$&   4.2$\pm$0.6 &  3.12$\pm$ 0.2 &  161$\pm$31  &  3.7$\pm$0.6&    3.19$\pm$0.2 &    162$\pm$31&   11.10&       11.04\\
        170& 1740  & 1.395 &  3.3$\pm0.2$&  12.8$\pm$2.0 &  2.62$\pm$ 0.1 &  368$\pm$44  & 19.5$\pm$1.9&    2.37$\pm$0.1 &    357$\pm$44&   12.30&       11.82\\
        433& 2147  & 1.395 &  5.3$\pm0.3$&   5.3$\pm$0.8 &  2.82$\pm$ 0.1 &  226$\pm$47  &  3.4$\pm$0.8&    3.12$\pm$0.1 &    232$\pm$47&   11.51&      10.95\\
        637&  --   & 1.397 &  3.2$\pm0.2$&   1.9$\pm$0.1 &  3.76$\pm$0.04 &  172$\pm$66  &  2.3$\pm$0.1&    3.62$\pm$0.04&    170$\pm$66&   10.82&      10.69\\
\hline
\end{tabular}
}
{$^a$ ID from \cite{ciocca17}. 
Structural parameters are those of \cite{ciocca17}.\\
$^b$ Redshift and velocity dispersion measurements are taken from \cite{beifiori17}.
Velocity dispersion values are scaled to the proper effective radii according to eq. \ref{eq:sigscale}.
}
\end{minipage}
\end{table*}

\section{Best fitting linear relations in case of correlated errors}
{ Given the correlated errors between variables in some of the relations considered,
%However, since <I$_e$> depends on R$_e$, the errors on these two quantities  
%are correlated among them.
%To properly take into account for this, 
we repeated the fitting using the code \texttt{linmix\_err} that applies the Bayesian method 
proposed by \cite{kelly07} to linear regression between data with correlated measurement errors.
Table \ref{tab:kelly} summarizes the relations thus obtained.
In all the case considered, correlated errors do not affect significantly the resulting
relation.}

\begin{table*}
\caption{\label{tab:kelly} Relations obtained adopting the Bayesian approach by 
\citet{kelly07} to account for correlated errors among variables.}
\centerline{
\begin{tabular}{ll}
%\hline
\hline
Relation& Reference\\
\hline
log<I$_e$>=(-1.48$\pm$0.12)log(R$_e$)+(3.90$\pm0.08$) & $z\simeq1.3$, spectroscopic sample (\S 4.3) \\
log<I$_e$>=(-1.42$\pm$0.09)log(R$_e$)+(3.81$\pm0.05$) &$z\simeq1.3$, photometric sample (\S 4.3) \\
log<I$_e$>=(-1.26$\pm$0.06)log(R$_e$)+(2.99$\pm0.03$) &$z=0$ sample  (\S 4.3) \\
\hline
log($M_{dyn}/L_B)$=(0.57$\pm$0.08)log($M_{dyn}$)-(6.2$\pm$0.9)& $z\simeq1.3$ sample (eq. \ref{eq:msul_z})\\
%log($M_{dyn}/L_B)$=(0.6$\pm$0.1)log($M_{dyn}/\lagle M_{dyn}\rangle$)+0.3$\pm$0.06& $z\simeq1.3$ sample\\
log($M_{dyn}/L_B)$=(0.32$\pm$0.03)log($M_{dyn}$)-(2.8$\pm$0.2) &$z=0$ sample (eq. \ref{eq:msul_0})\\
$\Delta${log}($M_{dyn}/L_B)$=(0.25$\pm$0.09)log($M_{dyn}$)+(3.4$\pm$0.9) & (eq. \ref{eq:msul_diff})\\
log($age(z_f)$)=(-0.3$\pm$0.1)log($M_{dyn}$)+(4.0$\pm$0.9)& (eq. \ref{eq:age_mdyn})\\
\hline
log($M_{dyn}/L_B)$=(2.1$\pm$0.6)log($\sigma_e$)-(4.7$\pm$1.3)& $z\simeq1.3$ sample (Fig. \ref{fig:MsuL})\\
%log($M_{dyn}/L_B)$=(0.6$\pm$0.1)log($M_{dyn}/\lagle M_{dyn}\rangle$)+0.3$\pm$0.06& $z\simeq1.3$ sample\\
log($M_{dyn}/L_B)$=(1.1$\pm$0.5)log($\sigma_e$)-(1.6$\pm$0.9) &$z=0$ sample (Fig. \ref{fig:MsuL})\\
\hline
\end{tabular}
}
\end{table*}

\section{Best fitting FP relations}
In this appendix we show the best-fitting FP relation for the local comparison 
samples (Fig. \ref{fig:fit_fpl}) and  the high-redshift sample (Fig. \ref{fig:fit_fph}). 
The best fitting coefficients $\alpha$, $\beta$ and $\gamma$ of eq. \ref{eq:fp}
have been determined using the code \texttt{lts\_planefit} 
\citep{cappellari13}. { Note that error bars in the figures should be tilted ellipses
rather than crosses due to the correlated errors among variables.
}

In Fig. \ref{fig:fp_sim}, the best-fitting FP relation obtained for one of the 100 simulated 
samples of spheroids at z$\sim$1.3 is shown as example. 
{ Simulated samples consist in varying the values of R$_e$ and L$_B$ of each galaxy by adding a shift 
$\delta$R$_e$ and $\delta$L$_B$ randomly chosen from gaussian distributions with sigmas 
dR$_e$ and dL$_B$ respectively, where dR$_e$ and dL$_B$
are the errors on the two parameters.
Then, for each galaxy, a new value of I$_e$ is derived from these perturbed values.}
Structural parameters considered are those obtained with r$^{1/4}$ profile.

\begin{figure*}
	\includegraphics[width=8.truecm]{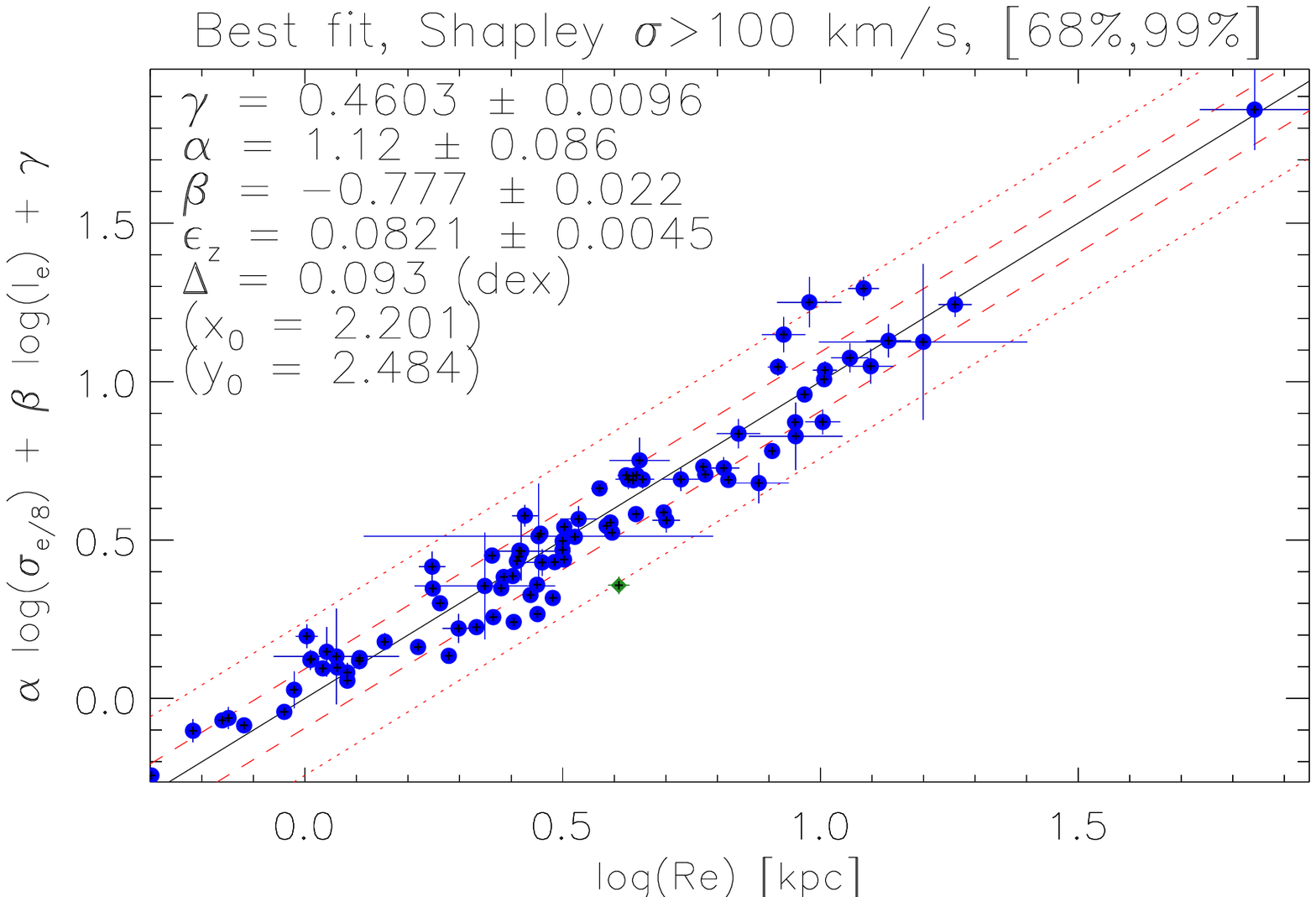}
	\includegraphics[width=8.truecm]{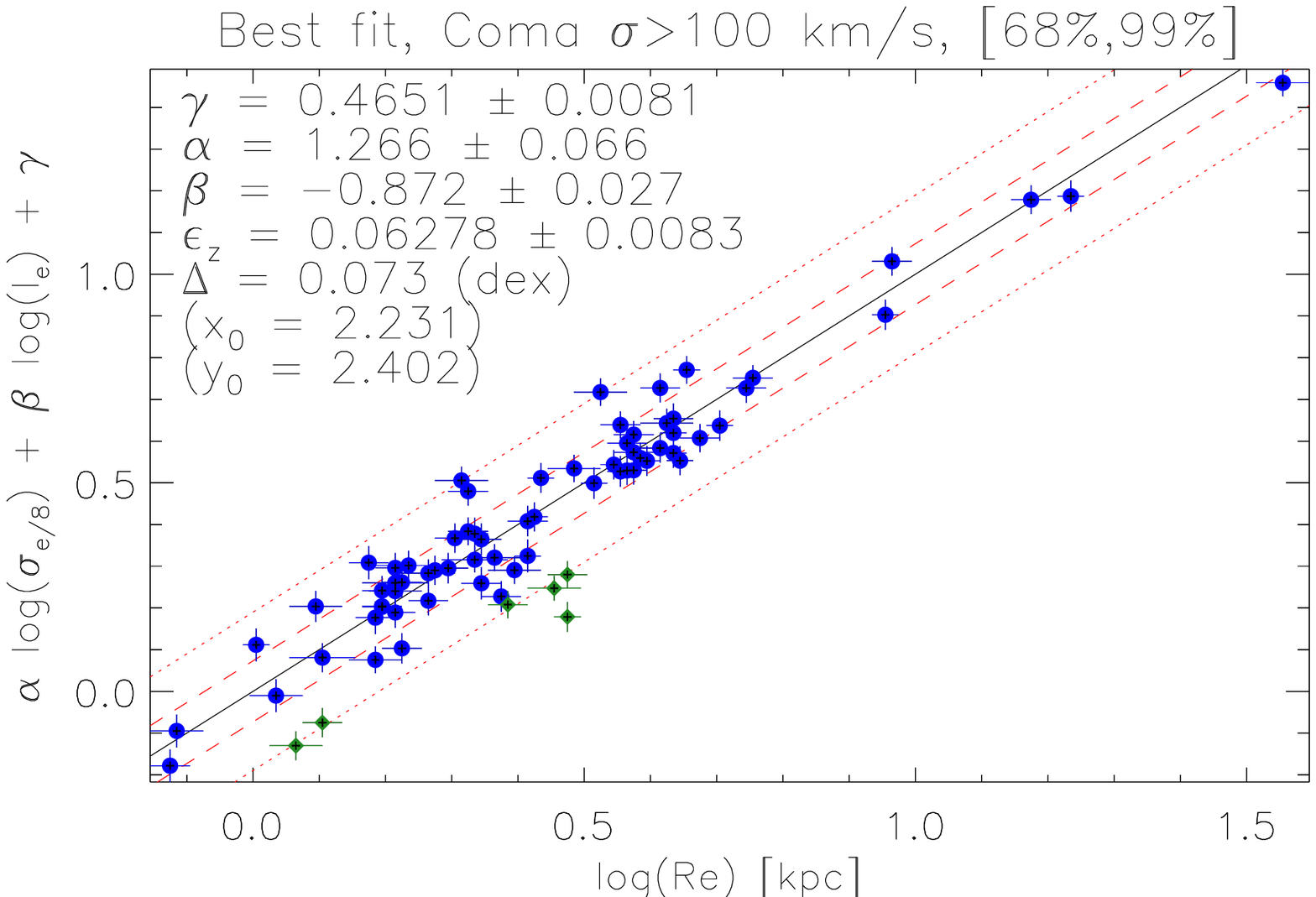}
\caption{\label{fig:fit_fpl} Fundamental Plane of local cluster
spheroidal galaxies. The edge-on view of 
the observed FP and best fitting relation 
$\rm{log}(Re)=\alpha\ log(\sigma_{e/8})+\beta\ log(I_e)+\gamma$ (eq. \ref{eq:fp})
are shown for the two local samples considered,
the Shapley sample (top panel) and the Coma sample (lower panel).
The best fitting values $\alpha\,\beta\,\gamma$ are shown at the top left
of each panel together with the corresponding observed scatter $\Delta$,
{ the intrinsic scatter $\epsilon_z$, and the pivot
 values $x_0$ and $y_0$ \citep[see][for details.]{cappellari13}}.
The dashed and the dotted (red) lines mark the regions enclosing 68\% (1$\sigma$) 
and 99\% (2.6$\sigma$) of the data, respectively. 
Green data points are the outliers excluded from the fit by the procedure 
\texttt{lts\_planefit}.
 }
\end{figure*}	

\begin{figure}
	\includegraphics[width=8.truecm]{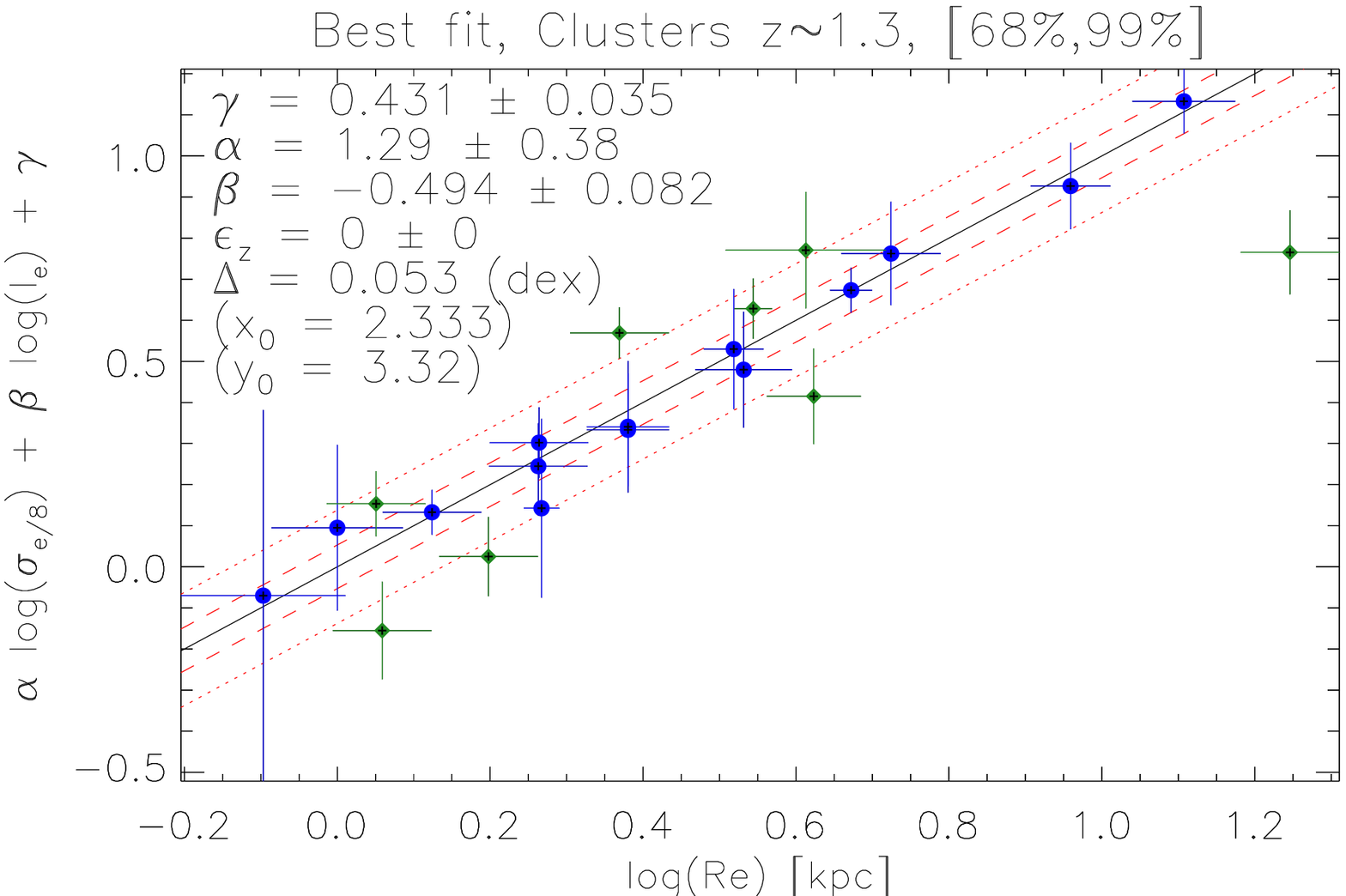}
\caption{\label{fig:fit_fph} Fundamental Plane of cluster spheroidal
galaxies at z$\sim$1.3. Same as Fig. \ref{fig:fit_fpl} but 
for the high-redshift sample composed of the 22 spheroids in cluster 
at $z\sim1.3$. Structural parameters are derived from Sersi\'c
profile fitting.
 }
\end{figure}	

\begin{figure}
	\includegraphics[width=8.truecm]{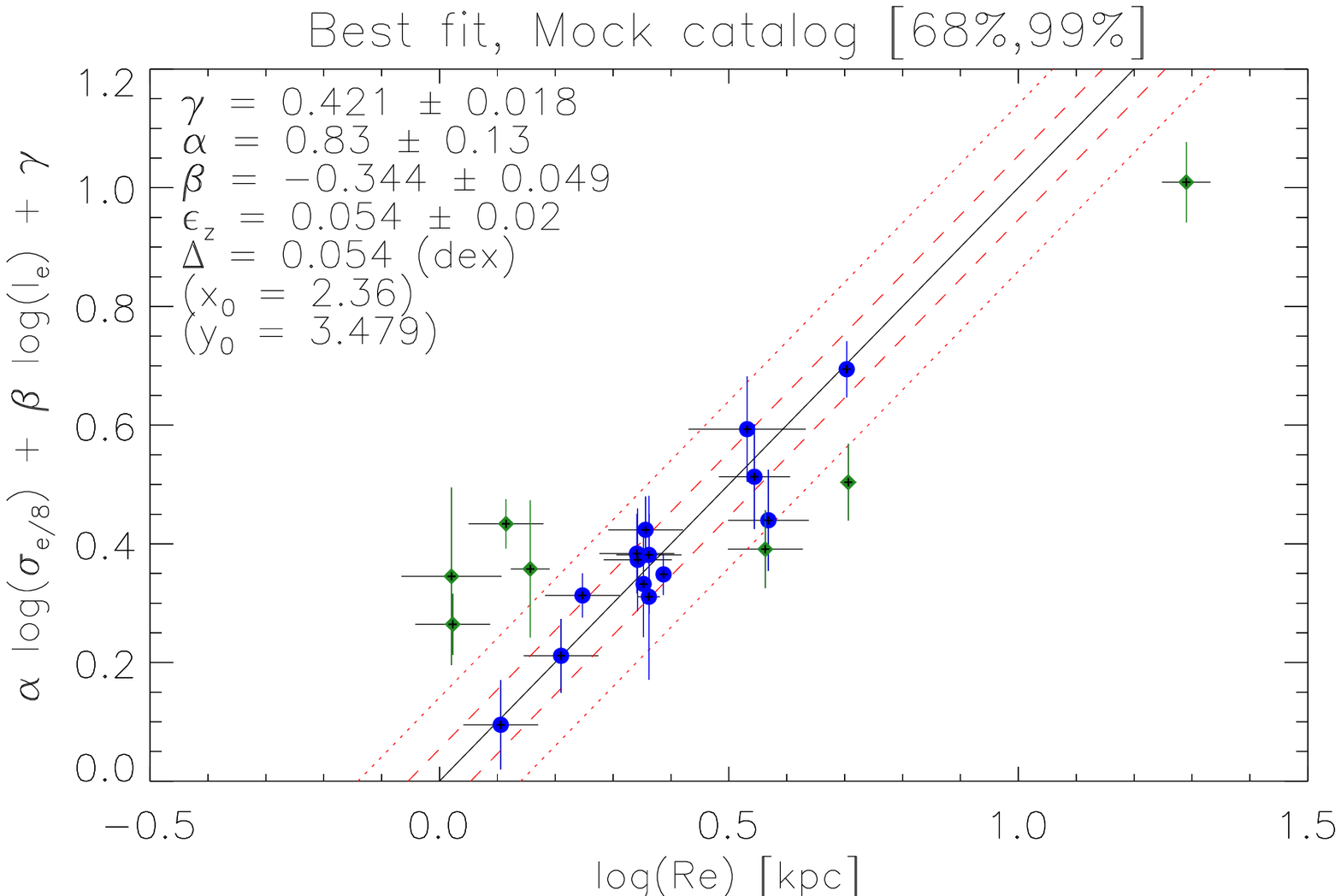}
\caption{\label{fig:fp_sim} Best-fitting Fundamental Plane 
obtained for one of the 100 simulated samples of spheroids at 
z$\sim$1.3. 
Structural parameters are those derived in case of r$^{1/4}$
profile.
 }
\end{figure}

\section{Size-redshift relation data}
In Fig. \ref{fig:re_z}, besides the sample of 55 spheroids in the three clusters
at $z\sim1.3$  \cite[red filled point;][S17]{saracco17}, the following data samples
were considered.
The early-type galaxies 
at $0.4<z<0.8$ of the sample of \citet[][S10]{saglia10}  
based on the ESO Distant Cluster Survey (EDisCS) dataset.
{Stellar masses were derived using a 'diet' Salpeter IMF
\citep[see][]{saglia10},
resulting similar to a Chabrier IMF.
Thus, we did not apply any correction to their masses.}
On the contrary, according to the recipe of \cite{longhetti09} 
\citep[see also][]{saracco14,tamburri14},
we scaled by a factor 1.7 the stellar masses of the early-type galaxies 
in groups and in cluster at $z\sim1.27$ selected by \citet[][R12]{raichoor11, raichoor12},
whose masses have been derived using a Salpeter IMF.
The same correction has been applied to the sample of red sequence galaxies 
selected by \citet[][A16]{andreon16} in 
clusters at 0.2$<$$z$$<$1.8 \citep[see e.g.][for the differences due to different selection
criteria]{tortorelli18}.   
We also plot the cluster early-type galaxies belonging to the 
sample of \citet[][D14]{delaye14} in the redshift range $0.8<z<1.45$ that have 
stellar masses based on Chabrier IMF, as for our data. 
All the size measurements are based on HST-ACS observations with the exception of
measurements at $z>1$ in the sample of \cite{andreon16} based on WFC3 images.
For the local Universe we report the median value of the Coma sample considered
in this work \citep[][starred symbol]{jorgensen95}
and the median value of cluster ellipticals from 
\cite{valentinuzzi10} \citep[see also][ open square]{poggianti13a}.

%%%%%%%%%%%%%%%%%%%%%%%%%%%%%%%%%%%%%%%%%%%%%%%%%%
% Don't change these lines
%\bsp	% typesetting comment
\label{lastpage}
\end{document}